\def\allblackversion{1}
\def\singlesupplement{1}

\documentclass[aps,prl,reprint,superscriptaddress,amsmath,amssymb,floatfix,footinbib]{revtex4-2}

\usepackage{graphicx}
\usepackage{tikz}
\usepackage{placeins}
\usepackage{float}
\usepackage{xcolor}
\ifdefined\allblackversion
  \usepackage[colorlinks=true,citecolor=blue,linkcolor=blue,urlcolor=blue]{hyperref}
\else
  \usepackage[colorlinks=true,citecolor=blue,linkcolor=blue,urlcolor=blue]{hyperref}
\fi
\hypersetup{
  pdftitle={A Single Spin Switches the Steady-State Phase of an Open Quantum System},
  pdfauthor={Jianwen Jie}
}

\newenvironment{revision}{\color{black}}{}
\colorlet{introblue}{blue}
\definecolor{introgreen}{RGB}{0,128,0}
\definecolor{introred}{RGB}{220,0,0}
\newcommand{\sentblack}[1]{{\color{black}#1}}
\newcommand{\smref}[2][app:mean-field-symmetry-limits]{%
  \hyperref[#1]{{\color{blue}#2}}}
\let\originalref\ref
\renewcommand{\ref}[1]{{\color{blue}\originalref{#1}}}
\let\originaleqref\eqref
\renewcommand{\eqref}[1]{{\color{blue}\originaleqref{#1}}}
\let\originalpageref\pageref
\renewcommand{\pageref}[1]{{\color{blue}\originalpageref{#1}}}
\ifdefined\allblackversion
  
  \newcommand{\sentblue}[1]{{\color{black}#1}}
  \newcommand{\sentgreen}[1]{{\color{black}#1}}
  \newcommand{\sentred}[1]{{\color{black}#1}}
\else
  
  \newcommand{\sentblue}[1]{{\color{introblue}#1}}
  \newcommand{\sentgreen}[1]{{\color{introgreen}#1}}
  \newcommand{\sentred}[1]{{\color{introred}#1}}
\fi
\newcommand{\vthirteen}[1]{#1}

\begin{document}
\title{A Single Spin Switches the Steady-State Phase of an Open Quantum System}
\author{Jianwen Jie}
\email{Jianwen.Jie1990@gmail.com}
\affiliation{College of Engineering Physics, Shenzhen Technology University, Shenzhen 518118, China}
\date{August 28, 2026}

\begin{abstract}
\sentblack{Changing a many-body system by one constituent is normally expected to produce only a
vanishing correction to an intensive observable.}
\sentblue{Here we show that, above a critical dissipation imbalance, adding or removing one spin at fixed intensive controls switches the steady state of a dissipative collective spin between a phase-averaged ring and a south-polar fixed point.}
\sentgreen{Unlike previous one-spin sensing through dark-state interference, parity here determines whether the Dicke ladder samples an interior zero of the nonlinear jump amplitude.}
\sentred{The integer-spin ladder samples this zero, making Dicke states with $m<0$
transient, whereas the half-integer ladder misses it and remains a single recurrent class.}
\sentblack{Exact finite-size steady states show that exponential competition between stationary weights amplifies this microscopic connectivity difference, yielding a first-order dissipative phase transition confined to the odd-\(N\) sequence.}
\sentblue{Multisector one-spin loading and state-unresolved removal
confirm the switch, while the exponentially increasing switching time
reflects the metastable isolation of the competing macroscopic basins.}
\sentgreen{Our results establish representation-lattice sampling as a route to phase control, suggesting parity-based detection of single-spin addition or removal.}
\end{abstract}

\maketitle
\enlargethispage{\baselineskip}

\begin{figure}[t]
\centering
\includegraphics[width=\columnwidth,trim=0 7.52bp 0 0,clip]{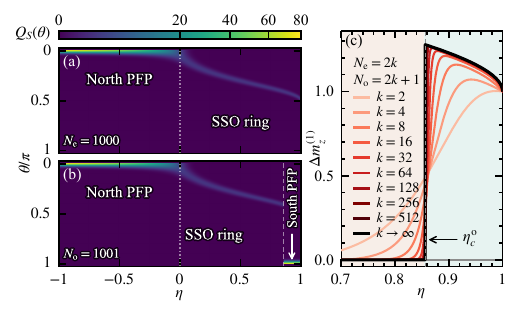}
\caption{\label{fig:parity-stationary-overview}
\sentblack{Finite-size parity comparison.}
\sentblue{(a,b) Normalized Husimi-$Q$ functions
$Q_S(\theta)$ for the adjacent sizes $N_{\mathrm{e}}=1000$ and $N_{\mathrm{o}}=1001$.}
\sentgreen{The dotted and dashed vertical lines mark $\eta=0$ and
$\eta_c^{\mathrm{o}}$, respectively.}
\sentred{(c) Adjacent-size magnetization contrast \(\Delta m_{z,k}^{(1)}=|m_{z,k}^{\mathrm{o}}-m_{z,k}^{\mathrm{e}}|\) for \(N_{\mathrm{e}}=2k\) and \(N_{\mathrm{o}}=2k+1\). The black curve is the thermodynamic-limit result.}}
\end{figure}

\begin{figure}[t]
\centering
\includegraphics[width=\columnwidth]{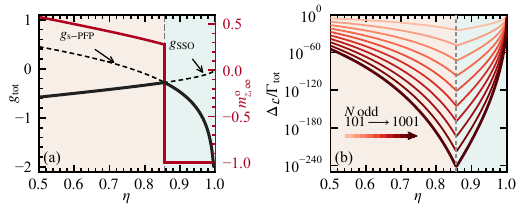}
\caption{\label{fig:boundary-condensation}
\sentblack{First-order DPT of the odd-\(N\) sequence.}
\sentblue{(a) Branch-resolved rate potentials $g_{\rm SSO}$ and
$g_{\rm s\text{-}PFP}$ (black dashed), their lower envelope
$g_{\rm tot}$ (black solid, left axis), and the conjugate order parameter---the
odd-sequence steady-state polarization $m_{z,\infty}^{\mathrm{o}}$ (red, right
axis).}
\sentgreen{(b) Liouvillian gap $\Delta_{\mathcal L}/\Gamma_{\rm tot}$
for odd-\(N=101,201,\ldots,1001\). Darker curves denote larger \(N\), and the
vertical dashed line marks $\eta_c^{\mathrm{o}}$.}}
\end{figure}

\paragraph{Introduction.---}\phantomsection\label{sec:introduction}
\begin{revision}
\setlength{\parskip}{0pt} 
\sentblack{Many-body systems are generally expected to approach a common
large-\(N\) limit, in which changing \(N\) by one modifies
intensive observables only by \(O(1/N)\)
\cite{Fisher1964,LiebLebowitz1972}.}
\sentblue{In open quantum systems, uniqueness within every finite irreducible
spin representation \cite{Evans1977,Frigerio1978,Yoshida2024}
further suggests that all size sequences approach the same
steady-state phase unless spontaneous symmetry breaking emerges
\cite{Minganti2018}.} \sentgreen{Yet macroscopic steady-state selection can depend not only on the local
stability of competing attractors \cite{AssafMeerson2017}, but also on the global connectivity of the
finite-size dynamics \cite{BaumgartnerNarnhofer2008}---a directed recurrent structure that can preserve the
parity of \(N\) in the thermodynamic limit.}
\sentred{This raises a fundamental question: can changing a single
constituent, while keeping all size-independent intensive controls fixed,
switch the system between macroscopically distinct steady-state phases?}


\sentgreen{Parity has qualitative---and sometimes macroscopic---effects in
closed many-body systems
\cite{LiebSchultzMattis1961,Haldane1983,Loss1992,Garg1993,
WernsdorferSessoli1999,MatveevLarkin1997,Maric2020Odd,Maric2022SciPost,
Shaikh2022Parity,Torre2022Odd}, including an exact symmetry-broken ground-state
doublet for odd-\(N\) with long-range order and conserved spin parity
\cite{Caleca2025}.}
\sentblack{Reservoir-engineered spin squeezing already exhibits a macroscopic
even--odd steady-state contrast caused by destructive dark-state interference
\cite{AgarwalPuri1990,Groszkowski2022}, which was proposed for detecting the
addition or removal of a single spin \cite{Groszkowski2022}. Related
representation-dependent interference and blockade also arise in finite-spin
synchronization \cite{Tan2022}.} \sentgreen{In the model studied below, previous work found that the parity-blind mean-field supports
two coexisting attractors: a finite-latitude self-sustained oscillation (SSO) and a
south-polar fixed point (s-PFP) \cite{Dutta2025}.} \sentred{Yet neither their local stability nor local Holstein--Primakoff expansions determine how the exact finite-$N$ steady state partitions weight between these macroscopically separated regions \cite{AssafMeerson2017}.}
\sentblack{Whether parity can reorganize Liouvillian recurrent support and thereby select distinct steady-state phases despite a unique steady state in every finite
irreducible spin representation remains unresolved.}

\newpage 
\begin{figure}[!b]
\centering
\vspace*{-7.25pt}
\includegraphics[width=\columnwidth]{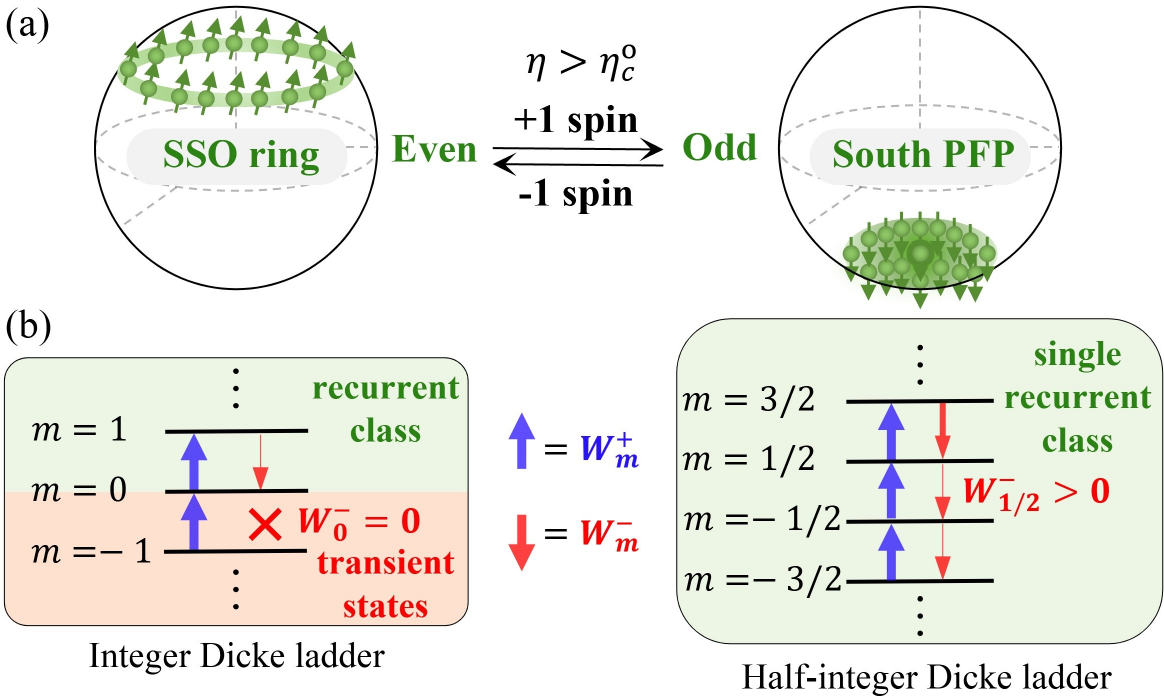}
\caption{\label{fig:parity-paradox}
\sentblack{{(a) Parity-selected macroscopic states.}}
\sentblue{At fixed
$\eta>\eta_c^{\mathrm{o}}$, adding or removing one spin switches between the
even-\(N\) SSO ring and the odd-\(N\) south-polar fixed point.}
\sentgreen{{(b) Parity-dependent Dicke-ladder connectivity.}}
\sentred{Blue (red) arrows denote \(W_m^+\) (\(W_m^-\)).}
\sentblack{For integer \(S\), the jump zero \(W_0^-=0\) makes \(m<0\) transient and leaves
\(m\geq0\) as the closed recurrent class, whereas for half-integer \(S\),
\(m=0\) is absent and \(W_{1/2}^->0\) connects the full ladder into a single
recurrent class.}}
\end{figure}

\sentblack{In this Letter, we show that one-spin sensitivity need not rely on
dark-state interference: above a critical dissipation imbalance, changing
\(N\) by one at fixed controls switches the steady state between the even-\(N\)
phase-averaged SSO ring and the odd-\(N\) s-PFP
[Fig.~\ref{fig:parity-paradox}(a)].}
\sentblue{The exact finite-size solution shows that the two sequences approach the same thermodynamic limit
up to the odd-sequence threshold \(\eta_c^{\mathrm{o}}\) and split above it
[Fig.~\ref{fig:parity-stationary-overview}].}
\sentgreen{The microscopic origin is a parity-dependent connectivity defect:
for \(S=N/2\), the even-\(N\) Dicke ladder contains \(m=0\), where the
ordered nonlinear-loss amplitude vanishes, deleting the \(0\to-1\)
transition, whereas the odd-\(N\) ladder misses this node and remains a
single recurrent class
[Fig.~\ref{fig:parity-paradox}(b)].}
\sentred{Only the odd-\(N\) ladder therefore retains both the SSO and s-PFP
contributions. Their exchange of exponential dominance produces a first-order
dissipative phase transition (DPT), signaled by a polarization jump and an
exponentially closing Liouvillian gap at rate coexistence
[Fig.~\ref{fig:boundary-condensation}].}
\sentblack{The switch survives multisector one-spin loading and
state-unresolved removal from the maximal sector. Quantum-jump trajectories
reveal rare-fluctuation-activated relaxation
[Fig.~\ref{fig:parity-quench-collapse}].}
\sentblue{These results establish discrete sampling of a state-dependent jump
zero as a mechanism whereby changing one constituent reorganizes Liouvillian
recurrence and selects a macroscopically distinct dissipative phase.}
\end{revision}

\begin{revision}
\paragraph{Model, symmetry and exact solution.---}\phantomsection\label{sec:model-exact-solution}
\sentblack{We consider \(N\) spin-\(1/2\) constituents with collective spin
\(\hat{\boldsymbol S}=\sum_{j=1}^N\hat{\boldsymbol\sigma}_j/2\), governed by the Lindblad equation \cite{Gorini1976,Lindblad1976}}
\begin{equation}
\dot{\hat\rho}=-i[\omega_0\hat S_z,\hat\rho]
+\mathcal D[\hat L_+]\hat\rho+\mathcal D[\hat L_-]\hat\rho.
\label{eq:lindblad}
\end{equation}
\sentblue{Here, \(\mathcal D[\hat L]\hat\rho=\hat L\hat\rho \hat L^\dagger-\{\hat L^\dagger \hat L,\hat\rho\}/2\),
while \(\hat L_+=\sqrt{2\Gamma_+/N}\,\hat S_+\) and
\(\hat L_-=\sqrt{8\Gamma_-/N^3}\,\hat S_-\hat S_z\) describe linear gain and nonlinear loss,
respectively \cite{Wang2026}.}
\sentblack{The strong symmetry
\(\hat{\boldsymbol S}^{\,2}\) decomposes the Hilbert space into invariant
total-spin sectors \cite{BucaProsen2012}.}
\sentblue{Within each sector, the Liouvillian acts identically on
all multiplicity copies, and its restriction to a single irreducible
spin-\(S\) representation has a unique finite-size steady state for
\(-1<\eta<1\); see
\hyperref[end:finite-uniqueness]{End Matter, Sec.~A} and
\smref[app:finite-uniqueness]{Supplemental Material (SM), Sec.~I}
~\cite{Supplemental}.} \sentgreen{Here, \(\eta=(\Gamma_--\Gamma_+)/\Gamma_{\rm tot}\) is the dissipation imbalance, with the total dissipation rate \(\Gamma_{\rm tot}=\Gamma_++\Gamma_-\).}

\sentblue{{In this work, we restrict the static analysis to the maximal
total-spin sector \(S=N/2\), whereas spin loading below retains all populated total-spin sectors.}}
\sentgreen{{Within this sector, previous work has shown that, for
\(0<\eta<1\), the large-\(S\) mean-field flow supports two
coexisting, parity-blind attractors: a SSO at
\(\theta_{\rm SSO}=\arccos z_\eta\), with
\(\phi(t)=\omega_0t+\phi_0\), and an s-PFP at
\(\theta=\pi\) \cite{Dutta2025},}} where \sentred{\(z_\eta=\sqrt{(1-\eta)/(1+\eta)}\), while
\((\theta,\phi)\) parametrize
\(\boldsymbol m=\langle\hat{\boldsymbol S}\rangle/S\).}
\sentblue{The driven system exhibits a transition from synchronization to a time crystal \cite{Wang2026}. Here we instead study steady-state selection in the undriven system.}

\sentred{The Liouvillian also has a weak $U(1)$ symmetry under
$\hat U_\varphi=e^{-i\varphi \hat S_z}$.}
\sentblack{Representation-reduced uniqueness then makes the steady
state $U(1)$ invariant and hence diagonal in the Dicke basis,
$\hat\rho_{\rm ss}^{(S)}=\sum_{m=-S}^{S}p_m^{(S)}
|S,m\rangle\langle S,m|$ \cite{BucaProsen2012,Nigro2020}.} \sentblue{The populations $p_m^{(S)}$ obey an exact closed birth-death equation, whose zero-current condition yields the recurrence
\((m+1)^2p_{m+1}^{(S)}=(Sz_\eta)^2p_m^{(S)}\); see \smref[app:population-derivation]{SM,
Sec.~II}~\cite{Supplemental}.}
\sentgreen{Hence the exact populations are}
\end{revision}
\begin{equation}
p_m^{(S)}=\frac{w_m}{\mathcal Z_{\rm tot}},~
w_m=\frac{(S z_\eta)^{2m}}{\Gamma^2(m+1)},~
\mathcal Z_{\rm tot}=\sum_{n=-S}^{S}w_n .
\label{eq:unified-population}
\end{equation}
\sentred{This solution covers both parities: for integer \(S\), the poles of \(\Gamma(m+1)\) set \(w_m=0\) for \(m<0\), whereas the half-integer-\(S\) support spans the full Dicke ladder.}
\begin{revision}

\sentblack{We visualize the exact steady state using the normalized Husimi-$Q$
function \cite{Roulet2018Smallest,Zhang2023TrappedIon},
$Q_{S}(\theta,\phi)=(2S+1)\langle\theta,\phi|\hat\rho_{\rm ss}^{(S)}
|\theta,\phi\rangle/(4\pi)$.} \sentblue{\(U(1)\) invariance makes \(Q_S\) independent of \(\phi\).
The explicit binomial form used in
Fig.~\ref{fig:parity-stationary-overview} is given in
\smref[app:population-derivation]{SM, Sec.~II}
~\cite{Supplemental}.}
\sentgreen{Figures~\ref{fig:parity-stationary-overview}(a,b) show the same
north-PFP/SSO structure for both parities below
\(\eta_c^{\mathrm{o}}\). Above it, even-\(N\) remains on the SSO ring while
odd-\(N\) develops an s-PFP peak.} \sentred{Figure~\ref{fig:parity-stationary-overview}(c) quantifies this effect: as \(N\) increases, the adjacent-size magnetization contrast develops an increasingly sharp onset above the threshold. We next determine whether it survives when the thermodynamic limit is taken separately along the even-\(N\) and odd-\(N\) subsequences.}
\end{revision}

\begin{revision}
\paragraph{Parity-selected thermodynamic limits.---}\phantomsection\label{sec:parity-limits}
\sentblack{For these paired sequences, define
$m_{z,\infty}^{\mathrm{e,o}}=
\lim_{k\to\infty}m_{z,k}^{\mathrm{e,o}}$.}
\sentblue{The large-$S$ asymptotics of the exact populations in
Eq.~\eqref{eq:unified-population} show that the two parity sequences
 have the same thermodynamic limit at and below $\eta_c^{\mathrm{o}}$:
$m_{z,\infty}^{\mathrm{e,o}}=1$ for $-1<\eta\leq0$ and
$m_{z,\infty}^{\mathrm{e,o}}=z_\eta$ for
 $0<\eta\leq\eta_c^{\mathrm{o}}$.}
\sentgreen{Above $\eta_c^{\mathrm{o}}$, however, the
parity-resolved thermodynamic limits split:}
\begin{equation}
(m_{z,\infty}^{\mathrm{e}},m_{z,\infty}^{\mathrm{o}})
=(z_\eta,-1),\qquad
\eta_c^{\mathrm{o}}<\eta<1.
\label{eq:parity-resolved-magnetization-main}
\end{equation}
\sentred{These limits and the odd-\(N\) threshold $\eta_c^{\mathrm{o}}$ are derived in
\smref[app:parity-thermodynamic-limits]{SM, Sec.~III}~\cite{Supplemental}.} 
\sentblack{Accordingly, the adjacent-size contrast $\Delta m_{z,k}^{(1)}$ tends to zero at and below
$\eta_c^{\mathrm{o}}$ and to $1+z_\eta$ above it as shown in Fig.~\ref{fig:parity-stationary-overview}(c).}
\sentblue{Therefore, despite a unique steady state within every finite irreducible
spin-\(S\) representation, the paired even-\(N\) and odd-\(N\)
sequences converge to different steady-state phases above the threshold.} 

\sentgreen{Both parity sequences nevertheless inherit the same bistable large-\(S\) mean-field flow \cite{Wang2026}.}
\sentblack{This apparent tension reflects the noncommutativity of the long-time and thermodynamic limits \footnote{\sentblack{Writing $s_{+,S}(t)=\langle\hat S_+(t)\rangle/S$, phase diffusion at finite $S$ gives $\lim_{S\to\infty}[\lim_{t\to\infty}s_{+,S}(t)]=0$, whereas taking $S\to\infty$ first leaves a phase-resolved periodic orbit, so $\nexists\lim_{t\to\infty}[\lim_{S\to\infty}s_{+,S}(t)]$.} \sentblue{See \smref[app:order-of-limits]{SM, Sec.~IV.A}~\cite{Supplemental}.}}.} \sentblue{This order-of-limits issue, however, does not cause the parity splitting, which is encoded in two ingredients: a parity-dependent recurrent structure and the large-deviation amplification of this microscopic connectivity difference into macroscopic steady-state selection.} 
\end{revision}

\sentgreen{The first ingredient is the parity-dependent recurrence shown in Fig.~\ref{fig:parity-paradox}(b).}
\sentred{For the ordered nonlinear loss \(\hat L_-\propto \hat S_-\hat S_z\), the downward
rate satisfies \(W_m^-\propto m^2\), and hence \(W_0^-=0\).}
\sentblack{For even-\(N\), the integer Dicke ladder samples this zero, deleting
the \(0\to-1\) edge while retaining the reverse pumping edge. Consequently,
\(m\geq0\) is recurrent and \(m<0\) is transient, consistent with \(p_m=0\)
for \(m<0\) in Eq.~\eqref{eq:unified-population}.}
\sentblue{For odd-\(N\), \(m=0\) is absent and \(W_{1/2}^->0\), leaving the
full ladder, including \(m=-S\), as a single recurrent chain.}
\sentgreen{However, this ingredient alone does not explain the macroscopic contrast.}

\sentred{The second ingredient is the exponential nonuniformity of the
steady-state weights in Eq.~\eqref{eq:unified-population}.}
\sentblack{On the odd Dicke ladder, the full recurrent chain supports
two macroscopically separated concentrations of steady-state weight: an
interior SSO peak near \(m\simeq Sz_\eta\) and an s-PFP boundary layer
near \(m=-S\), whereas the latter is absent from the
even-\(N\) recurrent class.}
\sentblue{Because the two integrated branch weights scale exponentially with \(S\), their crossing defines \(\eta_c^{\mathrm{o}}\) and exchanges the dominant contribution.}

\sentred{Local semiclassical treatments miss the splitting because, near the
SSO peak, the half-spacing offset between integer and half-integer lattices is
negligible relative to the \(O(\sqrt S)\) fluctuation width, yielding the same
mean-field and leading Holstein--Primakoff theories
\cite{Wang2026}.} \sentblue{Selection instead compares the loss node, SSO peak, and south
boundary, separated by \(O(S)\). These structures survive exact truncation
but are lost in an unbounded local bosonic expansion
\cite{KonigHucht2021}; see \smref[app:mean-field-attractors]{SM, Sec.~IV.B}
~\cite{Supplemental}.} \sentgreen{The parity switch is therefore a global large-deviation effect, not a
finite-order correction about one attractor
\cite{AssafMeerson2017}.}

\paragraph{First-order DPT along the odd-\(N\) sequence.---}\phantomsection\label{sec:odd-dpt}
\sentblack{For the odd-\(N\) sequence, the first-order character is encoded in the large-$S$
asymptotics of the exact normalization.}
\sentblue{We partition the total normalization $\mathcal Z_{\rm tot}$ in
Eq.~\eqref{eq:unified-population} as
\(\mathcal Z_{\rm tot}=\mathcal Z_{\rm SSO}
+\mathcal Z_{\rm s\text{-}PFP}+\mathcal Z_{\rm rest}\).}
\sentgreen{For a mesoscopic window \(\sqrt {S}\ll\ell_{S}\ll S\), the first two terms collect
the weights near the interior SSO peak and the s-PFP boundary,
respectively:
\(\mathcal Z_{\rm SSO}=\sum_{|m-Sz_\eta|\leq\ell_{S}}w_m\) and
\(\mathcal Z_{\rm s\text{-}PFP}
=\sum_{0\leq m+S\leq\ell_{S}}w_m\).}
\sentred{\(\mathcal Z_{\rm rest}\) does not modify the leading large-\(S\)
asymptotics of \(\mathcal Z_{\rm tot}\).}
\sentblack{We therefore define the total and branch-resolved rate potentials as}
\begin{equation}
g_\alpha=-\lim_{S\to\infty}
\ln\mathcal Z_\alpha/(2S),~\alpha\in\{\mathrm{tot},\mathrm{SSO},\mathrm{s\text{-}PFP}\}.
\label{eq:odd-rate-potential}
\end{equation}

\sentblue{Applying Stirling asymptotics about the interior SSO peak and the Gamma-function reflection identity at the south boundary gives \(g_{\rm SSO}=-z_\eta\) and \(g_{\rm s\text{-}PFP}=1+\ln z_\eta\), respectively.}
\sentgreen{Using the discrete Laplace principle, we obtain the total rate as
the smaller of the two branch rates, as derived and classified in
\smref[app:odd-global-selection]{SM, Sec.~III.D}~\cite{Supplemental}:}
\begin{equation}
\begin{aligned}
g_{\rm tot}=\min\{g_{\rm SSO},g_{\rm s\text{-}PFP}\}.
\end{aligned}
\label{eq:odd-rate-branches}
\end{equation}
\sentblack{Equating the two branch rates, we obtain
\(1+z_\eta+\ln z_\eta=0\), whose solution is \(z_c=W_0(e^{-1})\), where \(W_0\) is the principal Lambert branch.}
\sentblue{Therefore, the transition point is at
\(\eta_c^{\mathrm{o}}=(1-z_c^2)/(1+z_c^2)\simeq0.856\).}
\sentgreen{Although not an equilibrium free
 energy, \(g_{\rm tot}\) acts as a nonequilibrium steady-state rate potential: the branch with the smaller rate is exponentially dominant. Away from \(\eta_c^{\mathrm{o}}\), its conjugate
derivative generates the order parameter through}
\begin{equation}
m_{z,\infty}^{\mathrm{o}}
=-\frac{\partial g_{\rm tot}}{\partial\ln z_\eta}.
\label{eq:odd-polarization-from-potential}
\end{equation}
\sentred{At \(\eta_c^{\mathrm{o}}\), \(g_{\rm tot}\) remains continuous but becomes non-differentiable. This cusp makes \(m_{z,\infty}^{\mathrm{o}}\) jump from \(z_c\) to \(-1\), establishing the first-order DPT in Fig.~\ref{fig:boundary-condensation}(a).}

\sentblack{We next test the spectral signature of the transition and connect it to the
previously identified slow s-PFP mode \cite{Dutta2025} by defining, within the
maximal-spin representation, the Liouvillian gap as
\(\Delta_{\mathcal L}
=-\max_{\lambda\neq0}\operatorname{Re}\lambda\).} \sentblue{Figure~\ref{fig:boundary-condensation}(b) shows \(\Delta_{\mathcal L}\) obtained
by evaluating all coherence-order blocks: it remains positive for every
displayed finite odd-\(N\), consistent with finite-size uniqueness, but
develops a deepening minimum whose location approaches
\(\eta_c^{\mathrm{o}}\).} \sentgreen{At odd-sequence rate coexistence, the analytically controlled
population-transfer mode obeys
\(\Delta_{\rm pop}(\eta_c^{\mathrm{o}})
=\Gamma_{\rm tot}e^{-2z_cN+o(N)}\). The complete block-resolved spectra give
\(\Delta_{\mathcal L}=\Delta_{\rm pop}\) for every displayed size, providing an independent spectral
signature of the first-order DPT
\cite{Casteels2017,Minganti2018,Carmichael2015}; see
\hyperref[end:full-gap]{End Matter, Sec.~B} and
\smref[app:liouvillian-gap-method]{SM, Sec.~V}
~\cite{Supplemental}.}

\begin{revision}
\paragraph{One-spin loading and quantum-fluctuation-activated switching.---}\phantomsection\label{sec:loading}
\sentblack{Whereas the preceding static analysis uses the maximal
representation \(S_0=N_0/2\), physically appending one spin produces
\(\hat\rho^{(1)}(0^+)=
\hat\rho_{\rm ss}^{(S_0)}\otimes\hat\rho_{1/2}\),
with support on
\(\mathcal S_1=\{S_0+1/2,S_0-1/2\}\).}
\sentred{This raises the question whether the multisector dynamics
preserves or rounds the one-spin switch at fixed
\(\eta>\eta_c^{\mathrm{o}}\).}

\sentblack{We denote by \(\mathcal S_\ell\) the sectors occupied after the
\(\ell\)th loading and by \(q_S^{(\ell)}\) the exact weight of sector
\(S\in\mathcal S_\ell\).}
\sentgreen{Although loading generally generates inter-\(S\) coherences, we
prove that they are invisible to collective observables and remain decoupled
from the sector-reduced dynamics under subsequent independent loadings; see
Eq.~\eqref{eq:end-loading-recursion}.}
\sentblue{Consequently, between loadings, the sector weights
\(q_S^{(\ell)}\) are conserved, and every collective observable is the exact
sector-weighted sum in
Eq.~\eqref{eq:end-sector-weighted-collective-evolution}, with all occupied
sectors retained without total-spin-sector postselection.}

\sentgreen{To resolve transfer between the two macroscopic branches, we
denote by \(p_{m,S}^{(\ell)}(t_\ell)\) the normalized Dicke population in
sector \(S\) and partition each Dicke ladder into loading-dependent basins
\(\mathcal B_{\alpha,\ell}^{(S)}\), with
\(\alpha\in\{\mathrm{SSO},\mathrm{s\text{-}PFP}\}\).} \sentred{For even-to-odd loading, the basin boundary is the discrete
bottleneck separating the s-PFP and SSO regions, whereas for odd-to-even loading, it is \(m=0\), the entrance to the
recurrent SSO class.}
\sentblack{Using the sector-resolved population recursion and basin
construction detailed in
\smref[app:cg-multisector-dynamics]{SM, Sec.~VI.A}~\cite{Supplemental},
the total basin weights are}
\begin{equation}
P_\alpha^{(\ell)}(t_\ell)
=\sum_{S\in\mathcal S_\ell}q_S^{(\ell)}
\sum_{m\in\mathcal B_{\alpha,\ell}^{(S)}}p_{m,S}^{(\ell)}(t_\ell).
\label{eq:parity-quench-basin-weights}
\end{equation}
\sentgreen{Let \(T_S^{(\ell)}\) denote the mean first-passage time (MFPT) \cite{Redner2023FirstPassage,Menczel2026} of the sector-$S$ for interbranch transfer during loading step \(\ell\).}
\sentred{We use the MFPT reconstruction detailed in
\smref[app:loading-large-size-reconstruction]{SM, Sec.~VI.B}
~\cite{Supplemental} to approximate the initial-basin weight and define the
equal-weight time \(t_{50}^{(\ell)}\) at loading step \(\ell\) through}
\begingroup
\setlength{\belowdisplayskip}{1pt}
\setlength{\belowdisplayshortskip}{1pt}
\begin{equation}
P_{\alpha_\ell}^{(\ell)}(t_\ell)\simeq
\sum_{S\in\mathcal S_\ell}q_S^{(\ell)}e^{-t_\ell/T_S^{(\ell)}},~
P_{\alpha_\ell}^{(\ell)}\!\left(t_{50}^{(\ell)}\right)=1/2.
\label{eq:reconstructed-loading-t50}
\end{equation}
\endgroup
\sentblack{The construction allows an arbitrary independently prepared state of the added spin. Figure~\ref{fig:parity-quench-collapse}(a) starts from the SSO steady state at \(N_0=1024\) and loads unpolarized
spins, \(\hat\rho_{1/2}=\hat I_2/2\), for which the sector-reduced recursion
closes on the Dicke populations.}
\sentblue{The first loading, \(1024\to1025\), has
\(\alpha_1=\mathrm{SSO}\) and populates two half-integer sectors, both of
which relax to s-PFP-dominated steady states.}
\sentgreen{A second loading,
\(1025\to1026\), has
\(\alpha_2=\mathrm{s\text{-}PFP}\) and populates three integer sectors, all
of which relax back to the SSO basin, after which the protocol is reset.}
\sentred{Thus Fig.~\ref{fig:parity-quench-collapse}(a) shows that the
parity-selected steady-state switch survives multisector one-spin loading.}

\sentblack{Applying the same protocol to each even-\(N_0\),
Fig.~\ref{fig:parity-quench-collapse}(b) shows the paired equal-weight times
\(t_{50}^{(1)}\) for \(N_0\to N_0+1\) and \(t_{50}^{(2)}\) for
\(N_0+1\to N_0+2\), up to \(N_0=1024\).}
\sentblue{Their asymptotically linear large-\(N_0\) trends establish
exponential growth in both loadings, with distinct exponents given
by Eqs.~\eqref{eq:end-loading-t50-exponent} and
\eqref{eq:end-reverse-loading-t50-exponent}.}
\sentgreen{Direct propagation for \(N_0\leq40\) agrees with these
predictions on the exponential scale as shown in \hyperref[fig:end-loading-t50-validation]{Fig.~\ref*{fig:end-loading-t50-validation}} and
\smref[app:loading-switching-time]{SM, Sec.~VI.C}~\cite{Supplemental}.}
\sentred{For the first loading,
\(t_{50}^{(1)}=e^{o(N_0)}/\Delta_{{\rm pop},\min}^{(1)}\), identifying the
slowest occupied population mode as the asymptotic switching bottleneck
[see Eq.~\eqref{eq:end-loading-gap-time-scaling}].}

\begin{figure}[t]
\centering
\includegraphics[width=\columnwidth]{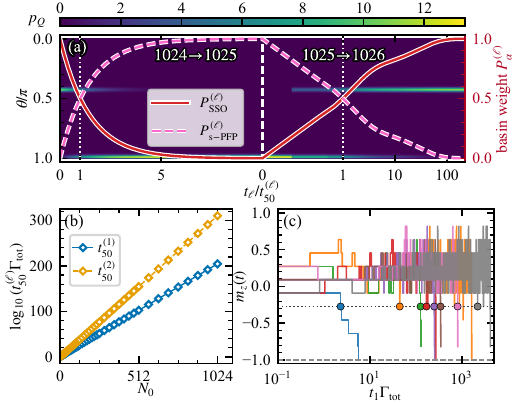}
\caption{\label{fig:parity-quench-collapse}
\sentblue{(a) Polar Husimi-\(Q\) density
\(p_Q(\theta,t_\ell)=2\pi\sin\theta\sum_{S\in\mathcal S_\ell}
q_S^{(\ell)}Q_S(\theta,t_\ell)\) and basin weights in the MFPT-based reconstruction of two successive loading stages; the dashed line separates their stage-local time axes.} \sentgreen{(b) Equal-weight times for the two loadings
versus their common even base size \(N_0\).}
\sentred{(c) Eight sector-resolved quantum-jump trajectories for \(10\to11\),
selected at equally spaced first-passage-time ranks from a 64-trajectory
ensemble. Color-matched circles mark first arrival at the common boundary
\(m_z^{\rm b}=-3/11\) (gray dotted), and the gray dashed line marks
\(m_z=-1\).}
\sentblack{All panels use \(\eta=0.9\).}}
\end{figure}

\sentblue{Figure~\ref{fig:parity-quench-collapse}(c) reveals the
trajectory-level origin of these exponentially long switching times.}
\sentgreen{Because one-spin loading changes intensive variables only by
\(O(1/N_0)\), the postloading mean-field state remains in the locally stable
SSO basin and cannot cross the basin boundary.}
\sentred{At finite \(N\), however, quantum jumps generate rare fluctuations that drive
individual trajectories across the basin boundary, yielding a broad
distribution of first-passage times
\cite{Macieszczak2016,Macieszczak2021,BrownMacieszczakJack2024}; see
\smref[app:sector-resolved-stochastic-dynamics]{SM, Sec.~VI.D} \cite{Supplemental}.}
\sentred{One-spin addition and removal are therefore opposite parity quenches
at fixed intensive controls, rather than parameter sweeps across the DPT; see
\smref[app:spin-removal-switch]{SM, Sec.~VI.E}~\cite{Supplemental}.}
\end{revision}

\begin{revision}
\paragraph{Generalized nonlinear jumps and zero-preserving robustness.---}
\phantomsection
\label{sec:generalized-jumps-robustness}
\sentblack{We now extend the mechanism from the specific nonlinear loss above
to collective-spin Lindbladians with closed Dicke-population dynamics and
ordered state-dependent \(q\)-step jumps
\(\hat L_{\pm,F}^{(q)}\propto
\hat S_\pm^{\,q}F(\hat S_z/S)\).}
\sentblue{Distinct particle-number subsequences can sample an isolated zero of
\(F\) differently and acquire different recurrent supports.}
\sentgreen{This kinematic distinction becomes thermodynamic phase selection
only when the surviving branches compete through exponential stationary
weights.}
\sentred{\(F(z_0)=0\) deletes the edge
\(m\to m\pm q\) when \(m/S=z_0\). If all active channels preserve the
residue classes, the Dicke graph splits into arithmetic sublattices and can
support \(N\bmod q\)-dependent selection.}
\sentblack{For \(F(z)=z(1+z^2)\), all signatures persist, with a shifted
critical point and switching steady-state exponent.}
\sentblue{By contrast, a weak linear loss that restores the deleted edge
removes the parity splitting at any fixed strength, although finite systems
retain it below a crossover that shrinks exponentially with size; see
\smref[app:general-mechanism]{SM, Sec.~VII}~\cite{Supplemental}.}
\end{revision}

\paragraph{Conclusion and discussion.---}\phantomsection\label{sec:conclusion}
\begin{revision}
\sentblack{We have established a mechanism by which particle-number parity selects
macroscopically distinct steady-state phases even though the mean-field flow is
parity blind and every finite irreducible spin representation has a unique steady
state.} \sentblue{The integer Dicke lattice samples the zero of the ordered nonlinear loss whereas the adjacent half-integer lattice avoids it, changing the recurrent support.} \sentgreen{The resulting global branch normalization produces a first-order DPT along the odd-\(N\) sequence.} \sentred{Unpolarized one-spin loading preserves this selection across all
populated total-spin sectors, while state-unresolved removal from the
maximal-spin representation realizes the reverse switch.} \sentblack{More broadly, these results show that discrete representation structure can remain thermodynamically relevant and control macroscopic steady-state selection in an open quantum system.}

\sentblack{Experimentally, the central challenge is to realize a collective jump
\(\hat S_-F(\hat S_z/S)\) with a Dicke-lattice-resolved zero. The specific
\(\hat S_-\hat S_z\) channel has not, to our knowledge, been implemented directly.}
\sentblue{A driven \(^{88}\mathrm{Sr}\) cavity-QED ensemble has nevertheless
demonstrated cavity-mediated collective radiation and an \(N\)-scaled
dissipation-induced superradiant transition for
\(N\sim10^3\!-\!10^4\) atoms \cite{Song2025}, establishing a suitable
large-\(N\) platform while leaving the \(S_z\)-dependent nonlinear factor to
be engineered.}
\sentgreen{Relevant ingredients have been demonstrated separately, including
state-selective ordered dissipation at spin 1 through polarization-controlled
optical pumping \cite{Laskar2020} and two-phonon damping in trapped-ion quantum
van der Pol oscillators through second-order red-sideband driving and rapid
spin reset \cite{Li2025QvdP,Liu2025QvdP}.}
\sentred{Related nonlinear loss also stabilizes driven-dissipative steady-state
phases in cavity-QED settings \cite{Shah2025TwoPhotonDicke}.}
\sentblack{Combined with collective reservoir engineering and effective-operator
elimination, these advances suggest generating the desired jump through a
rapidly damped ancillary mode dispersively coupled to \(S_z\)
\cite{Groszkowski2022,ReiterSorensen2012}.}
\sentblue{The switch can be detected from the order-one change in the
normalized polarization \(2\langle\hat S_z\rangle/N\) between adjacent sizes.}
\sentblack{This macroscopic polarization response suggests a route to detecting
single-spin addition or removal through collective readout, connecting our
mechanism to parity-based sensing \cite{Groszkowski2022} and to the broader
use of dissipative signal amplification in quantum metrology
\cite{Koppenhoefer2022SpinAmplifier}.}
\sentgreen{The required switching time, however, grows exponentially with
system size, reflecting the increasing dynamical isolation of the competing
macroscopic basins.}
\sentgreen{The resulting stability--accessibility tradeoff favors mesoscopic
ensembles, which retain an order-one parity contrast while switching within
experimentally accessible times.}

\end{revision}

\section*{Acknowledgments}
\sentblack{We thank Liang-Liang Wan for helpful discussions.}
\sentblue{This work was supported by the National Natural Science Foundation of China
under Grant No.~12575026, the Shenzhen Science and Technology Program under
Grant No.~JCYJ20250604145221028, and the Natural Science Foundation of Top
Talent of SZTU under Grant No.~GDRC202527.}

\paragraph*{Data availability.---}
\sentblack{The data supporting the findings of this article are publicly
available in Ref.~\cite{Jie2026FigureData}.}

\bibliography{references}

\clearpage
\section*{End Matter}

\paragraph{A. Representation-reduced finite-size uniqueness.---}
\setcounter{equation}{0}
\renewcommand{\theequation}{A\arabic{equation}}
\renewcommand{\theHequation}{endmatter.A\arabic{equation}}
\phantomsection
\label{end:finite-uniqueness}

\sentblack{Fix \(N\) and restrict the collective generator to an allowed
irreducible spin representation \(\mathcal R_S\), keeping the Kac
normalization fixed by \(N\).}
\sentblue{For \(-1<\eta<1\), the Dicke graph in every allowed \(\mathcal R_S\)
has a single closed irreducible class, with \(\mathcal R_0\) trivially one
dimensional. Within that class the weighted jumps connect all Dicke vectors
into one enclosure, all remaining states feed into it, and the diagonal
Hamiltonian creates no additional enclosure.}
\sentred{The finite-dimensional enclosure decomposition therefore rules out a
second stationary enclosure or equivalent enclosures carrying noiseless
coherence \cite{Frigerio1978,BaumgartnerNarnhofer2008,
AlbertJiang2014,Yoshida2024}, and hence}
\begin{equation}
\dim\ker\mathcal L_{N,S}=1,~
N<\infty,\quad -1<\eta<1.
\label{eq:end-finite-uniqueness}
\end{equation}
\sentred{This is a representation-reduced statement: with
\(\mathcal H_N=\bigoplus_S(\mathcal R_S\otimes\mathcal M_{S,N})\), collective
dynamics acts trivially on the multiplicity space and therefore}
\begin{equation}
\begin{aligned}
\ker\!(\left.\mathcal L_N\right|_{\mathcal R_S\otimes
\mathcal M_{S,N}})=\operatorname{span}\{\hat\rho_{\rm ss}^{(N,S)}\}
\otimes\mathcal B(\mathcal M_{S,N}).
\end{aligned}
\label{eq:end-isotypic-kernel}
\end{equation}
\sentblack{Multiplicity degrees of freedom are noiseless, while
\(\dim\mathcal M_{N/2,N}=1\); see
\smref[app:finite-uniqueness]{SM, Sec.~I}~\cite{Supplemental} for technical
details and the endpoint classification at \(\eta=\pm1\).}

\paragraph{B. Liouvillian gap at thermodynamic rate coexistence.---}
\setcounter{equation}{0}
\renewcommand{\theequation}{B\arabic{equation}}
\renewcommand{\theHequation}{endmatter.B\arabic{equation}}
\phantomsection
\label{end:full-gap}

\sentblack{Weak \(U(1)\) covariance decomposes operator space into invariant
coherence-order blocks}
\begin{equation}
\mathcal V_\nu=\operatorname{span}
\{|S,m\rangle\langle S,m-\nu|\},~
\dim\mathcal V_\nu=2S+1-|\nu|.
\label{eq:end-coherence-blocks}
\end{equation}
\sentblue{Writing \(\mathcal L_\nu=\mathcal L|_{\mathcal V_\nu}\) and
\(\Delta_\nu\) for its smallest nonzero decay rate, the full gap is exactly}
\begin{equation}
\Delta_{\mathcal L}=\min_{|\nu|\leq2S}\Delta_\nu,~
\Delta_\nu=-\max_{\substack{\lambda\in\operatorname{spec}\mathcal L_\nu\\
\lambda\neq0}}\operatorname{Re}\lambda.
\label{eq:end-block-gap-minimum}
\end{equation}
\sentgreen{The Hamiltonian shifts every eigenvalue in \(\mathcal V_\nu\) by
\(-i\nu\omega_0\) and therefore leaves its real part unchanged.}
\sentred{The \(\nu=0\) block closes exactly on the Dicke populations,
\(\dot{\boldsymbol p}=\mathsf K_{\rm pop}\boldsymbol p\). For the odd-\(N\)
chain at \(\eta=\eta_c^{\mathrm{o}}\), detailed balance makes \(\mathsf K_{\rm pop}\) similar
to a real symmetric Jacobi matrix, whose
first nonzero decay rate \(\Delta_{\rm pop}\) satisfies
\(-\Delta_{\rm pop}\in\operatorname{spec}\mathcal L\) and
\(\Delta_{\mathcal L}\leq\Delta_{\rm pop}\).}
\sentblue{At rate coexistence the negative-side rate function is minimized at
\(m_{\rm b}=-Sz_c+O(1)\). The resulting bottleneck conductance and the
one-dimensional Hardy constant obey
\cite{Miclo1999Hardy}}
\begin{equation}
\begin{aligned}
c_{m_{\rm b}}=\Gamma_{\rm tot}e^{-2z_cN+o(N)},&~B_N=\Gamma_{\rm tot}^{-1}e^{2z_cN+o(N)},\\
\frac{B_N}{2}\leq&\Delta_{\rm pop}^{-1}\leq4B_N.
\end{aligned}
\label{eq:end-hardy-scaling}
\end{equation}
\sentgreen{The analytic Hardy bound then gives the population-gap exponent}
\begin{equation}
\Delta_{\rm pop}(\eta_c^{\mathrm{o}})
=\Gamma_{\rm tot}e^{-2z_cN+o(N)}.
\label{eq:end-gap-exponent}
\end{equation}
\begin{figure}[t]
\centering
\includegraphics[width=0.6\columnwidth]{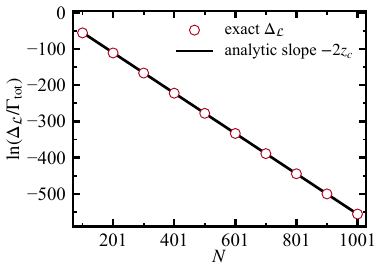}
\caption{\label{fig:end-gap-coexistence-scaling}
\sentblack{Full Liouvillian gaps at $\eta=\eta_c^{\mathrm{o}}$ for the odd sizes
used in Fig.~\ref{fig:boundary-condensation}(b) (symbols). For every symbol the
minimum lies in the population block. The solid guide has the analytic
population-gap slope $-2z_c$ from Eq.~\eqref{eq:end-gap-exponent}.}}
\end{figure}
\sentblue{Figure~\ref{fig:end-gap-coexistence-scaling} shows that, for every
displayed size, the block-resolved spectra place the leading nonpopulation
mode in the \(|\nu|=1\) phase-diffusion sector above the population mode,
establishing \(\Delta_{\mathcal L}=\Delta_{\rm pop}\) along the computed
sequence.}
\sentgreen{Equation~\eqref{eq:end-gap-exponent} analytically proves the
population-gap exponent but does not assume this equality beyond that
sequence.}
\sentred{The rate-function derivation, Jacobi--Sturm computation, Hardy bound,
and full finite-size block comparison are given in
\smref[app:liouvillian-gap-method]{SM, Sec.~V}~\cite{Supplemental}.}

\paragraph{C. Collective loading and switching.---}
\setcounter{equation}{0}
\renewcommand{\theequation}{C\arabic{equation}}
\renewcommand{\theHequation}{endmatter.C\arabic{equation}}
\phantomsection
\label{end:loading-closure}

\sentblack{For an arbitrary many-spin state, define the observable reduced
block
\(\hat{\bar\rho}_S=\operatorname{Tr}_{\mathcal M_{S,N}}
(\hat\Pi_S\hat\rho\hat\Pi_S)\), whose trace is the physical sector weight \(q_S\).}
\sentblue{At loading step \(\ell\), appending a spin in state
\(\hat\rho_{1/2}\) maps a parent sector \(S_{\rm p}\) to child sectors
\(S_{\rm c}=S_{\rm p}\pm1/2\), whose multiplicity copies occupy
orthogonal summands of \(\mathcal M_{S_{\rm c},N+1}\).}
\sentgreen{Orthogonality of the multiplicity-path embeddings removes
coherences between distinct parent sectors from the copy-space trace without
physically dephasing the state.}
\sentred{Let \(\hat C_{S_{\rm c}\leftarrow S_{\rm p}}\) be the
Clebsch--Gordan partial isometry defined by
\(\langle S_{\rm c},M|\hat C_{S_{\rm c}\leftarrow S_{\rm p}}
|S_{\rm p},m;\tfrac12,\sigma\rangle
=C^{S_{\rm c},M}_{S_{\rm p},m;\,1/2,\sigma}\). For each fixed child sector
\(S_{\rm c}\), summing the surviving contributions from all compatible parent
sectors gives the exact reduced loading recursion}
\begin{equation}
\hat{\bar\rho}_{S_{\rm c}}^{(\ell)}
=\sum_{\substack{S_{\rm p}\in\mathcal S_{\ell-1}\\
|S_{\rm c}-S_{\rm p}|=1/2}}
\hat C_{S_{\rm c}\leftarrow S_{\rm p}}
(\hat{\bar\rho}_{S_{\rm p}}^{(\ell-1)}\otimes\hat\rho_{1/2})
\hat C_{S_{\rm c}\leftarrow S_{\rm p}}^{\dagger}.
\label{eq:end-loading-recursion}
\end{equation}
\sentblue{Trace preservation within each sector conserves
\(q_S^{(\ell)}\) between loadings. With
\(\hat\rho_S^{(\ell)}(0^+)=
\hat{\bar\rho}_S^{(\ell)}(0^+)/q_S^{(\ell)}\), the resulting collective evolution is}
\begin{equation}
\langle\hat O(t_\ell)\rangle_\ell
=\sum_{S\in\mathcal S_\ell}q_S^{(\ell)}
\operatorname{Tr}\!\left[
\hat O_S e^{\mathcal L_{N_\ell,S}t_\ell}
\hat\rho_S^{(\ell)}(0^+)\right].
\label{eq:end-sector-weighted-collective-evolution}
\end{equation}
\sentred{Define the directly propagated equal-weight time \(t_{50,\mathrm{dir}}^{(\ell)}\) as the first crossing, from either direction, satisfying}
\nopagebreak[4]
\begin{equation}
P_{\mathrm{s\text{-}PFP}}^{(\ell)}
\bigl(t_{50,\mathrm{dir}}^{(\ell)}\bigr)=\frac12.
\label{eq:end-exact-loading-t50}
\end{equation}
\clearpage
\begin{figure}[H]
\centering
\includegraphics[width=0.6\columnwidth]{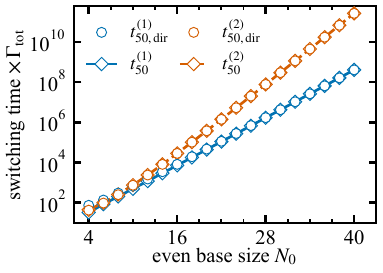}
\caption{\label{fig:end-loading-t50-validation}
\sentblack{Paired finite-size validation at \(\eta=0.90\): each independently
reset even base size \(N=4,6,\ldots,40\) defines
\(N\to N+1\to N+2\), with the second loading initialized from the first-stage
s-PFP state.}
\sentblue{Circles are the directly propagated \(t_{50,\mathrm{dir}}^{(1)}\) (blue) and
\(t_{50,\mathrm{dir}}^{(2)}\) (orange). Diamond curves are the corresponding
MFPT-based values \(t_{50}^{(1)}\) (solid) and
\(t_{50}^{(2)}\) (dashed).}}
\end{figure}
\sentgreen{Figure~\ref{fig:end-loading-t50-validation} shows that the MFPT
reconstruction reproduces the directly propagated equal-weight times for both
loading directions on the exponential scale over the accessible sizes.}
\sentblue{For the first loading, write \(S_1=(N_0+1)/2\).  The two occupied
sectors \(S\in\{S_1,S_1-1\}\) have the same leading barrier between the
positive mode and the bottleneck at
\(m_{\rm b}=-S_1z_\eta+O(1)\).}
\sentgreen{The exact birth--death MFPT, together with metastable spectral
separation inside the initial basin, then gives the same leading exponent for
every fixed survival quantile and for the finite mixture over sectors:}
\begin{equation}
\Gamma_{\rm tot}t_{50,\mathrm{dir}}^{(1)}
=\exp[2z_\eta N_0+o(N_0)].
\label{eq:end-loading-t50-exponent}
\end{equation}
\sentred{Combining the same sector-MFPT scaling with
Eq.~\eqref{eq:reconstructed-loading-t50} gives
\(\Gamma_{\rm tot}t_{50}^{(1)}
=\exp[2z_\eta N_0+o(N_0)]\), and hence
\(\ln[t_{50,\mathrm{dir}}^{(1)}/t_{50}^{(1)}]=o(N_0)\).}
\sentblack{The minimum participating population gap and equal-weight time
therefore satisfy}
\begin{align}
\Delta_{\mathrm{pop},\min}^{(\ell)}
=\min_{S\in\mathcal S_\ell}\Delta_{\mathrm{pop},S},
\quad
&\Delta_{\mathrm{pop},\min}^{(1)}
=\Gamma_{\rm tot}e^{-2z_\eta N_0+o(N_0)},
\notag\\
\Gamma_{\rm tot}t_{50}^{(1)}
&=
\frac{\Gamma_{\rm tot}}
{\Delta_{\mathrm{pop},\min}^{(1)}}e^{o(N_0)}.
\label{eq:end-loading-gap-time-scaling}
\end{align}
\sentblue{For the reverse loading, the transient negative half-ladder in each
of the three occupied integer-spin sectors has the common barrier action
\(I_{\rm rev}(z)=-\ln z-1+z\). The deleted \(0\to-1\) edge makes first arrival
at the recurrent nonnegative class irreversible.}
\sentgreen{The subextensive width of the inherited s-PFP layer and the
nonzero limiting weights of all three sectors therefore preserve this
large-deviation exponent in the physical mixture:}
\begin{equation}
\Gamma_{\rm tot}t_{50,\mathrm{dir}}^{(2)}
=\exp\!\left[N_0I_{\rm rev}(z_\eta)+o(N_0)\right].
\label{eq:end-reverse-loading-t50-exponent}
\end{equation}
\sentred{The MFPT-based calculation has the same exponent,
\(\ln[t_{50,\mathrm{dir}}^{(2)}/t_{50}^{(2)}]=o(N_0)\). All three sectors contribute to the subexponential prefactor of the
equal-weight crossing; their finite limiting time-scale ratios do not alter
the common large-deviation exponent.} \sentblack{These exponential laws fix the order of limits: at fixed finite
\(N_0\), taking \(t\to\infty\) reaches the parity-selected steady state,
whereas at fixed \(t\Gamma_{\rm tot}\), taking \(N_0\to\infty\) leaves the
postloading state in its initial basin with probability approaching one.}
\sentblack{Equation~\eqref{eq:end-loading-t50-exponent} is the finite-\(\eta\)
escape-time counterpart of the inverse gap in
Eq.~\eqref{eq:end-gap-exponent}: the former governs switching at fixed
\(\eta>\eta_c^{\mathrm{o}}\), while the latter governs relaxation at
rate coexistence, and their time-scale exponents coincide as
\(\eta\downarrow\eta_c^{\mathrm{o}}\).}
\sentblack{The exact recursions, MFPT construction, quantile proof, and
finite-size rate--gap validation are given in
\smref[app:intersector-coherence-closure]{SM, Sec.~VI} and
\smref[fig:app-rate-gap-validation]{Fig.~\ref*{fig:app-rate-gap-validation}}
~\cite{Supplemental}.}

\clearpage
\onecolumngrid
\setcounter{equation}{0}
\renewcommand{\theequation}{S\arabic{equation}}
\renewcommand{\theHequation}{supplement.S\arabic{equation}}
\setcounter{figure}{0}
\renewcommand{\thefigure}{S\arabic{figure}}
\renewcommand{\theHfigure}{supplement.S\arabic{figure}}
\providecommand{\vthirteen}[1]{#1}
\ifdefined\singlesupplement
  \newenvironment{smwidetable}{\begin{table}[t]}{\end{table}}
  \newenvironment{smwidefigure}{\begin{figure}}{\end{figure}}
  \newenvironment{smbranchfigure}{\begin{figure}[t]}{\end{figure}}
  \newenvironment{smfullfigure}[1][]{\begin{figure}[#1]}{\end{figure}}
  \newcommand{\smcompactfigurewidth}{0.68\textwidth}
  \newcommand{\smwidefigurewidth}{0.86\textwidth}
  \newcommand{\smfullplotwidth}{0.96\textwidth}
  \newcommand{\smrobustplotwidth}{0.96\textwidth}
  \newcommand{\smpairedfigurebarrier}{}
\else\ifdefined\allblackversion
  \newenvironment{smwidetable}{\begin{table*}[t]}{\end{table*}}
  \newenvironment{smwidefigure}{\begin{figure*}}{\end{figure*}}
  \newenvironment{smbranchfigure}{\begin{figure*}[t]}{\end{figure*}}
  \newenvironment{smfullfigure}[1][]{\begin{figure*}[#1]}{\end{figure*}}
  \newcommand{\smcompactfigurewidth}{0.52\textwidth}
  \newcommand{\smwidefigurewidth}{0.74\linewidth}
  \newcommand{\smfullplotwidth}{1.5\columnwidth}
  \newcommand{\smrobustplotwidth}{1.45\columnwidth}
  \newcommand{\smpairedfigurebarrier}{}
\else
  \newenvironment{smwidetable}{\begin{table*}[t]}{\end{table*}}
  \newenvironment{smwidefigure}{\begin{figure}}{\end{figure}}
  \newenvironment{smbranchfigure}{\begin{figure}[t]}{\end{figure}}
  \newenvironment{smfullfigure}[1][]{\begin{figure}[#1]}{\end{figure}}
  \newcommand{\smcompactfigurewidth}{0.66\columnwidth}
  \newcommand{\smwidefigurewidth}{0.75\columnwidth}
  \newcommand{\smfullplotwidth}{0.96\columnwidth}
  \newcommand{\smrobustplotwidth}{0.96\columnwidth}
  \newcommand{\smpairedfigurebarrier}{}
\fi\fi
\ifdefined\singlesupplement
  \newcommand{\smjoinrow}{\qquad}
  \newcommand{\smjoincontinuation}[1]{#1}
\else
  \newcommand{\smjoinrow}{\\}
  \newcommand{\smjoincontinuation}[1]{\\&#1}
\fi
\renewcommand{\topfraction}{0.95}
\renewcommand{\dbltopfraction}{0.95}
\renewcommand{\textfraction}{0.05}
\renewcommand{\floatpagefraction}{0.90}
\renewcommand{\dblfloatpagefraction}{0.90}
\setlength{\textfloatsep}{6pt plus 2pt minus 1pt}
\setlength{\dbltextfloatsep}{6pt plus 2pt minus 1pt}
\setlength{\intextsep}{6pt plus 2pt minus 1pt}
\setlength{\floatsep}{6pt plus 2pt minus 1pt}
\setlength{\abovedisplayskip}{2pt plus 1pt minus 1pt}
\setlength{\belowdisplayskip}{2pt plus 1pt minus 1pt}
\setlength{\abovedisplayshortskip}{0pt plus 1pt}
\setlength{\belowdisplayshortskip}{0pt plus 1pt}
\setlength{\jot}{2pt}
\newcommand{\smcompactlist}{%
  \setlength{\itemsep}{0pt}%
  \setlength{\parsep}{0pt}%
  \setlength{\topsep}{2pt}%
  \setlength{\partopsep}{0pt}}
\section*{SUPPLEMENTAL MATERIAL}
\label{app:mean-field-symmetry-limits}

\hyperref[app:finite-uniqueness]{Sections~I}--
\hyperref[app:parity-thermodynamic-limits]{III} establish the finite-size and
thermodynamic results. \hyperref[app:mean-field-finite-steady]{Secs.~IV}--
\hyperref[app:general-mechanism]{VII} address mean-field limits, the full
Liouvillian gap, physical loading, and generalized jumps.

\subsection{I. Symmetries and finite-size steady-state uniqueness}
\label{app:finite-uniqueness}

\subsubsection{\textbf{I.A. Strong symmetry and the relevant Hilbert space}}
\label{app:strong-symmetry-hilbert}

A Hermitian operator \(\hat Q\) generates a strong symmetry of a Lindblad
generator when \cite{BucaProsen2012}
\begin{equation}
[\hat Q,\hat H]=0,~
[\hat Q,\hat L_\mu]=[\hat Q,\hat L_\mu^\dagger]=0
\quad\text{for every }\mu .
\label{eq:app-strong-symmetry-definition}
\end{equation}
Its eigenspaces are then dynamically invariant symmetry sectors. In finite
dimension, every invariant symmetry sector supports at least one steady state, so a
nontrivial strong symmetry generally precludes unconditional uniqueness on the
full Hilbert space.

For \(N\) spin-\(1/2\) constituents, define
\(\hat{\boldsymbol S}=\sum_i\hat{\boldsymbol\sigma}^{(i)}/2\).  The model is
generated by \(\hat H=\omega_0\hat S_z\),
\(\hat L_+=\sqrt{2\Gamma_+/N}\,\hat S_+\), and
\(\hat L_-=\sqrt{8\Gamma_-/N^3}\,\hat S_-\hat S_z\), through
\(\dot{\hat\rho}=\mathcal L[\hat\rho]
=-i[\hat H,\hat\rho]+\sum_{\mu=\pm}\mathcal D[\hat L_\mu]\hat\rho\), with
\(\mathcal D[\hat L]\hat\rho=\hat L\hat\rho\hat L^\dagger
-\{\hat L^\dagger\hat L,\hat\rho\}/2\).
For this collective-spin model,
\begin{equation}
[\hat{\boldsymbol S}^{\,2},\hat H]=0,~
[\hat{\boldsymbol S}^{\,2},\hat L_\pm]=[\hat{\boldsymbol S}^{\,2},\hat L_\pm^\dagger]=0 .
\label{eq:app-strong-symmetry}
\end{equation}
Different total-spin sectors and their multiplicity spaces therefore do not
mix. More explicitly, the Hilbert space of \(N\) spins decomposes as
\begin{equation}
\hspace{-2em}\mathcal H_N=\bigoplus_S
 (\mathcal R_S\otimes\mathcal M_{S,N}), ~\hat S_a=\bigoplus_S
 (\hat S_a^{(S)}\otimes\hat I_{\mathcal M_{S,N}}),
 \label{eq:app-spin-multiplicity-decomposition}
\end{equation}
where \(a\in\{x,y,z\}\), \(\hat S_a^{(S)}\) acts on \(\mathcal R_S\), and
\(\mathcal M_{S,N}\) is the multiplicity space. The maximal sector has
\(S=N/2\) and \(\dim\mathcal M_{N/2,N}=1\).
The static calculations in
\hyperref[app:population-derivation]{Secs.~II}--%
\hyperref[app:liouvillian-gap-method]{V} specialize to this
permutation-symmetric representation, whereas the physical loading protocol in
\hyperref[app:physical-multisector-loading]{Sec.~VI} retains every populated
total-spin sector. The uniqueness result below applies
to the reduced Liouvillian on every irreducible representation
\(\mathcal R_S\), not to the full isotypic space
\(\mathcal R_S\otimes\mathcal M_{S,N}\).  Collective operators act trivially
on \(\mathcal M_{S,N}\), so multiplicity states remain noiseless and the full
\(2^N\)-dimensional problem is not unconditionally unique.

\subsubsection{\textbf{I.B. Weak \texorpdfstring{$U(1)$}{U(1)} covariance}}
\label{app:weak-u1-covariance}

Weak symmetry acts in Liouville space rather than partitioning Hilbert space.
For a unitary representation $\hat U_g$, define
$\mathcal U_g[\hat X]=\hat U_g\hat X\hat U_g^\dagger$. The dynamics is weakly symmetric, or
covariant, when
\begin{equation}
\begin{aligned}
\relax[\mathcal L,\mathcal U_g]=0,
~\text{i.e.,}~\mathcal L[\hat U_g\hat X\hat U_g^\dagger]
=\hat U_g\mathcal L[\hat X]\hat U_g^\dagger
\quad\text{for every }\hat X.
\end{aligned}
\label{eq:app-weak-symmetry-definition}
\end{equation}
This condition maps steady states to steady states, but by itself
neither proves uniqueness nor requires degeneracy.
For $\hat U_\varphi=e^{-i\varphi \hat S_z}$,
\begin{equation}
\hat U_\varphi \hat L_+\hat U_\varphi^\dagger=e^{-i\varphi}\hat L_+,
~
\hat U_\varphi \hat L_-\hat U_\varphi^\dagger=e^{i\varphi}\hat L_-.
\label{eq:app-jump-rotation}
\end{equation}
Because $\hat U_\varphi \hat H\hat U_\varphi^\dagger=\hat H$ and
$\mathcal D[e^{i\alpha}\hat L]=\mathcal D[\hat L]$, Eq.~\eqref{eq:app-weak-symmetry-definition}
holds. This is weak, not strong, $U(1)$ symmetry because
$[\hat U_\varphi,\hat L_\pm]\neq0$. Equivalently, its generator
$\mathcal Q[\hat X]=[\hat S_z,\hat X]$ commutes with the Liouvillian,
$[\mathcal L,\mathcal Q]=0$.
Within a fixed irreducible representation, the Dicke operators transform as
\begin{equation}
\mathcal U_\varphi[|S,m\rangle\langle S,n|]
=e^{-i\varphi(m-n)}|S,m\rangle\langle S,n|,
\label{eq:app-coherence-order}
\end{equation}
so Liouville space decomposes into coherence-order blocks $\nu=m-n$. The
$\nu=0$ block contains populations, while $\nu\neq0$ contains coherences. This
block decomposition does not conserve an individual Dicke index $m$: the
jumps still move populations along the Dicke ladder.
Once uniqueness on \(\mathcal R_S\) has been established, weak covariance
forces its unique representation-reduced state to inherit the symmetry,
\begin{equation}
\begin{aligned}
\hat U_\varphi\hat\rho_{\rm ss}^{(N,S)}\hat U_\varphi^\dagger=\hat\rho_{\rm ss}^{(N,S)},~
\hat\rho_{\rm ss}^{(N,S)}
=\sum_{m=-S}^{S}p_m^{(N,S)}|S,m\rangle\langle S,m|.
\end{aligned}
\label{eq:app-unique-u1-state}
\end{equation}
Here the labels \(N\) and \(S\) distinguish the physical system size,
which fixes the Kac normalization, from the total-spin representation,
which fixes the Dicke ladder.

\subsubsection{\textbf{I.C. Steady-state uniqueness in every irreducible spin representation}}
\label{app:finite-uniqueness-proof}

\hyperref[end:finite-uniqueness]{The Letter's End Matter, Sec.~A} gives
the open-interval uniqueness proof. Here we record the general-\(S\) rates
used below and classify the dissipative endpoints.

Fix the physical size \(N\) and restrict the collective generator to any
allowed irreducible representation \(\mathcal R_S\).  In this subsection,
\(S\) denotes the selected sector value, which need not equal \(N/2\).  The physical jump
normalization remains fixed by \(N/2\), while \(S\) specifies the Dicke
ladder.  On this representation,
\begin{equation}
\begin{aligned}
 \hat H_{N,S}=\omega_0\hat S_z^{(S)},~&\hat L_{+,N,S}=\sqrt{\frac{2\Gamma_+}{N}}\,\hat S_+^{(S)},~\hat L_{-,N,S}=\sqrt{\frac{8\Gamma_-}{N^3}}\,
 \hat S_-^{(S)}\hat S_z^{(S)}.
\end{aligned}
 \label{eq:app-general-S-generator}
\end{equation}
We parameterize the two rates by their total
\(\Gamma_{\rm tot}=\Gamma_++\Gamma_-\) and the dissipation imbalance
\(\eta=(\Gamma_--\Gamma_+)/\Gamma_{\rm tot}\).
We first consider \(\Gamma_{\rm tot}>0\) and the open interval
\(-1<\eta<1\), for which both jump channels are active.  The population rates along the
spin-\(S\) Dicke ladder are
\begin{align}
W_{m\to m+1}^{(N,S)}
 &=\left|\langle S,m+1|\hat L_{+,N,S}|S,m\rangle\right|^2=\frac{2\Gamma_+}{N}(S-m)(S+m+1),
 \quad -S\leq m<S,\notag\\
W_{m\to m-1}^{(N,S)}
 &=\left|\langle S,m-1|\hat L_{-,N,S}|S,m\rangle\right|^2=\frac{8\Gamma_-}{N^3}m^2(S+m)(S-m+1),
 \quad -S<m\leq S.
 \label{eq:app-general-S-rates}
\end{align}
All other off-diagonal population rates vanish. Their positive normalization
factors depend on the physical size \(N\) but do not change the directed
connectivity used in the proof of
Eq.~\eqref{eq:end-finite-uniqueness}.

At the dissipative endpoints one jump channel vanishes, so the open-interval
argument no longer applies. Table~\ref{tab:app-finite-uniqueness-summary}
gives the complete classification on one irreducible spin-\(S\)
representation, including the exceptional integer-\(S\) dark subspace at
\(\eta=1\).

\begin{smwidetable}
\caption{Finite-size steady states in one spin-\(S>0\) representation. Kernel
dimensions refer to \(\mathcal B(\mathcal R_S)\), excluding multiplicities.
Weak \(U(1)\) covariance holds throughout.  The one-dimensional \(S=0\)
representation is always unique.}
\label{tab:app-finite-uniqueness-summary}
\centering
\footnotesize
\begin{tabular}{p{0.15\textwidth}p{0.07\textwidth}p{0.28\textwidth}p{0.34\textwidth}p{0.07\textwidth}}
\hline\hline
Parameters & \(S\) parity & Closed class or dark structure & Time-independent normalized steady states & Kernel dim.\\
\hline
$\eta=-1$, finite $\omega_0$ & all \(S>0\) & Pump only. $|S,S\rangle$ is the unique closed dark state & Unique north-polar state $|S,S\rangle\langle S,S|$ & $1$\\
$-1<\eta<1$, finite $\omega_0$ & integer & $\{0,\ldots,S\}$ is the unique closed irreducible class. $m<0$ is transient & Unique Dicke-diagonal state supported on $m=0,\ldots,S$ & $1$\\
$-1<\eta<1$, finite $\omega_0$ & half-integer & The full half-integer ladder is the unique closed irreducible class & Unique Dicke-diagonal state supported on $m=-S,\ldots,S$ & $1$\\
$\eta=1$, finite $\omega_0$ & half-integer & $|S,-S\rangle$ is the unique closed dark state & Unique south-polar state $|S,-S\rangle\langle S,-S|$ & $1$\\
$\eta=1$, $\omega_0\neq0$ & integer & $|S,0\rangle$ and $|S,-S\rangle$ are dark. Their coherences rotate at $|\omega_0|S$ & Arbitrary probability mixtures of the two dark-state projectors & $2$\\
$\eta=1$, $\omega_0=0$ & integer & The two-dimensional dark subspace is decoherence free & Any normalized $2\times2$ density matrix on the dark subspace, including coherent superpositions & $4$\\
\hline\hline
\end{tabular}
\end{smwidetable}

The last two rows clarify the exceptional integer-$S$ endpoint.  At
$\eta=1$, one has $\Gamma_+=0$, so the only dissipative channel is
$\hat L_-\propto\hat S_-\hat S_z$.  For integer $S>0$, this jump annihilates
$|S,0\rangle$ at its internal $\hat S_z$ node and $|S,-S\rangle$ at the lower
Dicke-ladder endpoint. Their span is the two-dimensional dark subspace
\(\mathcal H_{\rm dark}=\operatorname{span}\{|S,0\rangle,|S,-S\rangle\}\).
The dissipator annihilates every operator supported on this subspace.  For
$\omega_0\neq0$, however, the Hamiltonian leaves the two dark-state
projectors stationary while their off-diagonal coherences rotate at frequency
$|\omega_0|S$. Hence only arbitrary probability mixtures are time independent
and $\dim\ker\mathcal L_{N,S}=2$.  When $\omega_0=0$, this relative phase
rotation also vanishes, and
\begin{equation}
\begin{aligned}
 \ker\mathcal L_{N,S}=\mathcal B(\mathcal H_{\rm dark})=\operatorname{span}\bigl(
 |S,0\rangle\langle S,0|,
 |S,-S\rangle\langle S,-S|, 
 |S,0\rangle\langle S,-S|,
 |S,-S\rangle\langle S,0|
 \bigr),~\dim\ker\mathcal L_{N,S}=4.
\end{aligned}
 \label{eq:app-integer-endpoint-kernel}
\end{equation}
Thus every positive, unit-trace $2\times2$ density matrix on
$\mathcal H_{\rm dark}$ is stationary, including coherent superpositions.
The kernel dimension counts four independent operators, not four isolated
physical steady states: the normalized steady states form a continuous convex
set. This endpoint degeneracy does not contradict
Eq.~\eqref{eq:end-finite-uniqueness}, whose uniqueness statement requires
$-1<\eta<1$.

The table makes explicit why weak symmetry cannot determine uniqueness: the
same weak $U(1)$ covariance accompanies kernel dimensions $1$, $2$, and $4$.
Only when the kernel is one dimensional does covariance require the unique
steady state itself to be $U(1)$ invariant.

\subsection{II. Exact finite-size steady state}
\label{app:population-derivation}
Specializing Eq.~\eqref{eq:app-unique-u1-state} to the maximal representation
\(S=N/2\), let
$\Lambda_{S}=\{-S,-S+1,\ldots,S\}$ denote its Dicke lattice, and write the
diagonal representation-reduced state as
$\hat\rho^{(S)}=\sum_{m\in\Lambda_{S}}p_m^{(S)}|S,m\rangle\langle S,m|$.
Here and below, explicit time arguments are suppressed. The stationary state
is denoted by $\hat\rho_{\rm ss}^{(S)}$.  Because both
\(\hat H=\omega_0\hat S_z\) and \(\hat\rho^{(S)}\) are diagonal in the Dicke
basis, the Hamiltonian does not affect the populations. Thus the population
dynamics is generated entirely by the dissipators.
Using the Dicke ladder identities
\begin{equation}
\begin{aligned}
\hat S_+|S,m\rangle
 &=\sqrt{(S-m)(S+m+1)}|S,m+1\rangle,\smjoinrow
\hat S_-|S,m\rangle
 &=\sqrt{(S+m)(S-m+1)}|S,m-1\rangle.
\end{aligned}
\label{eq:app-dicke-ladders}
\end{equation}
the linear pump gives the upward transition rate
\begin{equation}
\begin{aligned}
W_m^+&=W_{m\to m+1}
 =\left|\langle S,m+1|\hat L_+|S,m\rangle\right|^2
 \smjoincontinuation{=}
 \frac{\Gamma_+}{S}(S-m)(S+m+1),
 \quad -S\leq m<S.
\end{aligned}
\label{eq:app-upward-rate}
\end{equation}
For the ordered nonlinear loss, the rightmost $\hat S_z$ supplies the
additional amplitude (m), giving
\begin{equation}
\begin{aligned}
W_m^-&=W_{m\to m-1}
 =\left|\langle S,m-1|\hat L_-|S,m\rangle\right|^2
 \smjoincontinuation{=}
 \frac{\Gamma_-}{S^3}m^2(S+m)(S-m+1),
 \quad -S<m\leq S.
\end{aligned}
\label{eq:app-downward-rate}
\end{equation}
The internal factor $\hat S_z$ produces the factor $m^2$. Reversing the ordering to
$\hat S_z\hat S_-$ would replace $m$ in the amplitude by $m-1$ and move the deleted
bond, thereby changing the steady-state problem.

To derive the population equation, write the two jump actions as
$\hat L_\pm|S,n\rangle=a_n^\pm|S,n\mathbin{\pm}1\rangle$, with
$|a_n^\pm|^2=W_n^\pm$. For the diagonal state defined above, projection gives
\begin{equation}
\begin{aligned}
\langle S,m|\hat L_\pm\hat\rho^{(S)} \hat L_\pm^\dagger|S,m\rangle
 =W_{m\mp1}^\pm p_{m\mp1}^{(S)},~\frac12\langle S,m|\{\hat L_\pm^\dagger \hat L_\pm,\hat\rho^{(S)}\}|S,m\rangle
 =W_m^\pm p_m^{(S)}.
\end{aligned}
\label{eq:app-jump-projection}
\end{equation}
Recycling inflow and anticommutator outflow give
\begin{equation}
\begin{aligned}
\langle S,m|\mathcal D[\hat L_+]\hat\rho^{(S)}|S,m\rangle~ =W_{m-1}^+p_{m-1}^{(S)}-W_m^+p_m^{(S)},~
\langle S,m|\mathcal D[\hat L_-]\hat\rho^{(S)}|S,m\rangle~ =W_{m+1}^-p_{m+1}^{(S)}-W_m^-p_m^{(S)}.
\end{aligned}
\label{eq:app-channel-populations}
\end{equation}
Summing the channels gives
\begin{equation}
\dot p_m^{(S)}=W_{m-1}^+p_{m-1}^{(S)}+W_{m+1}^-p_{m+1}^{(S)}
-\bigl(W_m^++W_m^-\bigr)p_m^{(S)}.
\label{eq:app-birth-death}
\end{equation}
Here $p_n^{(S)}=0$ outside $\Lambda_{S}$, and $W_{S}^+=W_{-S}^-=0$.  Summing
Eq.~\eqref{eq:app-birth-death} over the chain and shifting the two inflow
indices gives $\sum_m\dot p_m^{(S)}=0$: every internal outflow is a neighboring
inflow, while neither endpoint leaks probability.
Because Eq.~\eqref{eq:app-birth-death} contains only the squared jump
amplitudes and no cross term between the two channels, the parity dependence
is not caused by destructive dark-state interference. It arises because the
integer Dicke lattice samples the loss zero $W_0^-=0$, whereas the
half-integer lattice does not.

Define the net current across the bond $m\to m+1$ as
\(I_m=W_m^+p_m^{(S)}-W_{m+1}^-p_{m+1}^{(S)}\).
The master equation becomes $\dot p_m^{(S)}=I_{m-1}-I_m$. At steady state,
$I_{m-1}=I_m$, so every bond current equals a constant $I_{\rm ss}$. It does
not by itself set that constant to zero.  The present Dicke lattice is a
finite open chain, however.  At the upper endpoint,
$I_{S}=p_S^{(S)}W_{S}^+=0$, since $W_{S}^+=0$ and no state exists beyond $S$.  The common
steady-state current must therefore vanish on every bond:
\begin{equation}
I_m=0,
~ p_m^{(S)}W_m^+=p_{m+1}^{(S)}W_{m+1}^-.
\label{eq:app-zero-current}
\end{equation}
Thus zero current follows from the steady-state condition together with the open-chain
boundary condition. It is not an additional detailed-balance assumption.
Substitution of Eqs.~\eqref{eq:app-upward-rate} and
\eqref{eq:app-downward-rate} gives
\begin{equation}
\begin{aligned}
&p_m^{(S)}\frac{\Gamma_+}{S}(S-m)(S+m+1)
\smjoincontinuation{=}
p_{m+1}^{(S)}\frac{\Gamma_-}{S^3}
(m+1)^2(S+m+1)(S-m).
\end{aligned}
\label{eq:app-rate-balance-expanded}
\end{equation}
Introduce the rate-ratio parameter
\(z_\eta=\sqrt{\Gamma_+/\Gamma_-}
=\sqrt{(1-\eta)/(1+\eta)}\) for \(-1<\eta<1\).
The two rates across the same bond contain the common ladder factor
$(S-m)(S+m+1)$, which cancels exactly.  This yields
\begin{equation}
\frac{p_{m+1}^{(S)}}{p_m^{(S)}}
=\frac{W_m^+}{W_{m+1}^-}
=\frac{S^2\Gamma_+}{(m+1)^2\Gamma_-}
=\frac{(S z_\eta)^2}{(m+1)^2}.
\label{eq:app-population-recursion}
\end{equation}
This ratio is used on the recurrent support where the denominator is nonzero.
On the integer negative half-ladder, use instead the equivalent zero-current
identity $(m+1)^2p_{m+1}^{(S)}=(Sz_\eta)^2p_m^{(S)}$;
no ratio of two vanishing stationary populations is implied.
For half-integer $S$, none of the allowed bonds encounters $m+1=0$.
Iterating Eq.~\eqref{eq:app-population-recursion} upward from $-S$ gives
\begin{equation}
\frac{p_m^{(S)}}{p_{-S}^{(S)}}
=(S z_\eta)^{2(m+S)}
\left[\frac{\Gamma(1-S)}
{\Gamma(m+1)}\right]^2,
~ S\in\mathbb Z+\tfrac12 .
\label{eq:app-half-integer-product}
\end{equation}
For integer $S$, the internal zero of $W_0^-$ removes the edge $0\to-1$.
The recurrent class is $m=0,1,\ldots,S$, and iteration from $m=0$ yields
\begin{equation}
p_m^{(S)}=p_0^{(S)}\frac{(S z_\eta)^{2m}}{(m!)^2},~ m\geq0;
~ p_m^{(S)}=0,~ m<0 .
\label{eq:app-integer-product}
\end{equation}
Both cases are represented by the same unnormalized weight
$(S z_\eta)^{2m}/\Gamma^2(m+1)$. For integer $S$, the reciprocal Gamma
function vanishes at every negative integer, removing the transient
half-chain, but remains finite at negative half-integers. Normalization gives
\begin{align}
w_m=\frac{(S z_\eta)^{2m}}{\Gamma^2(m+1)},~\mathcal Z_{\rm tot}=\sum_{m=-S}^{S}w_m,~p_m^{(S)}=\frac{w_m}{\mathcal Z_{\rm tot}},~
\hat\rho_{\rm ss}^{(S)}&=\sum_{m=-S}^{S}p_m^{(S)}|S,m\rangle\langle S,m|.
\label{eq:app-unified-population}
\end{align}
This proves the unified expression used in Eq.~\eqref{eq:unified-population}.

The finite-size Husimi-$Q$ function used in
Fig.~\ref{fig:parity-stationary-overview} is
$Q_S(\theta,\phi)=(2S+1)\langle\theta,\phi|
\hat\rho_{\rm ss}^{(S)}|\theta,\phi\rangle/(4\pi)$, normalized as
$\int d\Omega\,Q_S=1$, where
$|\theta,\phi\rangle=e^{-i\phi\hat S_z}e^{-i\theta\hat S_y}|S,S\rangle$
is a spin-coherent state.  Weak $U(1)$ invariance makes the exact steady state
Dicke diagonal and hence $Q_S$ independent of $\phi$.  Substituting
Eq.~\eqref{eq:app-unified-population} and expanding the coherent-state kernel
gives the explicit binomial form
\begin{equation}
\begin{aligned}
 Q_S(\theta)&=\frac{2S+1}{4\pi}
 \sum_{m=-S}^{S}p_m^{(S)}\binom{2S}{S+m}
 \smjoincontinuation{\times}
 u^{S+m}(1-u)^{S-m},
 ~ u=\cos^2\frac{\theta}{2}.
\end{aligned}
 \label{eq:app-husimi-binomial}
\end{equation}

\subsection{III. Parity-selected thermodynamic limits}
\label{app:parity-thermodynamic-limits}

\paragraph{\textbf{Limit convention for Sec.~III.---}}
All limits in this section are taken along the spin sequence displayed beneath
the corresponding limit.  We write $f_S=O(g_S)$ when
$|f_S/g_S|$ stays bounded, and $f_S=o(g_S)$ when
$f_S/g_S\to0$, along that sequence.

\subsubsection{\textbf{III.A. Common north-boundary limit of both parities}}
\label{app:north-boundary-limit}

With the boundary coordinate \(k=S-m\), the common north-polar limit reduces
to the width criterion
\begin{equation}
\begin{aligned}
\langle k\rangle_S,\sigma_{k,S}=
\begin{cases}
O(1),&-1<\eta<0,\\
O(\sqrt S),&\eta=0,
\end{cases}~
\langle k\rangle_S,\sigma_{k,S}=o(S).
\end{aligned}
\label{eq:app-north-width-criterion}
\end{equation}

\paragraph{\textbf{1) North maximum and exact boundary weights.---}}
For $-1<\eta<0$, one has $z_\eta>1$. Equation~\eqref{eq:app-population-recursion}
applies equally to the unnormalized weights and gives
$w_{m+1}/w_m\geq z_\eta^2>1$ on every bond of the recurrent Dicke chain,
because $|m+1|\leq S$. This statement applies to the nonnegative recurrent
chain for integer $S$ and to the full half-integer chain for
$S\in\mathbb Z+\tfrac12$.  Thus, for both parities, the unique global maximum
is the north endpoint $m=S$.

To resolve localization near the upper edge of the Dicke ladder, introduce
the boundary coordinate $k=S-m$.  It counts the number of Dicke steps
below $m=S$ and introduces no new degree of freedom: a state localized near
the upper edge keeps $k$ fixed even though $m$ and $S$ both grow.  Setting
$m=S-k$ with fixed $k\in\mathbb N_0$ and iterating
Eq.~\eqref{eq:app-population-recursion} downward gives
\begin{equation}
\frac{w_{S-k}}{w_{S}}
=z_\eta^{-2k}\prod_{j=0}^{k-1}(1-\frac{j}{S})^2,
~
\lim_{\substack{S\to\infty\\2S\in\mathbb N}}
\frac{w_{S-k}}{w_S}=z_\eta^{-2k}.
\label{eq:app-north-boundary-ratios}
\end{equation}
The condition $2S\in\mathbb N$ takes the limit through all physical integer-
and half-integer-spin values and therefore contains both parity subsequences.

\paragraph{\textbf{2) Fixed \(-1<\eta<0\): geometric boundary layer.---}}
For fixed negative \(\eta\), define
\(r=z_\eta^{-2}=(1+\eta)/(1-\eta)\in(0,1)\).
\paragraph{\textbf{2a) Normalization and limiting law.---}}
For integer \(S\), the recurrent support \(m=0,\ldots,S\) corresponds to
\(k=0,\ldots,S\), whereas for half-integer \(S\), the full ladder
\(m=-S,\ldots,S\) corresponds to \(k=0,\ldots,2S\).  The normalized
probability is therefore
\begin{equation}
p_{S-k}^{(S)}
=
\begin{cases}
\displaystyle
\frac{w_{S-k}/w_S}{\sum_{\ell=0}^{S}w_{S-\ell}/w_S},
&S\in\mathbb Z,\\[3mm]
\displaystyle
\frac{w_{S-k}/w_S}{\sum_{\ell=0}^{2S}w_{S-\ell}/w_S},
&S\in\mathbb Z+\tfrac12.
\end{cases}
\label{eq:app-north-normalized-finite-S}
\end{equation}
Throughout either recurrent support, the exact adjacent ratio obeys
\begin{equation}
\begin{aligned}
\frac{w_{S-(\ell+1)}}{w_{S-\ell}}=r\left(1-\frac{\ell}{S}\right)^2\leq r,~
0\leq\ell<
\begin{cases}
S,&S\in\mathbb Z,\\
2S,&S\in\mathbb Z+\tfrac12.
\end{cases}
\end{aligned}
\label{eq:app-north-adjacent-bound}
\end{equation}
Iteration of this bound, together with
Eq.~\eqref{eq:app-north-boundary-ratios}, gives, for every fixed \(\ell\),
\begin{equation}
0\leq\frac{w_{S-\ell}}{w_S}\leq r^\ell,
~
\lim_{\substack{S\to\infty\\2S\in\mathbb N}}
\frac{w_{S-\ell}}{w_S}=r^\ell.
\label{eq:app-north-pointwise-bound}
\end{equation}
To place both parity sequences on the common index set
\(\ell\in\mathbb N_0\), define the summand to be zero for \(\ell>S\) when
\(S\) is integer and for \(\ell>2S\) when \(S\) is half-integer.
Equation~\eqref{eq:app-north-pointwise-bound} then supplies the summable
dominating sequence \(r^\ell\), since
\(\sum_{\ell=0}^{\infty}r^\ell<\infty\).  The dominated convergence theorem
therefore gives
\begin{equation}
\begin{aligned}
\lim_{\substack{S\to\infty\\S\in\mathbb Z}}
\sum_{\ell=0}^{S}\frac{w_{S-\ell}}{w_S}=
\lim_{\substack{S\to\infty\\S\in\mathbb Z+\frac12}}
\sum_{\ell=0}^{2S}\frac{w_{S-\ell}}{w_S}=\sum_{\ell=0}^{\infty}r^\ell
=\frac{1}{1-r}.
\end{aligned}
\label{eq:app-north-normalizer-limit}
\end{equation}
Substituting Eq.~\eqref{eq:app-north-normalizer-limit} into
Eq.~\eqref{eq:app-north-normalized-finite-S} gives, for every fixed
$k\in\mathbb N_0$,
\begin{equation}
\lim_{\substack{S\to\infty\\2S\in\mathbb N}}p_{S-k}^{(S)}
=(1-z_\eta^{-2})z_\eta^{-2k}=(1-r)r^k=\mathcal G_k.
\label{eq:app-north-boundary-layer}
\end{equation}
Here, \(\mathcal G_k=(1-r)r^k\) denotes the normalized geometric
boundary-layer distribution on \(k\in\mathbb N_0\).

\paragraph{\textbf{2b) Moments and width of the geometric layer.---}}
For any function \(f\) for which the sum converges, its expectation under
this limiting law is
\(\langle f(k)\rangle_{\mathcal G}=\sum_{k=0}^{\infty}f(k)\mathcal G_k\).
Using \(\sum_{k=0}^{\infty}r^k=(1-r)^{-1}\), the first moment is
\begin{equation}
\begin{aligned}
\langle k\rangle_{\mathcal G}
&=(1-r)\sum_{k=0}^{\infty}k r^k=(1-r)r\frac{d}{dr}
\sum_{k=0}^{\infty}r^k
\smjoincontinuation{=}
(1-r)r\frac{d}{dr}\frac{1}{1-r}=\frac{r}{1-r}.
\end{aligned}
\label{eq:app-north-boundary-mean}
\end{equation}
Applying the same operator twice gives the second moment,
\begin{equation}
\begin{aligned}
\langle k^2\rangle_{\mathcal G}
&=(1-r)\sum_{k=0}^{\infty}k^2r^k
=(1-r)\left(r\frac{d}{dr}\right)^2\frac{1}{1-r}
\smjoincontinuation{=}
(1-r)r\frac{d}{dr}
\frac{r}{(1-r)^2}=\frac{r(1+r)}{(1-r)^2}.
\end{aligned}
\label{eq:app-north-boundary-second-moment}
\end{equation}
Therefore,
\begin{equation}
\begin{aligned}
\operatorname{Var}_{\mathcal G}(k)
&=\langle k^2\rangle_{\mathcal G}
-\langle k\rangle_{\mathcal G}^2
\smjoincontinuation{=}
\frac{r(1+r)}{(1-r)^2}
-\frac{r^2}{(1-r)^2}=\frac{r}{(1-r)^2}.
\end{aligned}
\label{eq:app-north-boundary-variance}
\end{equation}

To connect these limiting moments to the finite-\(S\) steady state, define
\(\langle f(k)\rangle_S=\sum_k f(k)p_{S-k}^{(S)}\), where the sum runs over the parity-dependent support specified in
Eq.~\eqref{eq:app-north-normalized-finite-S}.  Since the denominator in that
equation is at least one, Eqs.~\eqref{eq:app-north-adjacent-bound} and
\eqref{eq:app-north-pointwise-bound} imply
\(0\leq p_{S-k}^{(S)}\leq r^k\).  The sequences
\(k r^k\) and \(k^2r^k\) are summable, so dominated convergence gives
\begin{equation}
\begin{aligned}
\lim_{\substack{S\to\infty\\2S\in\mathbb N}}
\langle k^j\rangle_S&=\langle k^j\rangle_{\mathcal G},
~\lim_{\substack{S\to\infty\\2S\in\mathbb N}}
\operatorname{Var}_S(k)&=\operatorname{Var}_{\mathcal G}(k)\quad j=1,2.
\end{aligned}
\label{eq:app-north-finite-moment-limits}
\end{equation}
The explicit limiting values are given in
Eqs.~\eqref{eq:app-north-boundary-mean}--\eqref{eq:app-north-boundary-variance}.
Here, \(\langle k\rangle_S\) is the mean number of Dicke steps below the upper
edge, while \(\sigma_{k,S}=\sqrt{\operatorname{Var}_S(k)}\) is the microscopic
width of the boundary layer.  For each fixed \(-1<\eta<0\), both approach
finite, \(S\)-independent limits.  As \(\eta\to-1^+\), \(r\to0\) and the
distribution contracts onto the uppermost Dicke level. As
\(\eta\to0^-\), \(r\to1\) and both limiting moments diverge.  Since
\(m=S-k\), the normalized magnetization satisfies
\begin{equation}
1-\frac{\langle m\rangle_S}{S}=O(S^{-1}),~
\operatorname{Var}_S\!\left(\frac{m}{S}\right)=O(S^{-2}).
\label{eq:app-north-normalized-concentration}
\end{equation}
Thus, for every fixed \(-1<\eta<0\), the steady state becomes macroscopically
concentrated at \(m/S=1\) while retaining an $O(1)$ microscopic boundary width.

\paragraph{\textbf{3) The endpoint \(\eta=0\): half-Gaussian boundary layer.---}}
At \(\eta=0\), one has \(r=1\), so the limiting geometric law in
Eq.~\eqref{eq:app-north-boundary-layer} is no longer normalizable and the
$O(1)$ boundary-layer limit does not apply.  It is replaced by a broader
scaling regime.  Equation~\eqref{eq:app-north-boundary-ratios} also shows
directly that \(w_{S-k}/w_S\leq1\), so the north endpoint remains a maximum.
We first determine the nonnegative-half-ladder profile and then show that the
extra negative-\(m\) support for half-integer \(S\) is asymptotically
negligible.

\paragraph{\textbf{3a) Nonnegative half-ladder.---}}
Let \(k=o(S^{2/3})\).  Expanding the logarithm of the exact product in
Eq.~\eqref{eq:app-north-boundary-ratios} gives
\begin{equation}
\begin{aligned}
\ln\frac{w_{S-k}}{w_{S}}
=2\sum_{j=0}^{k-1}\ln\left(1-\frac{j}{S}\right)=-\frac{k(k-1)}{S}+O\!\left(\frac{k^3}{S^2}\right),\ln\frac{w_{S-k}}{w_{S}}\leq-\frac{k(k-1)}{S},
~ 0\leq k\leq\lfloor S\rfloor.
\end{aligned}
\label{eq:app-eta-zero-boundary-expansion}
\end{equation}
Thus the characteristic boundary scale is $k=O(\sqrt S)$, on which
\begin{equation}
\frac{w_{S-k}}{w_S}
=\exp\!\left(-\frac{k^2}{S}\right)[1+o(1)],
~ k=O(\sqrt S).
\label{eq:app-eta-zero-relative-weight}
\end{equation}
To normalize the nonnegative half-ladder, we must sum these
relative weights over \(0\leq k\leq\lfloor S\rfloor\).  Set
\(x_k=k/\sqrt S\). Adjacent values are separated by
\(\Delta x=1/\sqrt S\), while the upper endpoint diverges.  On every bounded
\(x_k\) interval, Eq.~\eqref{eq:app-eta-zero-relative-weight} supplies the
Gaussian integrand \(e^{-x_k^2}\).  It remains to show that the region
\(k\gg\sqrt S\) carries no finite fraction of the rescaled normalization.
For \(k\geq2\), the global bound in
Eq.~\eqref{eq:app-eta-zero-boundary-expansion} and
\(k(k-1)\geq k^2/2\) give
\(w_{S-k}/w_S\leq e^{-k^2/(2S)}\).  For fixed \(A>0\), define the rescaled
tail beyond \(A\sqrt S\) and use the monotonicity of \(e^{-x^2/2}\) to obtain
\begin{equation}
\begin{aligned}
R_S(A)
=\frac{1}{\sqrt S}
\sum_{k=\lceil A\sqrt S\rceil}^{\lfloor S\rfloor}
\frac{w_{S-k}}{w_S}\leq\frac{1}{\sqrt S}
\sum_{k=\lceil A\sqrt S\rceil}^{\infty}
e^{-k^2/(2S)}\leq\frac{1}{\sqrt S}
+\int_A^\infty e^{-x^2/2}\,dx.
\end{aligned}
\label{eq:app-eta-zero-tail-bound}
\end{equation}
The Gaussian tail on the right vanishes as \(A\to\infty\). Hence
\begin{equation}
\lim_{A\to\infty}\limsup_{S\to\infty}R_S(A)=0.
\label{eq:app-eta-zero-tail-vanishing}
\end{equation}
Thus Eq.~\eqref{eq:app-eta-zero-relative-weight} determines the Riemann sum
on every bounded \(x_k\) interval, while
Eq.~\eqref{eq:app-eta-zero-tail-vanishing} shows that the omitted large-
\(x_k\) region contributes asymptotically nothing.  Sending first
\(S\to\infty\) and then \(A\to\infty\) gives the half-Gaussian normalization:
\begin{equation}
\sum_{k=0}^{\lfloor S\rfloor}\frac{w_{S-k}}{w_S}
=\frac{\sqrt{\pi S}}2[1+o(1)].
\label{eq:app-eta-zero-positive-normalizer}
\end{equation}

\paragraph{\textbf{3b) Half-integer negative half-ladder.---}}
For half-integer \(S\), the recursion at \(z_\eta=1\) is strictly increasing
throughout the negative half-ladder, so \(w_{-1/2}\) is its largest weight.
Stirling's formula and \(\Gamma(1/2)=\sqrt\pi\) give
\begin{align}
\frac{w_{-1/2}}{w_S}
&=\frac{\Gamma^2(S+1)}{\pi S^{2S+1}},
~\frac{w_{-1/2}}{w_S}=2e^{-2S}[1+O(S^{-1})],\notag\\
0\leq\frac{\sum_{m<0}w_m}{w_S}
&\leq\left(S+\frac12\right)\frac{w_{-1/2}}{w_S},~
\frac{\sum_{m<0}w_m}{w_S}
=O(Se^{-2S})=o(\sqrt S).
\label{eq:app-eta-zero-negative-suppression}
\end{align}
The negative half-ladder is therefore exponentially negligible compared with
the positive-side normalization, which grows as \(\sqrt S\).

\paragraph{\textbf{4) Common macroscopic limit.---}}
The positive-side sum in Eq.~\eqref{eq:app-eta-zero-positive-normalizer} is the
complete normalization for integer \(S\).  For half-integer \(S\), the
negative-half-ladder contribution in
Eq.~\eqref{eq:app-eta-zero-negative-suppression} must be added, but it is
exponentially smaller than the positive-side contribution.  Thus both parity
sequences have the same leading total normalization:
\begin{equation}
\mathcal Z_{\rm tot}
=\frac{\sqrt{\pi S}}2w_S[1+o(1)].
\label{eq:app-eta-zero-total-normalizer}
\end{equation}

Using the probability definition in Eq.~\eqref{eq:app-unified-population},
divide the local relative weight in
Eq.~\eqref{eq:app-eta-zero-relative-weight} by the common total normalization
in Eq.~\eqref{eq:app-eta-zero-total-normalizer}.  Both parity sequences then
share the normalized half-Gaussian boundary profile
\begin{equation}
p_{S-k}^{(S)}
=\frac{2}{\sqrt{\pi S}}e^{-k^2/S}[1+o(1)],
~ k=O(\sqrt S).
\label{eq:app-eta-zero-half-gaussian}
\end{equation}
whose first two centered scales are
\begin{equation}
\langle k\rangle_S=\sqrt{\frac S\pi}+o(\sqrt S),~
\operatorname{Var}_S(k)=S\left(\frac12-\frac1\pi\right)+o(S).
\label{eq:app-eta-zero-boundary-moments}
\end{equation}
The mode and the mean occupy distinct finite-size locations.  At \(\eta=0\),
the exact product in Eq.~\eqref{eq:app-north-boundary-ratios} gives an exact
degeneracy of the two uppermost weights, whereas
Eq.~\eqref{eq:app-eta-zero-boundary-moments} places the mean
at a distance proportional to \(\sqrt S\) below the north boundary:
\begin{equation}
\begin{aligned}
w_S&=w_{S-1}>w_{S-k},
~ k=2,3,\ldots,\smjoinrow
\langle m\rangle_S&=S-\langle k\rangle_S,~
S-\langle m\rangle_S=\sqrt{\frac S\pi}+o(\sqrt S).
\end{aligned}
\label{eq:app-eta-zero-mode-mean-separation}
\end{equation}
Thus the two uppermost Dicke levels are modal, whereas the mean is displaced
from the north boundary on the \(\sqrt S\) boundary-layer scale.  Both this
displacement and its standard deviation are subextensive relative to \(S\), so
\begin{equation}
\lim_{\substack{S\to\infty\\S\in\mathbb Z}}
\frac{\langle\hat S_z\rangle_{\rm ss}}{S}
=
\lim_{\substack{S\to\infty\\S\in\mathbb Z+1/2}}
\frac{\langle\hat S_z\rangle_{\rm ss}}{S}
=1,
~ -1<\eta\leq0.
\label{eq:app-north-thermodynamic-limit}
\end{equation}
The finite displacement and width for fixed \(-1<\eta<0\), and their
\(O(\sqrt S)\) growth at \(\eta=0\), are both subextensive, establishing the
common north-polar limit in Eq.~\eqref{eq:app-north-thermodynamic-limit}.
The endpoint $\eta=-1$ is the pure-pumping case discussed separately in
\hyperref[app:finite-uniqueness]{Sec.~I}.

\subsubsection{\textbf{III.B. Common SSO peak of both parities}}
\label{app:positive-interior-saddle}

For fixed $0<\eta<1$, choose a mesoscopic width
$\sqrt S\ll\ell_S\ll S$.  The first inequality contains the full Gaussian
core, while the second keeps the window local in $x=m/S$ and disjoint from
branches separated by $O(S)$ levels.  Any $\ell_S=S^\alpha$ with
$1/2<\alpha<1$ works. We use $\ell_S=S^{3/4}$.  For half-integer $S$, the
SSO and south-polar fixed-point (s-PFP) windows, together with their remainder,
are defined on the Dicke lattice
$\mathcal L_S=\{-S,-S+1,\ldots,S\}$ by
\begin{equation}
\begin{aligned}
\mathcal Z_{\rm SSO}
 =\sum_{\substack{m\in\mathcal L_S\\|m-Sz_\eta|\leq\ell_S}}w_m,~\mathcal Z_{\rm s\text{-}PFP}
 =\sum_{\substack{m\in\mathcal L_S\\0\leq m+S\leq\ell_S}}w_m,~
\mathcal Z_{\rm rest}
 =\mathcal Z_{\rm tot}-\mathcal Z_{\rm SSO}-\mathcal Z_{\rm s\text{-}PFP}.
\end{aligned}
\label{eq:app-branch-window-decomposition}
\end{equation}
The two windows are disjoint for sufficiently large $S$.  The same
definition of \(\mathcal Z_{\rm SSO}\) applies to integer $S$, whereas the s-PFP
window then has zero weight.

\paragraph{\textbf{1) Mode location and local Gaussian profile.---}}
By Eq.~\eqref{eq:app-population-recursion}, the positive-$m$ weights grow for
$m+1<Sz_\eta$ and decrease for $m+1>Sz_\eta$.  Accordingly, the positive-$m$
mode $m_+(S)$ satisfies $m_+(S)=Sz_\eta+O(1)$.  To obtain its Gaussian width
and exponentially large branch weight, we convert the exact Gamma-function
weight into a smooth large-$S$ rate function.  At an allowed level $m=Sx$
with fixed $x>0$, one has
\(w_{Sx}=(Sz_\eta)^{2Sx}/\Gamma^2(Sx+1)\). Applying
\(\ln\Gamma(Sx+1)=Sx\ln(Sx)-Sx+\tfrac12\ln(2\pi Sx)+O(S^{-1})\)
then gives
\begin{equation}
\hspace{-2em}\ln w_{Sx}=2S\psi_+(x)+O(\ln S),~
\psi_+(x)=x\left[1+\ln\frac{z_\eta}{x}\right].
\label{eq:app-positive-rate-function}
\end{equation}
Since $\psi_+'(x)=\ln(z_\eta/x)$ and
$\psi_+''(x)=-1/x<0$, the unique positive-side maximum is
$x_+=z_\eta$, with $\psi_+(z_\eta)=z_\eta$.
A quadratic expansion around $m_+(S)$, with
$\delta m=m-m_+(S)$, gives
\begin{equation}
\frac{w_m}{w_{m_+(S)}}
=\exp\!\left[-\frac{\delta m^2}{Sz_\eta}\right][1+o(1)],
~ \delta m=O(\sqrt S).
\label{eq:app-sso-local-weight-ratio}
\end{equation}
\paragraph{\textbf{2) Discrete Laplace sum and integrated peak weight.---}}
We now sum the local weights without assuming that the approximation in
Eq.~\eqref{eq:app-sso-local-weight-ratio} is valid pointwise over the entire
mesoscopic window.  The calculation has four steps.

\begin{figure}[!htbp]
\centering
\begin{tikzpicture}[x=0.90cm,y=0.70cm,>=stealth]
  \node[anchor=west,font=\bfseries] at (0.05,1.55) {(a)};
  \draw[->] (0,0) -- (8.1,0) node[right] {$m$};
  \draw[dashed] (0.4,-0.35) -- (0.4,1.75);
  \draw[dashed] (7.7,-0.35) -- (7.7,1.75);
  \node[below] at (0.4,-0.35) {$Sz_\eta-\ell_S$};
  \node[below] at (7.7,-0.35) {$Sz_\eta+\ell_S$};
  \draw (3.0,-0.13) -- (3.0,0.13);
  \draw[fill=white] (3.0,0) circle (1.6pt);
  \node[below=2pt] at (3.0,0) {$Sz_\eta$};
  \draw (4.3,-0.13) -- (4.3,0.13);
  \fill (4.3,0) circle (1.7pt);
  \node[below=2pt] at (4.3,0) {$m_+(S)$};
  \draw (6.2,-0.13) -- (6.2,0.13);
  \fill (6.2,0) circle (1.7pt);
  \node[below=2pt] at (6.2,0) {$m$};
  \draw[<->] (3.0,0.72) -- (4.3,0.72)
    node[midway,above] {$d_S$};
  \draw[<->] (4.3,0.72) -- (6.2,0.72)
    node[midway,above] {$\delta m$};
  \draw[<->] (3.0,1.48) -- (6.2,1.48)
    node[midway,above] {$m-Sz_\eta=d_S+\delta m$};
  \node[align=center] at (4.05,-1.10)
    {original SSO window: $|m-Sz_\eta|\leq\ell_S$};

  \node[anchor=west,font=\bfseries] at (0.05,-1.72) {(b)};
  \node at (4.20,-1.70)
    {$u_{\delta m}=\delta m/\sqrt{Sz_\eta}$};
  \node at (4.20,-2.35)
    {$|\delta m|\leq A\sqrt{Sz_\eta}\Longleftrightarrow |u|\leq A$};
  \draw[->] (0.55,-3.35) -- (7.85,-3.35) node[right] {$u$};
  \draw[dashed] (0.90,-3.65) -- (0.90,-2.85);
  \draw[dashed] (7.20,-3.65) -- (7.20,-2.85);
  \node[below] at (0.90,-3.65) {$-A$};
  \node[below] at (7.20,-3.65) {$A$};
  \foreach \x in {1.20,1.75,2.30,2.85,3.40,3.95,4.50,5.05,5.60,6.15,6.70}
    {\fill (\x,-3.35) circle (1.25pt);}
  \node[below=2pt] at (3.95,-3.35) {$0~(m=m_+)$};
  \node[above=2pt] at (5.60,-3.35) {$u_{\delta m}$};
  \draw[<->] (5.60,-4.08) -- (6.15,-4.08)
    node[midway,below] {$\Delta u=1/\sqrt{Sz_\eta}$};
\end{tikzpicture}
\caption{\label{fig:app-sso-coordinate-recentering}
Coordinates for the SSO-window sum.  (a) Dashed lines bound the window around
$Sz_\eta$. Here $m_+(S)$ is the lattice mode and $\delta m$ its signed displacement
to $m$.  (b) Under $u=\delta m/\sqrt{Sz_\eta}$, the fixed core is $|u|\leq A$
and adjacent levels are separated by $\Delta u=1/\sqrt{Sz_\eta}$.}
\end{figure}
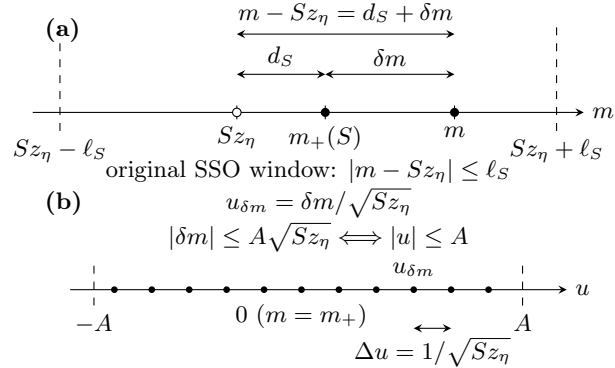

\textit{2a) Recenter the SSO window on the exact lattice mode.---}
Equation~\eqref{eq:app-sso-local-weight-ratio} describes the Gaussian weights
relative to the exact finite-$S$ lattice mode $m_+(S)$, whereas
$\mathcal Z_{\rm SSO}$ in Eq.~\eqref{eq:app-branch-window-decomposition} is
defined using a window centered at the continuous large-$S$ location
$Sz_\eta$.  To insert the local Gaussian weights into that exact window sum
without silently replacing one center by the other, we recenter the summation
coordinate.  Define the offset between the two centers by
$d_S=m_+(S)-Sz_\eta=O(1)$.  Since differences between allowed Dicke levels are
integers, the signed displacement $\delta m=m-m_+(S)$ belongs to $\mathbb Z$.
Figure~\ref{fig:app-sso-coordinate-recentering}(a) summarizes this recentering,
while panel (b) previews fixed-core scaling.

Following panel (a), the exact identity
$m-Sz_\eta=d_S+\delta m$ rewrites the original window in
Eq.~\eqref{eq:app-branch-window-decomposition} as
\begin{equation}
\begin{aligned}
\mathcal D_S
=\{\delta m\in\mathbb Z:|\delta m+d_S|\leq\ell_S\},~
\frac{\mathcal Z_{\rm SSO}}{w_{m_+(S)}}
=\sum_{\delta m\in\mathcal D_S}
\frac{w_{m_+(S)+\delta m}}{w_{m_+(S)}}.
\end{aligned}
\label{eq:app-sso-recentered-sum}
\end{equation}
The $O(1)$ shift $d_S$ changes each window endpoint by only finitely many
lattice sites and is negligible compared with both $\sqrt S$ and $\ell_S$.

\textit{2b) Evaluate a fixed Gaussian core.---}
Let $A>0$ be an arbitrary $S$-independent cutoff measured in units of the
Gaussian width.  The proof first takes $S\to\infty$ at fixed $A$ and only
afterward takes $A\to\infty$.  Because $\ell_S/\sqrt S\to\infty$, the
corresponding core $|\delta m|\leq A\sqrt{Sz_\eta}$ lies inside $\mathcal D_S$
for all sufficiently large $S$.

Panel (b) motivates the scaled coordinate
\begin{equation}
\begin{aligned}
u_{\delta m}&=\frac{\delta m}{\sqrt{Sz_\eta}},~
\Delta u=\frac{1}{\sqrt{Sz_\eta}},\smjoinrow
\mathcal U_S(A)
&=\{u_{\delta m}:\delta m\in\mathbb Z,~|u_{\delta m}|\leq A\}.
\end{aligned}
\label{eq:app-sso-scaled-core-coordinate}
\end{equation}
Adjacent values of $\delta m$ differ by one, so the scaled points
$u_{\delta m}$ form a uniform grid with mesh $\Delta u$.  In particular,
$\lim_{S\to\infty}\Delta u=0$, and the outermost grid points approach
$-A$ and $A$.

Next quantify the difference between the exact weight and its Gaussian
approximation by defining
\begin{equation}
\begin{aligned}
r_S(u_{\delta m})
&=e^{u_{\delta m}^2}
\frac{w_{m_+(S)+\delta m}}{w_{m_+(S)}}-1,\smjoinrow
\epsilon_S(A)&=\sup_{u\in\mathcal U_S(A)}|r_S(u)|,~
\lim_{S\to\infty}\epsilon_S(A)=0.
\end{aligned}
\label{eq:app-sso-core-relative-error}
\end{equation}
The first line is an exact definition: it rewrites the weight ratio as
$e^{-u_{\delta m}^2}[1+r_S(u_{\delta m})]$.  The final limit is precisely the
uniform $o(1)$ statement in Eq.~\eqref{eq:app-sso-local-weight-ratio} on the
fixed core $|u|\leq A$.

Substituting the preceding identity into the rescaled core weight $Q_S(A)$,
with $1/\sqrt{Sz_\eta}=\Delta u$, separates a Gaussian grid sum from its error:
\begin{equation}
\begin{aligned}
Q_S(A)=\frac{1}{\sqrt{Sz_\eta}}
\sum_{|\delta m|\leq A\sqrt{Sz_\eta}}
\frac{w_{m_+(S)+\delta m}}{w_{m_+(S)}},~
Q_S(A)
=\sum_{u\in\mathcal U_S(A)}\Delta u\,e^{-u^2}+R_S(A),~
R_S(A)
=\sum_{u\in\mathcal U_S(A)}\Delta u\,e^{-u^2}r_S(u).
\end{aligned}
\label{eq:app-sso-core-decomposition}
\end{equation}

We first show that $R_S(A)$ vanishes.  The number of points in the fixed core
and their total grid length satisfy
\begin{equation}
\begin{aligned}
N_S(A)&=|\mathcal U_S(A)|
=2\left\lfloor\frac{A}{\Delta u}\right\rfloor+1,\smjoinrow
N_S(A)\Delta u&\leq2A+\Delta u.
\end{aligned}
\label{eq:app-sso-core-grid-count}
\end{equation}
Applying the finite-sum triangle inequality
$|\sum_u a_u|\leq\sum_u|a_u|$ to
$a_u=\Delta u\,e^{-u^2}r_S(u)$ gives the first line below.  The subsequent
lines use $|r_S(u)|\leq\epsilon_S(A)$ and then $e^{-u^2}\leq1$:
\begin{equation}
\hspace{-2.5em}
\begin{aligned}
|R_S(A)|
\leq\sum_{u\in\mathcal U_S(A)}
\Delta u\,e^{-u^2}|r_S(u)|~
&\leq\epsilon_S(A)
\sum_{u\in\mathcal U_S(A)}\Delta u\,e^{-u^2}~
\leq\epsilon_S(A)
\sum_{u\in\mathcal U_S(A)}\Delta u
=\epsilon_S(A)N_S(A)\Delta u\\
&\leq(2A+\Delta u)\epsilon_S(A),~\lim_{S\to\infty}|R_S(A)|=0.
\end{aligned}
\label{eq:app-sso-core-error-bound}
\end{equation}
The last line follows because $A$ is fixed and both $\Delta u$ and
$\epsilon_S(A)$ have zero limits as $S\to\infty$.  Thus the growing number of
lattice points does not accumulate a finite relative error.

It remains only to evaluate the Gaussian part.  The function $e^{-u^2}$ is
continuous on $[-A,A]$, while the uniform grid has vanishing mesh and endpoints
approaching $-A$ and $A$.  Hence its defining Riemann-sum limit is
\begin{equation}
\begin{aligned}
\lim_{S\to\infty}
\sum_{u\in\mathcal U_S(A)}\Delta u\,e^{-u^2}
&=\int_{-A}^{A}e^{-u^2}\,du,\smjoinrow
\lim_{S\to\infty}Q_S(A)
&=\int_{-A}^{A}e^{-u^2}\,du.
\end{aligned}
\label{eq:app-sso-gaussian-core}
\end{equation}
The second line also uses the vanishing-error result in
Eq.~\eqref{eq:app-sso-core-error-bound}.

\textit{2c) Control the part of the window outside the fixed core.---}
The local formula alone does not control
$A\sqrt{Sz_\eta}<|\delta m|\leq\ell_S$.  Taylor expansion of the smooth rate
function in Eq.~\eqref{eq:app-positive-rate-function} about $x=z_\eta$, while
accounting for $m_+(S)-Sz_\eta=O(1)$, gives uniformly for
$|\delta m|\leq\ell_S=o(S)$,
\begin{equation}
\ln\frac{w_{m_+(S)+\delta m}}{w_{m_+(S)}}
=-\frac{\delta m^2}{Sz_\eta}
+O\!\left(\frac{|\delta m|}{S}
+\frac{|\delta m|^3}{S^2}\right).
\label{eq:app-sso-uniform-log-expansion}
\end{equation}
The cubic remainder relative to the quadratic term is at most
$O(\ell_S/S)=o(1)$.  The linear remainder is bounded and can be absorbed into
a constant prefactor.  Hence there exist $S$-independent constants $C,c>0$
such that
\begin{equation}
\frac{w_{m_+(S)+\delta m}}{w_{m_+(S)}}
\leq C\exp\!\left(-c\frac{\delta m^2}{S}\right),
~
\delta m\in\mathcal D_S,
\label{eq:app-sso-uniform-gaussian-bound}
\end{equation}
for all sufficiently large $S$.  Define the rescaled outer part of the SSO
window by
\begin{equation}
T_S(A)=\frac{1}{\sqrt{Sz_\eta}}
\sum_{\substack{\delta m\in\mathcal D_S\\
|\delta m|>A\sqrt{Sz_\eta}}}
\frac{w_{m_+(S)+\delta m}}{w_{m_+(S)}}.
\label{eq:app-sso-tail-definition}
\end{equation}
For every fixed finite $A$, the condition
$\lim_{S\to\infty}\ell_S/\sqrt S=\infty$ ensures that the core boundary
$A\sqrt{Sz_\eta}$ lies inside $\mathcal D_S$ for all sufficiently large $S$.
At any finite $S$, the sum in Eq.~\eqref{eq:app-sso-tail-definition} still ends
at the finite boundary of $\mathcal D_S$.  The integral comparison first
bounds it by a Gaussian integral with that finite scaled endpoint. Extending
the upper endpoint to infinity only enlarges this nonnegative upper bound.
The limits below are nested: $S\to\infty$ is taken first at fixed $A$, and only
then is $A\to\infty$ taken.  Thus no finite-$S$ window is required to contain
an infinite core.
Equation~\eqref{eq:app-sso-uniform-gaussian-bound} and an integral comparison
then give
\begin{equation}
\begin{aligned}
\limsup_{S\to\infty}T_S(A)
&\leq\frac{2C}{\sqrt{z_\eta}}
\int_{A\sqrt{z_\eta}}^\infty e^{-cu^2}\,du,\smjoinrow
\lim_{A\to\infty}\limsup_{S\to\infty}T_S(A)&=0.
\end{aligned}
\label{eq:app-sso-gaussian-tail}
\end{equation}
Thus the part of the mesoscopic window beyond any growing collection of
Gaussian widths contributes no finite fraction of the normalization.

\textit{2d) Combine the core and tail, and then take the logarithm.---}
First take $S\to\infty$ at fixed $A$ in
Eq.~\eqref{eq:app-sso-gaussian-core}. Then take $A\to\infty$ and use
Eq.~\eqref{eq:app-sso-gaussian-tail}.  This gives
\begin{equation}
\lim_{S\to\infty}
\frac{\mathcal Z_{\rm SSO}}
{w_{m_+(S)}\sqrt{Sz_\eta}}
=\int_{-\infty}^{\infty}e^{-u^2}\,du=\sqrt\pi.
\label{eq:app-sso-full-gaussian-sum}
\end{equation}
Moreover, $m_+(S)/S=z_\eta+O(S^{-1})$ and
$\psi_+(z_\eta)=z_\eta$, so Eq.~\eqref{eq:app-positive-rate-function} gives
\begin{equation}
\begin{aligned}
\ln w_{m_+(S)}
&=2S\psi_+\!\left(z_\eta+O(S^{-1})\right)+O(\ln S)
\smjoincontinuation{=}
2Sz_\eta+O(\ln S).
\end{aligned}
\label{eq:app-sso-mode-height}
\end{equation}
Including the Stirling prefactor refines this to
\begin{equation}
w_{m_+(S)}=\frac{e^{2Sz_\eta}}{2\pi Sz_\eta}[1+o(1)].
\label{eq:app-sso-mode-prefactor}
\end{equation}
Substitution into Eq.~\eqref{eq:app-sso-full-gaussian-sum} now yields
\begin{equation}
\begin{aligned}
\mathcal Z_{\rm SSO}
&=w_{m_+(S)}\sqrt{\pi Sz_\eta}[1+o(1)],\smjoinrow
\ln\mathcal Z_{\rm SSO}
&=\ln w_{m_+(S)}+\frac12\ln(\pi Sz_\eta)+o(1)
\smjoincontinuation{=}
2Sz_\eta+O(\ln S).
\end{aligned}
\label{eq:app-sso-normalization}
\end{equation}
\paragraph{\textbf{3) Normalized branch profile and thermodynamic limit.---}}
The branch-conditioned population
$p_{m\mid\mathrm{SSO}}^{(S)}=w_m/\mathcal Z_{\rm SSO}$ consequently satisfies
the following local Gaussian profile for $\delta m=O(\sqrt S)$:
\begin{equation}
p_{m\mid\mathrm{SSO}}^{(S)}
=\frac{1}{\sqrt{\pi Sz_\eta}}
\exp\!\left[-\frac{\delta m^2}{Sz_\eta}\right][1+o(1)].
\label{eq:app-ring-gaussian}
\end{equation}
Thus the normalized branch mean has the common integer- and
half-integer-sequence limit $z_\eta$, while
$\operatorname{Var}_{\rm SSO,S}(m)=Sz_\eta/2+o(S)$.
Its Dicke width is $O(\sqrt S)$, and its $U(1)$-invariant Husimi representation is the
phase-averaged steady-state counterpart of the finite-latitude SSO ring.
Its global weight in the half-integer-sequence steady state is determined in
\hyperref[app:odd-global-selection]{Sec.~III.D}.

\subsubsection{\textbf{III.C. Odd-parity south-boundary branch}}
\label{app:odd-south-boundary}

\paragraph{\textbf{1) Negative-side rate function and barrier.---}}
Only the half-integer sequence has nonzero stationary weight on a negative-$m$
branch, whereas for integer $S$, the corresponding reciprocal-Gamma zeros suppress
all levels below $m=0$.  For an allowed negative level of the half-integer
sequence, set $y=-m/S\in(0,1]$.  Then $Sy=-m$ is a positive half-integer, so
$|\sin(\pi Sy)|=1$.  Applying
\(\Gamma(1-Sy)\Gamma(Sy)=\pi/\sin(\pi Sy)\) to the reciprocal Gamma function
in Eq.~\eqref{eq:app-unified-population} therefore gives
\begin{equation}
w_{-Sy}
=\frac{(Sz_\eta)^{-2Sy}\Gamma^2(Sy)}{\pi^2}.
\label{eq:app-negative-reflected-weight}
\end{equation}
For fixed $y>0$, Stirling expansion at $Sy$ gives
\begin{equation}
\begin{aligned}
\ln w_{-Sy}
=-2Sy\ln(Sz_\eta)+2\ln\Gamma(Sy)-2\ln\pi~
=2Sy\left(\ln\frac{y}{z_\eta}-1\right)-\ln(Sy)~
\phantom{=}+\ln\frac{2}{\pi}+O((Sy)^{-1}).
\end{aligned}
\label{eq:app-negative-stirling-expanded}
\end{equation}
Consequently,
\begin{equation}
\hspace{-2.5em}
\begin{aligned}
\psi_-(y)=y\left(\ln\frac{y}{z_\eta}-1\right),~
\lim_{\substack{S\to\infty,~S\in\mathbb Z+1/2\\
m_S\in\mathcal L_S,~m_S<0,~-m_S/S\to y}}
\frac{1}{2S}\ln w_{m_S}=\psi_-(y).
\end{aligned}
\label{eq:app-negative-rate-function}
\end{equation}
The convergence is uniform on every compact interval
$y\in[\varepsilon,1]$ with fixed $\varepsilon>0$. No Stirling estimate at
$Sy=O(1)$ is being assumed.  The endpoint $y=0$ can be checked separately.
For an allowed sequence $m_S<0$ with $-m_S/S\to0$, set $t_S=-m_S$.  If $t_S$
remains bounded, Eq.~\eqref{eq:app-negative-reflected-weight} gives
$\ln w_{m_S}=O(\ln S)$. If $t_S$ diverges, Stirling's formula gives
$(2S)^{-1}\ln w_{m_S}=(t_S/S)[\ln(t_S/(Sz_\eta))-1]+o(1)$.  In both cases,
$\lim_{x\downarrow0}x\ln x=0$ implies
\begin{equation}
\lim_{\substack{S\to\infty,~S\in\mathbb Z+1/2\\
m_S\in\mathcal L_S,~m_S<0\\-m_S/S\to0}}
\frac{1}{2S}\ln w_{m_S}=0=\psi_-(0).
\label{eq:app-negative-zero-endpoint-rate}
\end{equation}

The derivatives
$\psi_-'(y)=\ln(y/z_\eta)$ and $\psi_-''(y)=1/y>0$ show that
$y=z_\eta$ is the unique minimum on $(0,1]$.  It is therefore the
large-deviation barrier on the negative half-ladder, not a competing maximum.
Because a convex function on $[0,1]$ can attain its maximum only at an
endpoint, and because
$\psi_-(0)=0<\psi_+(z_\eta)=z_\eta$, the only negative-side endpoint that can
compete exponentially with the positive SSO peak is $y=1$, namely $m=-S$.
Substituting $y=1$ into
Eq.~\eqref{eq:app-negative-stirling-expanded} gives its height including the
subexponential factor:
\begin{equation}
\hspace{-1em}
\begin{aligned}
\psi_-(1)&=-1-\ln z_\eta,\smjoinrow
\ln w_{-S}
&=-2S(1+\ln z_\eta)-\ln S+\ln\frac{2}{\pi}+O(S^{-1}).
\end{aligned}
\label{eq:app-south-height}
\end{equation}
This identifies the candidate south branch and its exponential weight, but it
does not yet determine whether that branch survives global normalization.
That comparison is deferred to \hyperref[app:odd-global-selection]{Sec.~III.D}.

\paragraph{\textbf{2) South-boundary normalization.---}}
To resolve the south boundary without reusing the north-boundary coordinate
of \hyperref[app:north-boundary-limit]{Sec.~III.A}, define $k_-=S+m\geq0$, the number of Dicke steps above
$m=-S$.  Since $k_-$ is integer, the south window contains
$0\leq k_-\leq K_S$, where $K_S=\lfloor\ell_S\rfloor$, $K_S\to\infty$,
and $K_S/S\to0$.
For $0\leq k_-\leq K_S$, iteration of the exact recursion gives the relative
boundary weight
\begin{equation}
\begin{aligned}
\mathcal R_{S,k_-}^{(-)}
=\frac{w_{-S+k_-}}{w_{-S}}
=\prod_{j=1}^{k_-}\frac{z_\eta^2}{(1-j/S)^2},~
\lim_{\substack{S\to\infty\\S\in\mathbb Z+1/2}}
\mathcal R_{S,k_-}^{(-)}&=z_\eta^{2k_-}
~\text{for every fixed }k_-\in\mathbb N_0.
\end{aligned}
\label{eq:app-south-boundary-ratios}
\end{equation}
The second line follows because every factor in the finite product has the
fixed-$k_-$ limit $z_\eta^2$.

To obtain a bound uniform over the growing window, choose a fixed $q$ with
$z_\eta^2<q<1$, and then choose $\delta\in(0,1)$ sufficiently small that
$z_\eta^2/(1-\delta)^2\leq q$.  The condition $K_S/S\to0$ ensures
$K_S/S\leq\delta$ for all
sufficiently large $S$.  Hence, for every $1\leq j\leq k_-\leq K_S$,
\begin{equation}
\begin{aligned}
\frac{z_\eta^2}{(1-j/S)^2}
&\leq\frac{z_\eta^2}{(1-K_S/S)^2}
\leq\frac{z_\eta^2}{(1-\delta)^2}\leq q,\smjoinrow
0\leq\mathcal R_{S,k_-}^{(-)}&\leq q^{k_-}.
\end{aligned}
\label{eq:app-south-uniform-geometric-bound}
\end{equation}
This is the summable, $S$-independent dominating sequence needed below.

To apply dominated convergence on a fixed index set, extend the truncated
relative weights to all $k_-\in\mathbb N_0$ by
\begin{equation}
\widetilde{\mathcal R}_{S,k_-}^{(-)}=
\begin{cases}
\mathcal R_{S,k_-}^{(-)},&0\leq k_-\leq K_S,\\
0,&k_->K_S.
\end{cases}
\label{eq:app-south-zero-extension}
\end{equation}
Because $K_S$ diverges, every fixed $k_-$ eventually lies inside the window.
Therefore the pointwise limit is $z_\eta^{2k_-}$, while
Eq.~\eqref{eq:app-south-uniform-geometric-bound} gives domination by
$q^{k_-}$.  Dominated convergence now yields
\begin{equation}
\begin{aligned}
&\frac{\mathcal Z_{\rm s\text{-}PFP}}{w_{-S}}=\sum_{k_-=0}^{\infty}\widetilde{\mathcal R}_{S,k_-}^{(-)},~
\lim_{\substack{S\to\infty\\S\in\mathbb Z+1/2}}
\frac{\mathcal Z_{\rm s\text{-}PFP}}{w_{-S}}
=\sum_{k_-=0}^{\infty}z_\eta^{2k_-}
=\frac{1}{1-z_\eta^2},~\\
&\mathcal Z_{\rm s\text{-}PFP}
=\frac{w_{-S}}{1-z_\eta^2}[1+o(1)],~\ln\mathcal Z_{\rm s\text{-}PFP}
=-2S(1+\ln z_\eta)-\ln S\phantom{=}+\ln\frac{2}{\pi(1-z_\eta^2)}+o(1).
\end{aligned}
\label{eq:app-pfp-normalization}
\end{equation}
The last line combines the converged geometric sum with the endpoint
asymptotics in Eq.~\eqref{eq:app-south-height}.

\paragraph{\textbf{3) Conditional law and macroscopic limit.---}}
Conditioned on the south window, the exact finite-$S$ distribution can be
placed on the fixed set $\mathbb N_0$ using the zero extension in
Eq.~\eqref{eq:app-south-zero-extension}:
\begin{equation}
p_{k_-\mid\mathrm{s\text{-}PFP}}^{(S)}
=\frac{\widetilde{\mathcal R}_{S,k_-}^{(-)}}
{\sum_{j=0}^{\infty}\widetilde{\mathcal R}_{S,j}^{(-)}}.
\label{eq:app-south-conditional-finite}
\end{equation}
For $0\leq k_-\leq K_S$, this definition is exactly
$w_{-S+k_-}/\mathcal Z_{\rm s\text{-}PFP}$, whereas for $k_->K_S$, it is zero.
It is therefore normalized for every finite $S$.  The pointwise convergence of
the numerator and the normalization limit in
Eq.~\eqref{eq:app-pfp-normalization} give, for every fixed
$k_-\in\mathbb N_0$,
\begin{equation}
\lim_{\substack{S\to\infty\\S\in\mathbb Z+1/2}}
p_{k_-\mid\mathrm{s\text{-}PFP}}^{(S)}
=\mathcal H_{k_-}=(1-z_\eta^2)z_\eta^{2k_-}.
\label{eq:app-south-boundary-layer}
\end{equation}
The finite-$S$ denominator in Eq.~\eqref{eq:app-south-conditional-finite} is at
least its $k_-=0$ term, which equals one.  Hence
$p_{k_-\mid\mathrm{s\text{-}PFP}}^{(S)}\leq q^{k_-}$ for all sufficiently
large $S$.  Since $\sum_{k_-=0}^{\infty}k_-^r q^{k_-}<\infty$ for $r=1,2$,
dominated convergence gives
\begin{equation}
\begin{aligned}
\lim_{\substack{S\to\infty\\S\in\mathbb Z+1/2}}
\langle k_-\rangle_{\rm s\text{-}PFP,S}
=\frac{z_\eta^2}{1-z_\eta^2},~
\lim_{\substack{S\to\infty\\S\in\mathbb Z+1/2}}
\operatorname{Var}_{\rm s\text{-}PFP,S}(k_-)
=\frac{z_\eta^2}{(1-z_\eta^2)^2}.
\end{aligned}
\label{eq:app-south-boundary-moments}
\end{equation}
Finally, $m=-S+k_-$ implies
\begin{equation}
\begin{aligned}
\lim_{\substack{S\to\infty\\S\in\mathbb Z+1/2}}
\frac{\langle m\rangle_{\rm s\text{-}PFP,S}}{S}
=-1,~
\lim_{\substack{S\to\infty\\S\in\mathbb Z+1/2}}
\operatorname{Var}_{\rm s\text{-}PFP,S}(m)
=\frac{z_\eta^2}{(1-z_\eta^2)^2}.
\end{aligned}
\label{eq:app-south-macroscopic-moments}
\end{equation}
The collapse to the south pole also holds in probability.  For every fixed
$\varepsilon>0$, Markov's inequality gives
\begin{equation}
\begin{aligned}
\Pr_{\rm s\text{-}PFP}^{(S)}
\left(\left|\frac{m}{S}+1\right|>\varepsilon\right)=\Pr_{\rm s\text{-}PFP}^{(S)}(k_->\varepsilon S)\leq\frac{\langle k_-\rangle_{\rm s\text{-}PFP,S}}{\varepsilon S},~
\lim_{\substack{S\to\infty\\S\in\mathbb Z+1/2}}
\Pr_{\rm s\text{-}PFP}^{(S)}
\left(\left|\frac{m}{S}+1\right|>\varepsilon\right)&=0.
\end{aligned}
\label{eq:app-south-probability-concentration}
\end{equation}
Thus this branch has $O(1)$ Dicke width and converges conditionally to the
south pole, in contrast to the $O(\sqrt S)$ width of the interior SSO branch.
Its stationary weight is determined by the global comparison in
\hyperref[app:odd-global-selection]{Sec.~III.D}.

\subsubsection{\textbf{III.D. Odd-parity branch selection and the exact threshold}}
\label{app:odd-global-selection}

\paragraph{\textbf{1) Exhaustion of the two branch windows.---}}
We now compare the three contributions defined in
Eq.~\eqref{eq:app-branch-window-decomposition}.  The SSO tail immediately
outside its window is suppressed relative to
Eq.~\eqref{eq:app-sso-normalization} by
$O(\exp[-c\ell_S^2/S])$ for some $c>0$.
For the south branch, choose a fixed
$\delta>0$ small enough that $z_\eta^2/(1-\delta)^2<q<1$, where $q$ is the
constant introduced in \hyperref[app:odd-south-boundary]{Sec.~III.C}.  The recursion makes the local tail
$\ell_S<k_-\leq\delta S$ an $O(q^{\ell_S})$ fraction of the south-window
normalization.  Both local tails vanish because
$\ell_S/\sqrt S$ and $\ell_S$ both diverge.

It remains to control the region macroscopically separated from both windows.
On the positive side, $\psi_+$ is strictly concave with its unique maximum at
$x=z_\eta$, whereas on the negative side, the convex function $\psi_-$ can maximize
only at $y=0$ or $y=1$, and $\psi_-(0)=0<\psi_+(z_\eta)$.  Hence any closed
region separated from the SSO peak and south endpoint has a strictly lower
large-deviation height than the larger of the two branch heights.  Combining
the local-tail and macroscopic-region bounds gives
\begin{equation}
\begin{aligned}
\mathcal Z_{\rm rest}
&=o(\mathcal Z_{\rm SSO}+\mathcal Z_{\rm s\text{-}PFP}),\smjoinrow
\mathcal Z_{\rm tot}
&=(\mathcal Z_{\rm SSO}+\mathcal Z_{\rm s\text{-}PFP})[1+o(1)].
\end{aligned}
\label{eq:app-rest-negligible}
\end{equation}
Although the local tails need not have distinct rates on the $S$ scale, their
integrated weights vanish. Equation~\eqref{eq:app-rest-negligible} therefore rules
out finite stationary weight in any third region.

\paragraph{\textbf{2) Competing exponential rates.---}}
Define the branch-resolved stationary rate potentials by
\begin{equation}
\begin{aligned}
g_{\rm SSO}
 =-\lim_{\substack{S\to\infty\\S\in\mathbb Z+1/2}}
\frac{\ln\mathcal Z_{\rm SSO}}{2S}=-z_\eta,~
g_{\rm s\text{-}PFP}
 =-\lim_{\substack{S\to\infty\\S\in\mathbb Z+1/2}}
\frac{\ln\mathcal Z_{\rm s\text{-}PFP}}{2S}=1+\ln z_\eta.
\end{aligned}
\label{eq:app-competing-branch-rates}
\end{equation}
The SSO value is unchanged along the integer-$S$ sequence, whereas the half-integer
restriction in Eq.~\eqref{eq:app-competing-branch-rates} places both quantities
on the odd sequence where they compete.
The discrete Laplace principle yields
\begin{equation}
g_{\rm tot}
=-\lim_{\substack{S\to\infty\\S\in\mathbb Z+1/2}}
\frac{\ln\mathcal Z_{\rm tot}}{2S}
=\min\!\left(g_{\rm SSO},g_{\rm s\text{-}PFP}\right).
\label{eq:app-odd-scaled-normalizer}
\end{equation}
The difference between the two branch rates is
\begin{equation}
\Delta g(\eta)=g_{\rm SSO}-g_{\rm s\text{-}PFP}
=-\bigl(1+z_\eta+\ln z_\eta\bigr).
\label{eq:app-intensive-weight-imbalance}
\end{equation}

\paragraph{\textbf{3) Normalized branch selection.---}}
To connect this rate comparison to the normalized steady state, define for
half-integer $S$ the finite-size branch probabilities
\(B_{\alpha}^{(S)}=\mathcal Z_{\alpha}/\mathcal Z_{\rm tot}\), with
\(\alpha\in\{\mathrm{SSO},\mathrm{s\text{-}PFP}\}\).
Their ratio satisfies
\begin{equation}
\frac{B_{\rm SSO}^{(S)}}{B_{\rm s\text{-}PFP}^{(S)}}
=\frac{\mathcal Z_{\rm SSO}}{\mathcal Z_{\rm s\text{-}PFP}}
=\exp[-2S\Delta g(\eta)+O(\ln S)].
\label{eq:app-branch-log-odds}
\end{equation}
Equations~\eqref{eq:app-rest-negligible} and
\eqref{eq:app-branch-log-odds} then imply
\begin{equation}
\begin{aligned}
\Delta g<0:\quad
\lim_{\substack{S\to\infty\\S\in\mathbb Z+1/2}}
(B_{\rm SSO}^{(S)},B_{\rm s\text{-}PFP}^{(S)})=(1,0),~
\Delta g>0:\quad
\lim_{\substack{S\to\infty\\S\in\mathbb Z+1/2}}
(B_{\rm SSO}^{(S)},B_{\rm s\text{-}PFP}^{(S)})=(0,1).
\end{aligned}
\label{eq:app-branch-probability-limits}
\end{equation}
Thus the sign of $\Delta g$ identifies the globally dominant branch.  This
comparison cannot be replaced by a local stability
analysis: the two peaks remain separated by $O(S)$ Dicke sites, and an expansion
around the SSO peak contains no information about the exponentially weighted
south endpoint.

\paragraph{\textbf{4) Exact threshold, critical selection, and equal-weight window.---}}
The thermodynamic rate-coexistence point is
\begin{equation}
z_c=-1-\ln z_c=W_0(e^{-1})\simeq0.2785,
~
\eta_c^{\mathrm{o}}=\frac{1-z_c^2}{1+z_c^2}\simeq0.8561.
\label{eq:app-odd-critical}
\end{equation}
Here $W_0$ denotes the principal branch of the Lambert $W$ function. This
notation is distinct from the transition rates $W_m^\pm$.
Displayed derived values are rounded.
At $\eta=\eta_c^{\mathrm{o}}$, the exponential rates are equal, while the
prefactors in Eqs.~\eqref{eq:app-sso-mode-prefactor},
\eqref{eq:app-sso-normalization}, and \eqref{eq:app-pfp-normalization} yield
\begin{equation}
\begin{aligned}
\frac{\mathcal Z_{\rm SSO}}{\mathcal Z_{\rm s\text{-}PFP}}
&=C_c\sqrt S[1+o(1)],\smjoinrow
C_c&=\frac{\sqrt\pi(1-z_c^2)}{4\sqrt{z_c}}
\simeq0.7746.
\end{aligned}
\label{eq:app-critical-prefactor-selection}
\end{equation}
Thus $B_{\rm SSO}^{(S)}\to1$ and $B_{\rm s\text{-}PFP}^{(S)}\to0$: at
coexistence, the polynomial prefactor selects the SSO branch. At finite $S$,
equal branch weight occurs at
\begin{equation}
\eta_{\rm eq}(S)=\eta_c^{\mathrm{o}}
+\frac{\ln S}{4S\Delta g'(\eta_c^{\mathrm{o}})}+O(S^{-1}),
\label{eq:app-finite-size-equal-weight-shift}
\end{equation}
where
$\Delta g'(\eta_c^{\mathrm{o}})=(1+z_c)/(1-(\eta_c^{\mathrm{o}})^2)>0$.
Hence $\eta_{\rm eq}(S)>\eta_c^{\mathrm{o}}$ for sufficiently large $S$.
A prescribed nontrivial branch mixture requires tuning inside this shrinking
equal-weight window.  For fixed $\eta$, the SSO branch is selected for
$0<\eta\leq\eta_c^{\mathrm{o}}$, whereas the s-PFP branch is selected for
$\eta_c^{\mathrm{o}}<\eta<1$.

\paragraph{\textbf{5) Selected odd-sequence thermodynamic state.---}}
Combining the branch-probability limits with
the conditional means obtained in Secs.~III.B and III.C gives
\begin{equation}
\lim_{\substack{S\to\infty\\S\in\mathbb Z+1/2}}
\frac{\langle \hat S_z\rangle_{\rm ss}}{S}
=\begin{cases}
 z_\eta,&0<\eta\leq\eta_c^{\mathrm{o}},\\
-1,&\eta_c^{\mathrm{o}}<\eta<1.
\end{cases}
\label{eq:app-odd-order-parameter}
\end{equation}
The equality case follows from Eq.~\eqref{eq:app-critical-prefactor-selection}.
The remainder bound then ensures that exactly one branch carries unit weight
at every fixed $0<\eta<1$.

\paragraph{\textbf{6) First-order classification.---}}
For every finite half-integer $S$, $\mathcal Z_{\rm tot}$ is a finite sum of
strictly positive analytic weights. The nonanalyticity appears only when the
thermodynamic limit turns the logarithm of the branch sum into the lower
envelope in Eq.~\eqref{eq:app-odd-scaled-normalizer}. Because
$w_m\propto z_\eta^{2m}$, the exact normalization obeys
\begin{equation}
\frac{\partial\ln\mathcal Z_{\rm tot}}{\partial\ln z_\eta}
=2\langle m\rangle,
~
m_{z,\infty}^{\mathrm{o}}
=-\frac{\partial g_{\rm tot}}{\partial\ln z_\eta},~
\eta\neq\eta_c^{\mathrm{o}}.
\label{eq:app-potential-polarization-conjugacy}
\end{equation}
At $\eta_c^{\mathrm{o}}$, the two analytic branch rates cross with unequal
slopes, so $g_{\rm tot}$ has a cusp. The order parameter equals $z_c$ at the
crossing and approaches $-1$ immediately above it, giving a polarization
discontinuity of $1+z_c$.

\subsubsection{\textbf{III.E. Parity-resolved thermodynamic-limit comparison}}
\label{app:husimi-thermodynamic-limits}

\paragraph{\textbf{Husimi-measure limits.---}}
The magnetization limits in the Letter specify only a first moment. Applying the population
limits established above to the exact finite-size Husimi representation in
Eq.~\eqref{eq:app-husimi-binomial} determines the complete macroscopic
phase-space measures.  Write
$x=\cos\theta$ and define
\begin{equation}
\begin{aligned}
\mu_{Q,S}^{a}(d\Omega)&=Q_S^{a}(\theta,\phi)\,d\Omega,\smjoinrow
\mu_{\rm ring,\eta}(d\Omega)
&=\frac{\delta(x-z_\eta)}{2\pi}\,dx\,d\phi,~
a\in\{\mathrm{e},\mathrm{o}\}.
\end{aligned}
\label{eq:app-husimi-measure-definitions}
\end{equation}
The measures $\mu_{Q,S}^{a}$ are normalized because
$d\Omega=\sin\theta\,d\theta\,d\phi$ and
$\int d\Omega\,Q_S^{a}=1$.  The weak limits are
\begin{equation}
\begin{aligned}
\mu_{Q,S}^{\mathrm{e}}
\Longrightarrow
\begin{cases}
\delta_{\rm north}(d\Omega),&-1<\eta\leq0,\\
\mu_{\rm ring,\eta}(d\Omega),&0<\eta<1,
\end{cases}~
\mu_{Q,S}^{\mathrm{o}}
\Longrightarrow
\begin{cases}
\delta_{\rm north}(d\Omega),&-1<\eta\leq0,\\
\mu_{\rm ring,\eta}(d\Omega),&0<\eta\leq\eta_c^{\mathrm{o}},\\
\delta_{\rm south}(d\Omega),&\eta_c^{\mathrm{o}}<\eta<1.
\end{cases}
\end{aligned}
\label{eq:app-husimi-weak-limits}
\end{equation}
Here $\delta_{\rm north}$ and $\delta_{\rm south}$ are unit point masses at
$\theta=0$ and $\theta=\pi$, respectively.  To see why these limits follow,
note that the binomial coherent-state kernel in
Eq.~\eqref{eq:app-husimi-binomial} has Dicke width $O(\sqrt S)$.
Convolving it with the SSO population of the same width therefore produces an
angular width $O(S^{-1/2})$ around $x=z_\eta$, whereas either
boundary layer has $S-|m|=o(S)$ and therefore collapses to the corresponding pole.
Thus the SSO becomes a phase-averaged ring, while the two PFP branches become
point masses at the corresponding poles.  These weak limits establish the
full phase-space counterpart of the parity-dependent magnetization selection.

\subsection{IV. Mean-field attractors, finite-size steady states, and
noncommuting limits}
\label{app:mean-field-finite-steady}

\subsubsection{\textbf{IV.A. Noncommuting long-time and thermodynamic limits}}
\label{app:order-of-limits}

Large-$S$ fixed-time factorization gives
\begin{equation}
\dot\theta=\frac{\Gamma_{\rm tot}}{2}\sin\theta
\big[(1+\eta)\cos^2\theta-(1-\eta)\big],
~ \dot\phi=\omega_0.
\label{eq:app-spherical-flow}
\end{equation}
For the SSO, $\cos\theta_{\rm SSO}=z_\eta$ and the azimuthal phase rotates at
$\omega_0$.

Let $\lambda_{\rm phase}(S)$ denote the leading Liouvillian eigenvalue in the
$\nu=\pm1$ coherence-order blocks associated with the first azimuthal harmonic of
the SSO. At every finite $S$, its real part defines the phase-diffusion rate
$D_\phi(S)=-\operatorname{Re}\lambda_{\rm phase}(S)>0$, with
\(\tau_\phi(S)=D_\phi^{-1}(S)=|\operatorname{Re}\lambda_{\rm phase}(S)|^{-1}\).
This sector-resolved decay rate equals the full Liouvillian gap only if the
phase mode is the globally slowest nonstationary mode.
Taking $t\to\infty$ first restores weak $U(1)$ symmetry. For the
even sequence at $0<\eta<1$, or the odd sequence at
$0<\eta\leq\eta_c^{\mathrm{o}}$, the subsequent large-$S$ limit is
\begin{equation}
\lim_{S\to\infty}Q_S(\theta,\phi)=
\frac{\delta(\theta-\theta_{\rm SSO})}{2\pi\sin\theta_{\rm SSO}},
\qquad \langle\hat S_+\rangle_{\rm ss}^{(S)}=0,
\qquad \cos\theta_{\rm SSO}=z_\eta.
\label{eq:app-static-ring}
\end{equation}
The limit is understood weakly as a probability measure. It is stationary,
not a state following a particular point around the classical orbit.
For the odd sequence at $\eta>\eta_c^{\mathrm{o}}$, the true stationary
measure instead converges to the south pole as in
Eq.~\eqref{eq:app-husimi-weak-limits}; a ring there is only metastable.

For an initial state localized on the SSO with a resolved phase and
$\omega_0\ne0$, if $S\to\infty$ is taken first, the phase-diffusion rate vanishes and
$\tau_\phi$ diverges. The initial phase remains resolved, and the subsequent
orbit has $m_z(t)=z_\eta$, $\theta(t)=\theta_{\rm SSO}$, and
$\phi(t)=\omega_0t+\phi_0$.
The noncommutativity is explicit in the phase-sensitive intensive observable
$s_{+,S}(t)=\langle\hat S_+(t)\rangle/S$. Defining
$s_{+,\infty}(t)=\lim_{S\to\infty}s_{+,S}(t)$, we obtain
\begin{equation}
\begin{aligned}
\lim_{S\to\infty}\left[\lim_{t\to\infty}s_{+,S}(t)\right]=0,~
s_{+,\infty}(t)=\sin\theta_{\rm SSO}\,
e^{i(\omega_0t+\phi_0)},~
\lim_{t\to\infty}s_{+,\infty}(t)\quad\text{does not exist}.
\end{aligned}
\label{eq:app-noncommuting-limits}
\end{equation}
Thus taking the long-time limit first gives the stationary Husimi measure in
Eq.~\eqref{eq:app-static-ring}, whereas taking $S\to\infty$ first gives the
phase-resolved orbit above, whose
pointwise long-time limit does not exist. The two limits are therefore
noninterchangeable without comparing density matrices on Hilbert spaces whose
dimensions vary with $S$.
For large finite half-integer $S$ at $\eta>\eta_c^{\mathrm{o}}$, starting
in the SSO basin, the time windows instead distinguish phase diffusion from
the much slower basin transfer:
\begin{equation}
\begin{aligned}
\tau_{\rm amp}\ll t\ll\tau_\phi(S)&:
\quad \text{phase-resolved SSO},\smjoinrow
\tau_\phi(S)\ll t\ll\tau_{\rm sw}(S)&:
\quad \text{phase-averaged metastable SSO},\\
t\gg\tau_{\rm sw}(S)&:
\quad \text{south-polar steady state}.
\end{aligned}
\label{eq:app-time-windows}
\end{equation}

In the stationary-ring regimes specified above, phase averaging gives the
limiting ring after $t\gg\tau_\phi$ without a subsequent dominant south branch.
This noncommutativity differs from the even--odd inequivalence in
\hyperref[eq:parity-resolved-magnetization-main]{the main text}.  The former
compares the $t$ and $N$ limits within one model sequence and originates from
a closing phase-diffusion rate, whereas the latter compares two steady-state size
sequences after $t\to\infty$ and originates from the Dicke-graph topology.

\subsubsection{\textbf{IV.B. Exact truncated Holstein--Primakoff map and the
limitation of a local expansion}}
\label{app:mean-field-attractors}

The distinction can be made explicit without abandoning bosonic variables.
Identify $|S,m\rangle$ with the number state $|n=S-m\rangle_{\rm b}$ and
restrict the bosonic occupation to $0\leq n\leq2S$.  On this truncated space,
the exact Holstein--Primakoff representation is
\begin{equation}
 \hat S_z=S-\hat n,
 ~
 \hat S_+=\sqrt{2S-\hat n}\,\hat a,
 ~
 \hat S_-=\hat a^\dagger\sqrt{2S-\hat n}.
 \label{eq:app-exact-truncated-hp}
\end{equation}
It reproduces the Dicke matrix elements and the finite cutoff at $n=2S$
exactly \cite{KonigHucht2021}, while retaining the half-spacing displacement
between integer- and half-integer-spin lattices.  In particular, the ordered
nonlinear loss becomes
\begin{equation}
 \hat L_-
 =\sqrt{\frac{\Gamma_-}{S^3}}\,
 \hat a^\dagger\sqrt{2S-\hat n}\,(S-\hat n),
 \label{eq:app-hp-ordered-loss}
\end{equation}
so its internal node is $n_0=S$ and the south boundary is $n_{\rm s}=2S$.
Because $n$ is integer valued, the node $n_0=S$ is an allowed occupation for
integer $S$ but lies halfway between two allowed occupations for half-integer
$S$.

For the SSO, $m_*=Sz_\eta+O(1)$ and
$n_*=S(1-z_\eta)+O(1)$. Writing $\hat n=n_*+\delta\hat n$ gives
$n_0-n_*=Sz_\eta+O(1)$ and $n_{\rm s}-n_*=S(1+z_\eta)+O(1)$.
Both distances are $O(S)$. By contrast, the SSO Gaussian profile in
Eq.~\eqref{eq:app-ring-gaussian} gives
$\operatorname{Var}_{\rm SSO,S}(n)=
\operatorname{Var}_{\rm SSO,S}(m)=Sz_\eta/2+o(S)$ and hence
quantum-jump-induced finite-size fluctuations of typical scale
$\delta n=O(\sqrt S)$. Consequently, a finite-order expansion in
$\delta n/S=O(S^{-1/2})$
resolves the same local Gaussian fluctuations for
the two parities: their half-spacing offset is negligible on this scale.
The zero of the factor
$S-\hat n$ lies at $\delta n=Sz_\eta+O(1)$, outside the domain of that local expansion, and the
usual unbounded fluctuation Fock space also discards the physical endpoint
$n=2S$.  Such an expansion can therefore determine intrabasin fluctuations
and relaxation but cannot compare the stationary normalization of the SSO
with the loss node and south boundary.

\subsection{V. Full Liouvillian gap: coherence-block resolution and
population-mode asymptotics}
\label{app:liouvillian-gap-method}

\subsubsection{\textbf{V.A. Full spectrum and coherence-order blocks}}
\label{app:gap-coherence-blocks}

Using Eqs.~\eqref{eq:end-coherence-blocks} and
\eqref{eq:end-block-gap-minimum}, we evaluate the coherence-order blocks
separately rather than forming one dense $(2S+1)^2$-dimensional matrix.

Define the signed Dicke-state amplitudes
\begin{equation}
\begin{aligned}
 a_m^+&=\sqrt{\frac{\Gamma_+}{S}(S-m)(S+m+1)},\smjoinrow
 a_m^-&=\sqrt{\frac{\Gamma_-}{S^3}}\,
 m\sqrt{(S+m)(S-m+1)}.
\end{aligned}
\label{eq:app-signed-jump-amplitudes}
\end{equation}
The amplitudes are set to zero whenever either the initial or final Dicke
level lies outside the ladder. In the ordered-loss amplitude \(a_m^-\), the
factor \(m\) retains its sign. This will matter for coherences even though the
population transition rate depends only on \(|a_m^-|^2\).

Use the operator basis
\(\hat E_m^{(\nu)}=|S,m\rangle\langle S,m-\nu|\), with both magnetic
indices restricted to the Dicke ladder. The pump term raises both indices,
the loss term lowers both, and diagonal anticommutators make
$\mathcal L_\nu$ tridiagonal:
\begin{equation}
\hspace{-1.5em}
\begin{aligned}
\mathcal L_\nu[\hat E_m^{(\nu)}]
={}a_m^+a_{m-\nu}^+\hat E_{m+1}^{(\nu)}+a_m^-a_{m-\nu}^-\hat E_{m-1}^{(\nu)}-\frac12\sum_{\sigma=\pm}
\bigl(|a_m^\sigma|^2+|a_{m-\nu}^\sigma|^2\bigr)
\hat E_m^{(\nu)}-i\nu\omega_0\hat E_m^{(\nu)}.
\end{aligned}
\label{eq:app-general-coherence-block}
\end{equation}
In particular, the ordered-loss recycling matrix element is proportional to
\(m(m-\nu)\). Replacing it by the positive square root of two population
rates would erase this sign and give an incorrect coherence block. At
\(\nu=0\), Eq.~\eqref{eq:app-general-coherence-block} reduces exactly to the
population generator below. Hermiticity preservation makes the spectra of
the \(\nu\) and \(-\nu\) blocks complex conjugates, so only \(\nu>0\) must be
diagonalized independently.

\subsubsection{\textbf{V.B. Population Jacobi reduction and full-block gap
identification}}
\label{app:population-jacobi}

\vthirteen{The $\nu=0$ block closes on the Dicke populations governed by
Eqs.~\eqref{eq:app-birth-death}--\eqref{eq:app-downward-rate}. Here $\nu=0$
denotes coherence order, not the level $m=0$.  Collect them into
$\mathbf p=(p_{-S},\ldots,p_S)^{\mathsf T}$.}
\vthirteen{The population generator $\mathsf K_{\rm pop}$ is defined by
$\dot{\mathbf p}=\mathsf K_{\rm pop}\mathbf p$, with
$\mathbf p(t)=e^{\mathsf K_{\rm pop}t}\mathbf p(0)$. In the column-vector
convention used here, its only nonzero matrix elements are}
\begin{equation}
\begin{aligned}
(\mathsf K_{\rm pop})_{m,m}&=-(W_m^++W_m^-),\smjoinrow
(\mathsf K_{\rm pop})_{m+1,m}&=W_m^+,~(\mathsf K_{\rm pop})_{m-1,m}=W_m^-.
\end{aligned}
\label{eq:app-population-generator-elements}
\end{equation}
\vthirteen{The nonnegative off-diagonal entries are transition rates, while
each column sums to zero so that $\sum_m p_m$ is conserved.}

\vthirteen{Let $\pi_m=p_{m,\rm ss}^{(S)}$. For the odd-\(N\) chains used in the
gap calculation and $0\leq\eta<1$, all $\pi_m$ are positive, and the
zero-current identity in Eq.~\eqref{eq:app-zero-current} makes the population
chain reversible.}
\vthirteen{For even-\(N\), ordering the Dicke basis as the transient set
$\mathcal T=\{-S,\ldots,-1\}$ followed by the closed recurrent class
$\mathcal R=\{0,\ldots,S\}$ makes $\mathsf K_{\rm pop}$ block triangular.
The one-way edge $-1\to0$ appears only in the off-diagonal block, so the full
population spectrum is the union of the spectra of the transient and recurrent
diagonal blocks. Each diagonal block is separately symmetrizable: the recurrent
block uses its stationary weights, whereas the transient block uses positive
unnormalized reversible weights for its internal edges. Restriction to
$\mathcal R$ is sufficient for the stationary state, but the full population
gap must be minimized over both blocks. No such decomposition is needed for
the odd-\(N\) scan.}

\vthirteen{For the odd chain, or for either diagonal block of the even chain,
let $\mathcal I$ denote its ordered state set and let $\pi_m>0$ be the
corresponding reversible weights. Define
$D_{\mathcal I}=\operatorname{diag}_{m\in\mathcal I}(\sqrt{\pi_m})$.}
\vthirteen{The similarity transformation
$\widetilde{\mathsf K}_{\mathcal I}=D_{\mathcal I}^{-1}
\mathsf K_{\mathcal I}D_{\mathcal I}$ has the general matrix element}
\begin{equation}
(\widetilde{\mathsf K}_{\mathcal I})_{m,n}
 =\frac{\sqrt{\pi_n}}{\sqrt{\pi_m}}(\mathsf K_{\mathcal I})_{m,n},
 ~m,n\in\mathcal I.
\label{eq:app-jacobi-element-general}
\end{equation}
\vthirteen{Detailed balance leaves the diagonal unchanged and gives}
\begin{equation}
\begin{aligned}
(\widetilde{\mathsf K}_{\mathcal I})_{m,m}
 =-(W_m^++W_m^-),~(\widetilde{\mathsf K}_{\mathcal I})_{m,m+1}=(\widetilde{\mathsf K}_{\mathcal I})_{m+1,m}=\sqrt{W_m^+W_{m+1}^-}.
\end{aligned}
\label{eq:app-jacobi-population-generator}
\end{equation}
\vthirteen{Hence $\widetilde{\mathsf K}_{\mathcal I}$ is a real symmetric
tridiagonal, or Jacobi, matrix.}
\vthirteen{Similarity preserves all eigenvalues. For even-\(N\), the direct sum
of the two symmetric blocks is therefore isospectral to the full
block-triangular population generator.  The Jacobi reduction is exact.}
\vthirteen{If the eigenvalues are ordered as
$0=\lambda_0>\lambda_1\geq\lambda_2\geq\cdots$, the exact finite-size
population gap is $\Delta_{\rm pop}=-\lambda_1$.}
\vthirteen{Because $\mathsf K_{\rm pop}$ represents $\mathcal L_0$ exactly,
$-\Delta_{\rm pop}\in\operatorname{spec}\mathcal L$ and
$\Delta_{\mathcal L}\leq\Delta_{\rm pop}$, with equality precisely when
$\nu=0$ minimizes Eq.~\eqref{eq:end-block-gap-minimum}.  The longest population
relaxation time is of order $\Delta_{\rm pop}^{-1}$.}

\vthirteen{Set
$m_j=-S+j$, $r_j=W_{m_j}^++W_{m_j}^-$, and
$a_j^2=W_{m_j}^+W_{m_j+1}^-$.}
\vthirteen{Here $\lambda$ is the spectral parameter of the Jacobi population
generator $\widetilde{\mathsf K}_{\rm pop}$, equivalently of the Liouvillian
restricted to the diagonal $\nu=0$ operator subspace.}
\vthirteen{Let $\widetilde{\mathsf K}_{\rm pop}^{(j)}$ be the leading
$j\times j$ principal block of $\widetilde{\mathsf K}_{\rm pop}$ and define
$\mathcal P_j(\lambda)=\det[\widetilde{\mathsf K}_{\rm pop}^{(j)}-\lambda I_j]$,
whose tridiagonal structure gives the continuant recurrence}
\begin{equation}
\begin{aligned}
\mathcal P_0=1,~\mathcal P_1=-r_0-\lambda,~
\mathcal P_{j+1}=(-r_j-\lambda)\mathcal P_j-a_{j-1}^2\mathcal P_{j-1}.
\end{aligned}
\label{eq:app-gap-characteristic-recurrence}
\end{equation}
\vthirteen{For the full matrix, the roots of
$\mathcal P_{N+1}(\lambda)=\det(\widetilde{\mathsf K}_{\rm pop}-\lambda I)$
give the exact population spectrum, with the root below $\lambda_0=0$ equal to
$-\Delta_{\rm pop}$.  The recurrence is an exact implicit finite-$N$ solution,
although its nonuniform coefficients yield no known elementary expression for
that root.}

\phantomsection
\label{app:gap-sturm-computation}
\vthirteen{Direct double-precision diagonalization cannot reliably separate
the exponentially small population eigenvalue from the exact stationary
root. We therefore obtain the leading $\nu=0$ eigenvalue from the symmetric
Jacobi matrix using arbitrary-precision Sturm bisection.}
For every displayed $N$ and sampled $\eta$, the spectrum of every
coherence-order block is evaluated, and the full Liouvillian gap is obtained
directly from Eq.~\eqref{eq:end-block-gap-minimum}. Figure~
\ref{fig:boundary-condensation}(b) shows the resulting block-resolved
finite-size spectrum.

\label{app:gap-full-liouvillian-identification}
\vthirteen{Near rate coexistence, for the sizes computed, the block-resolved
calculation identifies the leading nonpopulation mode in the
$|\nu|=1$ phase-diffusion sector. Over the computed
sequence, its closing is consistent with algebraic scaling in $N$, and it
remains above the exponentially small population-switching mode near
rate coexistence.}
The spectral minimum approaches $\eta_c^{\mathrm{o}}$ within the scan
resolution, although at finite $S$ it need not coincide with the equal-weight
point $\eta_{\rm eq}(S)$ defined by
equal branch probabilities and expanded in
Eq.~\eqref{eq:app-finite-size-equal-weight-shift}. The full-block calculation
therefore establishes the finite-size gap identification, while the following
subsection fixes the asymptotic rate-coexistence exponent of the population
mode.

\subsubsection{\textbf{V.C. Analytic population-gap scaling at rate coexistence}}
\label{app:gap-asymptotic-scaling}

The finite-$N$ population gap has no simple closed form, but its leading exponential
scaling follows analytically from the reversible chain.  Define the edge
conductance as $c_m=\pi_mW_m^+=\pi_{m+1}W_{m+1}^-$.
The equality of the two directed stationary fluxes follows from detailed
balance.  Thus $c_m$ is the conductance of the undirected edge
$m\leftrightarrow m+1$: a small $c_m$ is a weak link, with edge resistance
$c_m^{-1}$.
For two intervals $A$ and $B$ separated along the one-dimensional ladder,
their total series resistance and effective interbasin conductance are
\begin{equation}
R(A,B)=\sum_{m:A\to B}\frac{1}{c_m},~
C_{AB}=R(A,B)^{-1}.
\label{eq:app-resistance-capacity}
\end{equation}
Because the Dicke ladder is one dimensional, all intervening edges are in
series, so their resistances add.  The effective conductance $C_{AB}$ is the
stationary flux supported by a unit potential difference between the two
intervals and is known as the capacity in reversible-chain potential theory
$\operatorname{cap}(A,B)$.
If $\mu_A=\sum_{m\in A}\pi_m$ and
$\mu_B=\sum_{m\in B}\pi_m$, the two-basin reduction has decay rate
\begin{equation}
\Delta_{\rm eff}=C_{AB}
(\mu_A^{-1}+\mu_B^{-1}).
\label{eq:app-effective-two-basin-gap}
\end{equation}
Indeed, potential theory gives the coarse-grained rates
$k_{A\to B}=C_{AB}/\mu_A$ and
$k_{B\to A}=C_{AB}/\mu_B$. The only nonzero eigenvalue of
the resulting two-state generator is
$-(k_{A\to B}+k_{B\to A})=-\Delta_{\rm eff}$.
At rate coexistence, take $A$ and $B$ to be the mesoscopic SSO and south-polar
windows defined in Eq.~\eqref{eq:app-branch-window-decomposition}.  The two
branch normalizations have the same leading exponential rate, so
\begin{equation}
\ln\mathcal Z_{\rm tot}=2Sz_c+o(S),
~
\mu_A^{-1}+\mu_B^{-1}=e^{o(N)}.
\label{eq:app-coexistence-basin-masses}
\end{equation}
The second relation allows for the unequal algebraic prefactors in
Eqs.~\eqref{eq:app-sso-normalization} and
\eqref{eq:app-pfp-normalization}. They do not affect the exponential rate.

The negative-side rate function in
Eq.~\eqref{eq:app-negative-rate-function} obeys
$\psi_-'(y)=\ln(y/z_c)$ and $\psi_-''(y)=1/y>0$ at rate coexistence.  Its unique
minimum is therefore at $y=z_c$, so the nearest allowed Dicke level gives the
bottleneck $m_{\rm b}=-Sz_c+O(1)$, where
\begin{equation}
\ln w_{m_{\rm b}}=-2Sz_c+o(S).
\label{eq:app-barrier-weight-exponent}
\end{equation}
After division by the total normalization, the stationary bottleneck weight
therefore satisfies
$\pi_{m_{\rm b}}=\exp[-4Sz_c+o(S)]$.  The
transition rate multiplying this weight is only polynomial in $S$, and hence
\begin{equation}
c_{m_{\rm b}}
=\Gamma_{\rm tot}\exp[-2z_cN+o(N)].
\label{eq:app-bottleneck-conductance-exponent}
\end{equation}
All other edges have larger conductance on the exponential scale, so
Eqs.~\eqref{eq:app-resistance-capacity} and
\eqref{eq:app-bottleneck-conductance-exponent} give
\begin{equation}
C_{AB}
=\Gamma_{\rm tot}\exp[-2z_cN+o(N)].
\label{eq:app-capacity-exponent}
\end{equation}
Combining this result with
Eq.~\eqref{eq:app-coexistence-basin-masses} shows that the effective two-basin
decay rate has the same exponent.  To transfer this result to the exact
population gap, note first that reversibility gives the variational formula
\begin{equation}
\Delta_{\rm pop}
=\inf_{f\not={\rm const}}
\frac{\displaystyle\sum_{m=-S}^{S-1}c_m(f_{m+1}-f_m)^2}
{\displaystyle\sum_{m=-S}^{S}\pi_m(f_m-\langle f\rangle_\pi)^2}.
\label{eq:app-population-gap-variational}
\end{equation}
The numerator is the Dirichlet form and the denominator is the stationary
variance.  Applying the discrete Hardy inequality to this Rayleigh quotient
gives a sharp estimate of its inverse up to universal constants
\cite{Miclo1999Hardy}.  For a pivot $i$ on the Dicke ladder, define
\begin{align}
B_+(i)=\max_{x>i}\left(\sum_{m=x}^{S}\pi_m\right)
\left(\sum_{m=i}^{x-1}\frac{1}{c_m}\right),~
B_-(i)=\max_{x<i}\left(\sum_{m=-S}^{x}\pi_m\right)
\left(\sum_{m=x}^{i-1}\frac{1}{c_m}\right),~
B_N=\min_i\max\bigl(B_+(i),B_-(i)\bigr).
\label{eq:app-hardy-constant}
\end{align}
The inverse population gap, or Poincar\'e relaxation constant, obeys the
universal one-dimensional bounds
\begin{equation}
 \frac{B_N}{2}\leq \Delta_{\rm pop}^{-1}\leq4B_N.
 \label{eq:app-hardy-gap-bounds}
\end{equation}
At rate coexistence, both basin masses are $e^{o(N)}$, while the bottleneck is the
largest edge resistance and satisfies
$c_{m_{\rm b}}^{-1}=\Gamma_{\rm tot}^{-1}e^{2z_cN+o(N)}$.
For any pivot, at least one of $B_+$ and $B_-$ contains a path from a basin
of $e^{o(N)}$ mass across this bottleneck.  Conversely,
every resistance sum contains only $O(N)$ edges,
and no edge has a larger exponential resistance.
Therefore
\begin{equation}
B_N=\Gamma_{\rm tot}^{-1}e^{2z_cN+o(N)}.
\label{eq:app-hardy-constant-exponent}
\end{equation}
Equations~\eqref{eq:app-hardy-gap-bounds} and
\eqref{eq:app-hardy-constant-exponent} prove
Eq.~\eqref{eq:end-gap-exponent}. This determines the thermodynamic exponential
rate, not the complete finite-size prefactor. The full-block computation gives
$\Delta_{\mathcal L}=\Delta_{\rm pop}$ at rate coexistence for every displayed
$N=101,201,\ldots,1001$. Extending this equality analytically beyond the
computed sequence would require a lower bound on every $\nu\neq0$ block.

Gap closing is not, by itself, exclusive to the odd-sequence branch exchange.
For even \(N\), the population spectrum also contains the transient negative
half-ladder block identified in \hyperref[app:population-jacobi]{Sec.~V.B}.
The same one-dimensional escape calculation used for the reverse loading in
\hyperref[app:loading-switching-time]{Sec.~VI.C} gives its principal decay
rate as
\(\Gamma_{\rm tot}e^{-NI_{\rm rev}(z_\eta)+o(N)}\), where
\(I_{\rm rev}(z)=-\ln z-1+z\). At rate coexistence,
\(-\ln z_c=1+z_c\) implies \(I_{\rm rev}(z_c)=2z_c\), so this transient even
block has the same exponential scale as Eq.~\eqref{eq:end-gap-exponent}.
The gap therefore diagnoses exponentially slow relaxation, whereas the
odd-sequence first-order DPT is established by the stationary branch-rate
exchange and polarization jump.

\subsection{VI. Physical one-spin operations and switching}
\label{app:physical-multisector-loading}

\subsubsection{\textbf{VI.A. Exact collective closure and physical multisector loading}}
\label{app:intersector-coherence-closure}

Let \(N_0\) denote the even preloading particle number and define
\(N_\ell=N_0+\ell\) after loading step \(\ell\).
We write \(\mathcal L_{N_\ell,S}\) for the representation-reduced Liouvillian
in sector \(S\).  Step labels on reduced states and weights are suppressed
below until they are needed explicitly.

\paragraph{\textbf{The sectors in Fig.~\ref{fig:parity-quench-collapse}(a).---}}
Before the first loading, \(N_0=1024\) and the state lies in the maximal-spin
sector \(S_0=N_0/2=512\).  Adding one spin-\(1/2\) gives
\(512\otimes\tfrac12=512.5\oplus511.5\).
For the state used in Fig.~\ref{fig:parity-quench-collapse}(a), these two
half-integer sectors have exact weights
\(513/1025\simeq0.5005\) and \(512/1025\simeq0.4995\), respectively.  The
postloading state is not replaced by a classical mixture: in the coupled basis
it may contain an off-diagonal block
\(\hat\rho_{512.5,511.5}\).  During the ensuing \(N_1=1025\) evolution, however,
the collective Hamiltonian and jumps preserve \(S\). The two diagonal blocks
therefore evolve independently, while the off-diagonal block evolves
separately and cannot change their populations.

The second loading makes the only potentially ambiguous case explicit:
\(512.5\otimes\tfrac12=513\oplus512\) and
\(511.5\otimes\tfrac12=512\oplus511\).
Thus the \(N_2=1026\) state occupies \(S=513,512,511\), with the exact weights
\(1027/4100,1/2,1023/4100\) used in
Fig.~\ref{fig:parity-quench-collapse}(a).
The middle sector \(S=512\) receives two parent paths, one from \(S=512.5\) and
one from \(S=511.5\). These paths are not identified or projected onto
one another.  They occupy two orthogonal copies of the same spin-512
representation.  A collective operator acts identically on the two copies,
so their mutual coherence has zero contribution to the collective signal.
The polar Husimi-$Q$ density and the SSO/s-PFP basin weights plotted in
Fig.~\ref{fig:parity-quench-collapse}(a) are therefore obtained by retaining
both sectors after the first loading and all three sectors after the second.
No maximal-spin-sector postselection is used.

\paragraph{\textbf{Why the reduction is exact.---}}
The required structure can be stated without resolving the exponentially large
many-spin basis.  Using the isotypic decomposition in
Eq.~\eqref{eq:app-spin-multiplicity-decomposition}, \(\mathcal R_S\) is the usual Dicke ladder
\(\bigl(|S,m\rangle\bigr)_{m=-S}^{S}\), whereas \(\mathcal M_{S,N}\) only labels the
different orthogonal copies of that ladder.  In the concrete second loading
above, the two \(S=512\) copies carry the parent labels \(512.5\) and \(511.5\).
A collective Hamiltonian, jump operator, or observable does not act on this
copy label:
\begin{equation}
\begin{aligned}
 \hat H=\bigoplus_S(\hat H_S\otimes\hat I_{\mathcal M_{S,N}}),~
 \hat L_\mu=\bigoplus_S(\hat L_{\mu,S}\otimes\hat I_{\mathcal M_{S,N}}),~
 \hat O=\bigoplus_S(\hat O_S\otimes\hat I_{\mathcal M_{S,N}}).
\end{aligned}
 \label{eq:app-collective-isotypic-action}
\end{equation}
We may therefore discard only the unobserved copy label, not any occupied
total-spin sector, and define
\begin{equation}
\begin{aligned}
 \hat{\bar\rho}_S=\operatorname{Tr}_{\mathcal M_{S,N}}
 \!\left(\hat\Pi_S\hat\rho\hat\Pi_S\right),
~
 q_S=\operatorname{Tr}_{\mathcal R_S}\hat{\bar\rho}_S,
 ~ \hat\rho_S=\frac{\hat{\bar\rho}_S}{q_S}\quad(q_S>0).
\end{aligned}
 \label{eq:app-sector-reduced-density}
\end{equation}
During fixed-\(N\) evolution, collective observables obey
\begin{equation}
 \langle\hat O\rangle
 =\sum_S\operatorname{Tr}_{\mathcal R_S}(\hat O_S\hat{\bar\rho}_S),
~
 \frac{d\hat{\bar\rho}_S}{dt}
 =\mathcal L_{N_\ell,S}[\hat{\bar\rho}_S].
 \label{eq:app-reduced-collective-closure}
\end{equation}
The first identity follows from
\(\hat\Pi_S\hat O\hat\Pi_{\widetilde S}=0\) for \(S\neq\widetilde S\).  The second follows from
strong total-spin symmetry: an off-diagonal operator block
\(\hat\rho_{S\widetilde S}=\hat\Pi_S\hat\rho\hat\Pi_{\widetilde S}\) evolves within its own
\((S,\widetilde S)\) block and never feeds a diagonal reduced state \(\hat{\bar\rho}_S\).

Restoring the loading-step labels and integrating
Eq.~\eqref{eq:app-reduced-collective-closure} gives
Eq.~\eqref{eq:end-sector-weighted-collective-evolution}.

It remains to check that another loading event does not convert a previously
invisible coherence into a collective population.  Let \(\hat\rho_{1/2}\) be the
state of the appended spin.  At loading step \(\ell\), denote a parent sector
by \(S_{\rm p}\) and a child sector by
\(S_{\rm c}=S_{\rm p}\pm\tfrac12\).  Their Clebsch--Gordan coupling is
represented by the partial isometry
\(\hat C_{S_{\rm c}\leftarrow S_{\rm p}}:
\mathcal R_{S_{\rm p}}\otimes\mathbb C^2\to\mathcal R_{S_{\rm c}}\), with
\(\langle S_{\rm c},M|\hat C_{S_{\rm c}\leftarrow S_{\rm p}}
|S_{\rm p},m;\tfrac12,\sigma\rangle
=C^{S_{\rm c},M}_{S_{\rm p},m;\,1/2,\sigma}\) for \(M=m+\sigma\).
Equivalently,
\begin{equation}
\begin{aligned}
 \hat C_{S_{\rm c}\leftarrow S_{\rm p}}
 ={}\sum_{m=-S_{\rm p}}^{S_{\rm p}}
 \sum_{\sigma=\pm1/2}
 C^{S_{\rm c},m+\sigma}_{S_{\rm p},m;\,1/2,\sigma}~
 \times |S_{\rm c},m+\sigma\rangle
 \langle S_{\rm p},m|\langle\sigma|,
\end{aligned}
 \label{eq:app-cg-partial-isometry-expansion}
\end{equation}
where coefficients outside the child Dicke ladder vanish.  Unitarity of the
Clebsch--Gordan transformation gives
\begin{equation}
\begin{aligned}
 \hat C_{S_{\rm c}\leftarrow S_{\rm p}}
 \hat C_{S_{\rm c}\leftarrow S_{\rm p}}^\dagger
 =\hat I_{\mathcal R_{S_{\rm c}}},~
 \sum_{S_{\rm c}=S_{\rm p}\pm1/2}
 \hat C_{S_{\rm c}\leftarrow S_{\rm p}}^\dagger
 \hat C_{S_{\rm c}\leftarrow S_{\rm p}}
 =\hat I_{\mathcal R_{S_{\rm p}}\otimes\mathbb C^2}.
\end{aligned}
 \label{eq:app-cg-completeness}
\end{equation}
Thus each child map is completely positive, and their sum is trace
preserving for a fixed parent sector.  The corresponding reduced child map is
\(\Phi_{S_{\rm c}\leftarrow S_{\rm p}}^{(\rho_{1/2})}[\hat X]
=\hat C_{S_{\rm c}\leftarrow S_{\rm p}}
(\hat X\otimes\hat\rho_{1/2})
\hat C_{S_{\rm c}\leftarrow S_{\rm p}}^\dagger\).
The two parent multiplicity spaces enter the multiplicity space of the new
\(S_{\rm c}\) sector as the orthogonal direct sum
\(\mathcal M_{S_{\rm c},N+1}\cong
\mathcal M_{S_{\rm c}-1/2,N}\oplus\mathcal M_{S_{\rm c}+1/2,N}\).
Embedding the parent copy space with
\(\hat E_{S_{\rm c}\leftarrow S_{\rm p}}\) gives
\begin{equation}
 \hat E_{S_{\rm c}\leftarrow\widetilde S_{\rm p}}^\dagger
 \hat E_{S_{\rm c}\leftarrow S_{\rm p}}
 =\delta_{\widetilde S_{\rm p},S_{\rm p}}
 \hat I_{\mathcal M_{S_{\rm p},N}}.
 \label{eq:app-parent-embedding-orthogonality}
\end{equation}
For clarity, propagate an arbitrary input coherence
\(\hat\rho_{S_{\rm p}\widetilde S_{\rm p}}\) through the two parent paths and
then trace only the copy label.  The result is
\begin{equation}
\hspace{-1.5em}
\begin{aligned}
 \operatorname{Tr}_{\mathcal M_{S_{\rm c},N+1}}
 \left[(\hat C_{S_{\rm c}\leftarrow S_{\rm p}}
 \otimes\hat E_{S_{\rm c}\leftarrow S_{\rm p}})
 (\hat\rho_{S_{\rm p}\widetilde S_{\rm p}}
 \otimes\hat\rho_{1/2}) (\hat C_{S_{\rm c}\leftarrow\widetilde S_{\rm p}}^\dagger
 \otimes\hat E_{S_{\rm c}\leftarrow\widetilde S_{\rm p}}^\dagger)
 \right]
 =\delta_{S_{\rm p},\widetilde S_{\rm p}}\,
 \hat C_{S_{\rm c}\leftarrow S_{\rm p}}
 \bigl(\hat{\bar\rho}_{S_{\rm p}}\otimes\hat\rho_{1/2}\bigr)
 \hat C_{S_{\rm c}\leftarrow S_{\rm p}}^\dagger .
 \end{aligned}
 \label{eq:app-coherence-partial-trace-zero}
\end{equation}
The Kronecker delta is the entire point: if
\(S_{\rm p}\neq\widetilde S_{\rm p}\), the coherence ends between orthogonal
copies of \(S_{\rm c}\) and its copy-space trace is exactly zero.
For Fig.~\ref{fig:parity-quench-collapse}(a), this is precisely what happens to
the coherence between the two routes
\(512.5\to512\) and \(511.5\to512\).
For each fixed child sector \(S_{\rm c}\), summing the surviving contributions
from all compatible parents gives Eq.~\eqref{eq:end-loading-recursion}.
Repeating the same argument after every loading step combines all paths ending
at \(S_{\rm c}\) into the correct weight
\(q_{S_{\rm c}}^{(\ell)}=
\operatorname{Tr}_{\mathcal R_{S_{\rm c}}}
\hat{\bar\rho}_{S_{\rm c}}^{(\ell)}\), while preserving their orthogonal copy
labels in the full state.

Noncollective sector-mixing noise lies outside the protocol considered here.

For the unpolarized loading used in the Letter, every parent block is Dicke
diagonal.  Writing
\(\bar p_{m,S_{\rm p}}^{(\ell-1)}=
q_{S_{\rm p}}^{(\ell-1)}p_{m,S_{\rm p}}^{(\ell-1)}\), and defining the child
populations analogously, the exact matrix recursion reduces to
\begin{equation}
 \bar p_{M,S_{\rm c}}^{(\ell)}
 =\frac12\sum_{\substack{S_{\rm p}\in\mathcal S_{\ell-1}\\
 |S_{\rm c}-S_{\rm p}|=1/2}}
 \sum_{\sigma=\pm1/2}
 \left|C^{S_{\rm c},M}_{S_{\rm p},M-\sigma;\,1/2,\sigma}\right|^2
 \bar p_{M-\sigma,S_{\rm p}}^{(\ell-1)}.
 \label{eq:app-exact-reduced-loading-populations}
\end{equation}
Terms outside an allowed Dicke ladder are absent.  The plotted calculation
uses \(q_{S_{\rm c}}^{(\ell)}=\sum_M\bar p_{M,S_{\rm c}}^{(\ell)}\) and
\(p_{M,S_{\rm c}}^{(\ell)}=
\bar p_{M,S_{\rm c}}^{(\ell)}/q_{S_{\rm c}}^{(\ell)}\), retaining every
occupied \(S_{\rm c}\).

\phantomsection
\label{app:cg-multisector-dynamics}

\paragraph{\textbf{Explicit Clebsch--Gordan coefficients for unpolarized loading.---}}
For the unpolarized loading used in the Letter,
$\hat\rho_{1/2}=\hat I_2/2$, the Clebsch--Gordan amplitudes are
\begin{equation}
\begin{aligned}
|S,m\rangle|\!\uparrow\rangle={}
\sqrt{\frac{S+m+1}{2S+1}}\,
 |S+\tfrac12,m+\tfrac12\rangle-\sqrt{\frac{S-m}{2S+1}}\,
 |S-\tfrac12,m+\tfrac12\rangle,\\
|S,m\rangle|\!\downarrow\rangle={}
\sqrt{\frac{S-m+1}{2S+1}}\,
 |S+\tfrac12,m-\tfrac12\rangle+\sqrt{\frac{S+m}{2S+1}}\,
 |S-\tfrac12,m-\tfrac12\rangle.
\end{aligned}
\label{eq:app-cg-loading}
\end{equation}
Squaring these amplitudes and summing the appended-spin components gives
Eq.~\eqref{eq:app-exact-reduced-loading-populations}, including additive
contributions from orthogonal parent paths ending in the same $S$.

\paragraph{\textbf{Exact sector rates and loading-dependent basin partitions.---}}
Within a fixed loading step, the diagonal block in sector $S$ is an exact
birth--death chain.  Its only nonzero transition rates are
\begin{equation}
\begin{aligned}
W_{m\to m+1}^{(S,\ell)}=\frac{2\Gamma_+}{N_\ell}(S-m)(S+m+1),~
W_{m\to m-1}^{(S,\ell)}=\frac{8\Gamma_-}{N_\ell^3}m^2(S+m)(S-m+1).
\end{aligned}
\label{eq:app-loading-trajectory-rates}
\end{equation}
Here $N_\ell/2$ fixes the physical collective normalization, whereas $S$
fixes the Dicke-ladder matrix elements.  For a half-integer sector in the
first loading, the exact stationary ratio is
$w_{m+1,S}/w_{m,S}=W_{m\to m+1}^{(S,\ell)}/W_{m+1\to m}^{(S,\ell)}
=\bigl[N_\ell z_\eta/(2(m+1))\bigr]^2$.
We define $m_{\rm b}^{(S)}$ as the minimum of $w_{m,S}$ between the south
boundary and the positive-$m$ SSO mode.  Away from boundary truncation it is
the allowed Dicke level satisfying
$|m_{\rm b}^{(S)}|\geq N_\ell z_\eta/2\geq|m_{\rm b}^{(S)}+1|$,
so $2m_{\rm b}^{(S)}/N_\ell\to-z_\eta$ with only one-lattice-spacing
rounding.  Assigning the boundary level itself to the s-PFP basin gives the
first-loading partition.  For the second loading, the deleted $0\to-1$ edge
makes $m<0$ transient and $m\geq0$ the unique recurrent SSO class.  The two
loading-dependent partitions are therefore
\begin{align}
\mathcal B_{\mathrm{s\text{-}PFP},1}^{(S)}
 &=\{m:-S\leq m\leq m_{\rm b}^{(S)}\},~
\mathcal B_{\mathrm{SSO},1}^{(S)}
 =\{m:m_{\rm b}^{(S)}<m\leq S\}
 \label{eq:app-loading-basin-partition}\\
\mathcal B_{\mathrm{s\text{-}PFP},2}^{(S)}
 &=\{m:-S\leq m<0\},~
\mathcal B_{\mathrm{SSO},2}^{(S)}
 =\{m:0\leq m\leq S\}.\notag
\end{align}
Changing the ownership of the first-loading bottleneck level changes only a
finite-size threshold convention, not the large-size switching exponent.

\paragraph{\textbf{Normalized sector dynamics and basin probabilities.---}}
Between loading events the weights $q_S^{(\ell)}$ are conserved. Each
normalized child block obeys
\begin{equation}
\begin{aligned}
 \hat\rho_S^{(\ell)}(t_\ell)
 =e^{\mathcal L_{N_\ell,S}t_\ell}\hat\rho_S^{(\ell)}(0^+),~
 p_{m,S}^{(\ell)}(t_\ell)
 =\langle S,m|\hat\rho_S^{(\ell)}(t_\ell)|S,m\rangle.
\end{aligned}
 \label{eq:app-loaded-sector-deterministic-dynamics}
\end{equation}
where $\sum_{m=-S}^{S}p_{m,S}^{(\ell)}(t_\ell)=1$.  Using the loading- and sector-dependent
partition in Eq.~\eqref{eq:app-loading-basin-partition}, the normalized probability for
sector $S$ to occupy basin
$\alpha\in\{\mathrm{SSO},\mathrm{s\text{-}PFP}\}$ is
\begin{equation}
 P_{\alpha,S}^{(\ell)}(t_\ell)
 =\sum_{m\in\mathcal B_{\alpha,\ell}^{(S)}}p_{m,S}^{(\ell)}(t_\ell),
 ~
 \sum_\alpha P_{\alpha,S}^{(\ell)}(t_\ell)=1.
 \label{eq:app-sector-conditioned-basin-probability}
\end{equation}
Its contribution to the ensemble is
$q_S^{(\ell)}P_{\alpha,S}^{(\ell)}(t_\ell)$. Summing these contributions gives
Eq.~\eqref{eq:parity-quench-basin-weights}. The polarization follows from
Eq.~\eqref{eq:end-sector-weighted-collective-evolution} with
\(\hat O=2\hat S_z/N_\ell\).
Multiplicity copies are already combined in
$q_S^{(\ell)}p_{m,S}^{(\ell)}$. No extra degeneracy factor is inserted.

\subsubsection{\textbf{VI.B. MFPT-based description of
Fig.~\ref{fig:parity-quench-collapse}(a)}}
\label{app:loading-large-size-reconstruction}

All inputs to Fig.~\ref{fig:parity-quench-collapse}(a) are exact. Only its
$N_0=1024$ time dependence uses the single-exponential MFPT reconstruction
defined and validated in \hyperref[app:loading-switching-time]{Sec.~VI.C}.

\paragraph{\textbf{First loading: SSO-to-s-PFP MFPT calculation.---}}
At $N_0=1024$, Eq.~\eqref{eq:app-exact-reduced-loading-populations} maps the
exact maximal-sector steady populations, with
$\hat\rho_{1/2}=\hat I_2/2$, to $p_{m,S}^{(1)}(0^+)$ and $q_S^{(1)}$ in
$S\in\{512.5,511.5\}$.  The normalized stationary restriction to the basin
in Eq.~\eqref{eq:app-loading-basin-partition} is
\begin{equation}
 \pi_{m,S}^{\rm PFP}
 =\frac{w_{m,S}\,
 \mathbf 1_{\mathcal B_{\mathrm{s\text{-}PFP},1}^{(S)}}(m)}
 {\displaystyle\sum_{r\in\mathcal B_{\mathrm{s\text{-}PFP},1}^{(S)}}w_{r,S}}.
 \label{eq:app-conditional-pfp-distribution}
\end{equation}
Let $T_{S,\to{\rm s}}$ be the exact MFPT from the positive stationary mode to
the south endpoint, namely $T_S^{\rm s}$ in
Eq.~\eqref{eq:app-log-mfpt-construction}.  The first-stage MFPT calculation
uses
\begin{equation}
\begin{aligned}
 \Sigma_S^{(1)}(t_1)&=e^{-t_1/T_{S,\to{\rm s}}},\smjoinrow
 \sum_{S\in\{512.5,511.5\}}q_S^{(1)}
 \Sigma_S^{(1)}(t_{50}^{(1)})&=\frac12,
\end{aligned}
 \label{eq:app-first-loading-survival}
\end{equation}
and
\begin{equation}
 p_{m,S}^{(1)}(t_1)
 =\Sigma_S^{(1)}(t_1)p_{m,S}^{(1)}(0^+)
 +[1-\Sigma_S^{(1)}(t_1)]\pi_{m,S}^{\rm PFP}.
 \label{eq:app-first-loading-reconstruction}
\end{equation}
Thus the plotted first-stage basin weights are
\begin{equation}
\begin{aligned}
 P_{\rm SSO}^{(1)}(t_1)
 &=\sum_S q_S^{(1)}\Sigma_S^{(1)}(t_1),\smjoinrow
 P_{\mathrm{s\text{-}PFP}}^{(1)}(t_1)
 &=1-P_{\rm SSO}^{(1)}(t_1).
\end{aligned}
 \label{eq:app-first-loading-basin-reconstruction}
\end{equation}
Equations~\eqref{eq:app-first-loading-survival}--
\eqref{eq:app-first-loading-basin-reconstruction} retain the interbasin
transfer. Exponentially small wrong-basin weights and faster intrabasin modes
enter only finite-size prefactors.

\paragraph{\textbf{Second loading: s-PFP-to-SSO MFPT calculation.---}}
After the first MFPT-based switch, the second loading acts on
$q_S^{(1)}\pi_{m,S}^{\rm PFP}$ rather than a finite-time snapshot of
Eq.~\eqref{eq:app-first-loading-reconstruction}.  The exact loading recursion
produces the three integer-sector states $p_{m,S}^{(2)}(0^+)$ and their weights.
Their MFPTs $T_{S,\to0}=T_S^0$ are evaluated from the exact backward recurrence
in Eq.~\eqref{eq:app-reverse-loading-mfpt}.

Let $\pi_{m,S}^{(2)}$ be the exact normalized stationary population in the
corresponding integer sector.  The second-stage MFPT calculation is
\begin{align}
 \Sigma_S^{(2)}(t_2)=e^{-t_2/T_{S,\to0}},
 ~ \sum_{S\in\{513,512,511\}}q_S^{(2)}
 \Sigma_S^{(2)}(t_{50}^{(2)})=\frac12,~
 p_{m,S}^{(2)}(t_2)=\Sigma_S^{(2)}(t_2)p_{m,S}^{(2)}(0^+)+[1-\Sigma_S^{(2)}(t_2)]\pi_{m,S}^{(2)}.
 \label{eq:app-second-loading-survival}
\end{align}
Accordingly,
\begin{equation}
\begin{aligned}
 P_{\mathrm{s\text{-}PFP}}^{(2)}(t_2)
 =\sum_S q_S^{(2)}\Sigma_S^{(2)}(t_2),P_{\rm SSO}^{(2)}(t_2)
 =1-P_{\mathrm{s\text{-}PFP}}^{(2)}(t_2).
\end{aligned}
 \label{eq:app-second-loading-basin-reconstruction}
\end{equation}

\paragraph{\textbf{Husimi mixture and display-time map.---}}
For normalized $p_{m,S}$, let $Q_S(\theta;p)$ denote
Eq.~\eqref{eq:app-husimi-binomial} with $p_m^{(S)}$ replaced by $p_{m,S}$.
The two basin-conditioned endpoint functions are
\begin{align}
Q_{S,\mathrm{s\text{-}PFP}}(\theta)\equiv Q_S\!\left(\theta;\pi_{m,S}^{\rm PFP}\right),~S\in\{512.5,511.5\};~
Q_{S,\mathrm{SSO}}(\theta)\equiv Q_S\!\left(\theta;\pi_{m,S}^{(2)}\right),~S\in\{513,512,511\}.
 \label{eq:app-loading-basin-husimi-functions}
\end{align}
Here $\pi_{m,S}^{\rm PFP}$ is defined in
Eq.~\eqref{eq:app-conditional-pfp-distribution}, and $\pi_{m,S}^{(2)}$ is the
integer-sector SSO steady state.  We plot the normalized polar density
\(p_{Q,S}(\theta;p)=2\pi\sin\theta\,Q_S(\theta;p)\), with
\(\int_0^\pi p_{Q,S}(\theta;p)\,d\theta=1\).
The physical mixture is
\begin{equation}
 p_Q^{(\ell)}(\theta,t_\ell)
 =\sum_{S\in\mathcal S_\ell}q_S^{(\ell)}
 p_{Q,S}\!\left(\theta;p_{m,S}^{(\ell)}(t_\ell)\right).
 \label{eq:app-loading-husimi-mixture}
\end{equation}
At $\eta=0.90$,
$\log_{10}(\Gamma_{\rm tot}t_{50}^{(1)})\simeq204.67$ and
$\log_{10}(\Gamma_{\rm tot}t_{50}^{(2)})\simeq310.47$. We therefore
evaluate the reconstruction in the log domain, relative to the largest time
scale.

Panel (a) uses the stage-local variables
$u_\ell=t_\ell/t_{50}^{(\ell)}$: the first stage is linear over
$0\leq u_1\leq10$, while $0\leq u_2\leq220$ is monotonically compressed to
the same visual width.  The white dashed line separates stages, and the white
dotted lines mark $u_1=u_2=1$. The two segments do not share an absolute-time
axis.

\paragraph{\textbf{Sector-resolved relaxation in the physical protocol.---}}
Figure~\ref{fig:app-sector-resolved-loading} resolves the large-size basin
weights into $q_S^{(\ell)}P_{\alpha,S}^{(\ell)}$.  The spread of sector time
scales in the second loading broadens the reverse crossover, but every sector
returns to the SSO branch.

\begin{smfullfigure}[!htbp]
\centering
\includegraphics[width=0.8\columnwidth]{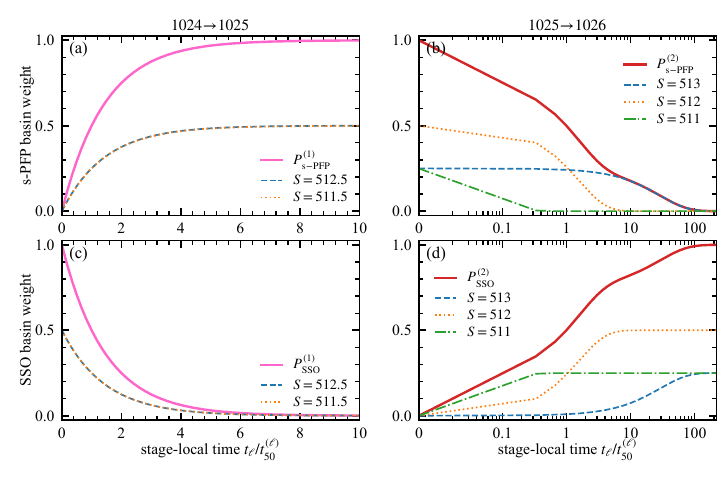}
\caption{\label{fig:app-sector-resolved-loading}
Sector decomposition underlying Fig.~\ref{fig:parity-quench-collapse}(a).
Solid magenta ($1024\to1025$) and red ($1025\to1026$) curves show the total
basin weights. Non-solid curves show $q_S^{(\ell)}P_{\alpha,S}^{(\ell)}$.
Panels (a,b) give
s-PFP weights and (c,d) SSO weights for the two loadings. Panels (b,d) use a
symmetric-logarithmic time axis.}
\end{smfullfigure}

\subsubsection{\textbf{VI.C. Switching-time construction, validation, and
asymptotic exponents}}
\label{app:loading-switching-time}

\paragraph{\textbf{Equal-weight time and metastable two-basin reduction.---}}
The direct and MFPT-based equal-weight times are defined in
Eqs.~\eqref{eq:end-exact-loading-t50} and
\eqref{eq:reconstructed-loading-t50}. Here $T_S^{(1)}=T_S^{\rm s}$ and
$T_S^{(2)}=T_S^0$.

The comparison between the equal-weight time and population gap follows from
the metastable two-basin reduction. Let $P_B(t_\ell)$ be the s-PFP weight and
$k_{A\to B},k_{B\to A}$ the effective transfer rates:
\begin{equation}
\begin{aligned}
 \dot P_B&=k_{A\to B}(1-P_B)-k_{B\to A}P_B,
 \smjoinrow
 \Delta_{2{\rm b}}&=k_{A\to B}+k_{B\to A},~P_{B,\rm ss}=\frac{k_{A\to B}}{\Delta_{2{\rm b}}}.
\end{aligned}
\label{eq:app-two-basin-kinetics}
\end{equation}
For an initially SSO-localized state, $P_B(0)=0$, this gives
\begin{equation}
\begin{aligned}
 P_B(t_\ell)&=P_{B,\rm ss}\bigl(1-e^{-\Delta_{2{\rm b}}t_\ell}\bigr),\smjoinrow
 t_{50,2{\rm b}}&=-\frac{1}{\Delta_{2{\rm b}}}
 \ln\!\left(1-\frac{1}{2P_{B,\rm ss}}\right),
\end{aligned}
 \label{eq:app-t50-two-basin}
\end{equation}
provided $P_{B,\rm ss}>1/2$. Hence the commonly used relation
$t_{50,2{\rm b}}\simeq(\ln2)/\Delta_{2{\rm b}}$ is exact only in the strongly biased,
effectively unidirectional limit $P_{B,\rm ss}\to1$. After spin loading, the
basin probability, allowing either loading direction, becomes
\begin{equation}
 P_B^{(\ell)}(t_\ell)=\sum_{S\in\mathcal S_\ell}q_S^{(\ell)}
 \left\{P_{B,\rm ss}^{(S)}+
 [P_{B,S}^{(\ell)}(0)-P_{B,\rm ss}^{(S)}]
 e^{-\Delta_{2{\rm b},S}t_\ell}\right\}.
 \label{eq:app-multisector-two-basin}
\end{equation}
There is therefore no exact single-gap identity at finite size.  The minimum
participating population gap
$\Delta_{{\rm pop},\min}^{(\ell)}=\min_{S\in\mathcal S_\ell}\Delta_{{\rm pop},S}$
always controls the longest tail, but it controls a fixed basin-weight
quantile only if the sectors at that exponential scale carry sufficient total
weight.  This is the metastable interbranch-transfer mechanism associated
with low-lying Liouvillian modes
\cite{Casteels2017,Macieszczak2016,Macieszczak2021}.  The paired protocol
shows both possibilities: $\Delta_{{\rm pop},\min}^{(1)}$ fixes the first-loading
exponent, whereas the second-loading median is controlled by a faster sector.

\paragraph{\textbf{First-passage definition and sector MFPTs.---}}
For a trajectory $m_t$ of the sector-$S$ birth--death chain after loading
step $\ell$, let $\mu_S^{(\ell)}$ be its initial distribution and
$\mathcal A_S^{(\ell)}$ the prescribed target set.  The first-passage time is
the stopping time
\(\tau_S^{(\ell)}=\inf\{t\geq0:m_t\in\mathcal A_S^{(\ell)}\}\).
Only the first entry is recorded, irrespective of any subsequent recrossing
\cite{Redner2023FirstPassage}.  Its survival probability, first-passage-time
density, and mean first-passage time are
\begin{equation}
\begin{aligned}
 \Sigma_S^{(\ell)}(t_\ell)=\Pr_{\mu_S^{(\ell)}}[\tau_S^{(\ell)}>t_\ell],~
 f_S^{(\ell)}(t_\ell)&=-\partial_{t_\ell}\Sigma_S^{(\ell)}(t_\ell),~
 T_S^{(\ell)}=\mathbb E_{\mu_S^{(\ell)}}[\tau_S^{(\ell)}]
 =\int_0^\infty\Sigma_S^{(\ell)}(t_\ell)\,dt_\ell.
\end{aligned}
 \label{eq:app-mfpt-definition}
\end{equation}
For a fixed initial level $m$, write
$\mathcal T_{m,S}^{(\ell)}=
\mathbb E[\tau_S^{(\ell)}\mid m_0=m]$.
The backward equation and absorbing target condition are
\begin{equation}
\begin{aligned}
 -1=W_{m\to m+1}^{(S,\ell)}
 \bigl(\mathcal T_{m+1,S}^{(\ell)}-\mathcal T_{m,S}^{(\ell)}\bigr)~
 +W_{m\to m-1}^{(S,\ell)}
 \bigl(\mathcal T_{m-1,S}^{(\ell)}-\mathcal T_{m,S}^{(\ell)}\bigr),~
 \mathcal T_{m,S}^{(\ell)}=0,
 ~m\in\mathcal A_S^{(\ell)},
\end{aligned}
 \label{eq:app-mfpt-backward-equation}
\end{equation}
with nonexistent boundary rates set to zero.

The MFPT calculation uses
$\mu_S^{(1)}(m)=\delta_{m,m_{+,S}}$ and
$\mathcal A_S^{(1)}=\{-S\}$ for the forward loading, and
$\mu_S^{(2)}(m)=\delta_{m,-S}$ and
$\mathcal A_S^{(2)}=\{0\}$ for the reverse loading.  Thus
$T_S^{(1)}=T_S^{\rm s}$ and $T_S^{(2)}=T_S^0$.
The circles in Fig.~\ref{fig:parity-quench-collapse}(c) instead record first
entry at the basin boundary.  That boundary-hitting time is a distinct
first-passage observable. As shown below, replacing the south endpoint by the
basin boundary changes only subexponential factors and leaves the large-size
switching exponent unchanged.

For the large-size SSO-to-s-PFP scan in
Fig.~\ref{fig:parity-quench-collapse}(b), let $m_{+,S}$ be the positive-$m$
mode of $w_{m,S}$. The exact birth--death MFPT from that mode to the south
endpoint is
\begin{equation}
 \hspace{-1.5em}
 T_S^{\rm s}=
 \sum_{k=-S+1}^{m_{+,S}}
 \frac{\sum_{r=k}^{S}w_{r,S}}
 {W_{k\to k-1}^{(S,1)}w_{k,S}},
~
 \sum_S q_S^{(1)}e^{-t_{50}^{(1)}/T_S^{\rm s}}=\frac12.
 \label{eq:app-log-mfpt-construction}
\end{equation}
The unknown normalization of $w_{m,S}$ cancels in every summand. We evaluate
the nested sums and the multisector half-survival equation in the log domain,
which prevents overflow when the MFPTs are exponentially large. The
\(\mathrm{dir}\) subscript marks the exact threshold time from direct
propagation.

For the second, reverse s-PFP-to-SSO switch of the paired protocol, every
postloading sector has integer $S$, and the deleted $0\to-1$ edge makes $m<0$
the transient side of the loading-induced boundary.  We use the exact MFPT
$T_S^0$ from the south endpoint to $m=0$.  Defining the successive MFPT
differences $D_m=T_m-T_{m+1}$ gives the stable recursion
\begin{equation}
\begin{aligned}
 D_{-S}&=\frac{1}{W_{-S\to-S+1}^{(S,2)}},\smjoinrow
 D_m&=\frac{1+W_{m\to m-1}^{(S,2)}D_{m-1}}
 {W_{m\to m+1}^{(S,2)}}\quad(-S<m<0),\smjoinrow
 T_S^0&=\sum_{m=-S}^{-1}D_m.
\end{aligned}
 \label{eq:app-reverse-loading-mfpt}
\end{equation}
The reverse-loading estimate is then fixed by
$\sum_S q_S^{(2)}e^{-t_{50}^{(2)}/T_S^0}=1/2$. Both loading directions are
evaluated in the log domain.

\paragraph{\textbf{Size range for panel (b) and paired direct spectral
validation.---}}
Both curves in Fig.~\ref{fig:parity-quench-collapse}(b) are obtained from the
MFPT calculation rather than direct large-matrix propagation, using every
even base size from $4$ to $40$ and a progressively coarser grid through
$N=1024$.
For each base $N$, we compute the exact maximal-sector steady state, apply the
first loading, and solve Eq.~\eqref{eq:app-log-mfpt-construction} for
$t_{50}^{(1)}$. The conditional first-stage s-PFP distributions then seed the
second loading and Eq.~\eqref{eq:app-reverse-loading-mfpt} gives
$t_{50}^{(2)}$.  Both times are plotted at the common base $N$, so each
blue--orange pair is one inherited $N\to N+1\to N+2$ protocol, ending with
$1024\to1025\to1026$.  The ordinate is
$\log_{10}(\Gamma_{\rm tot}t_{50}^{(\ell)})$.

For the paired validation in Fig.~\ref{fig:end-loading-t50-validation}, we
directly propagate the same inherited protocols for $N=4,6,\ldots,40$,
resetting only between base sizes.  In each occupied child sector, let
$\mathsf K_S=V_S\Lambda_SV_S^{-1}$ be the exact finite population generator
and let $\boldsymbol b_S$ indicate the s-PFP side: it selects
$\mathcal B_{\mathrm{s\text{-}PFP},1}^{(S)}$ for half-integer $S$ and
$\mathcal B_{\mathrm{s\text{-}PFP},2}^{(S)}$ for integer $S$.  The directly
propagated physical basin weight is
\begin{equation}
 P_{\mathrm{s\text{-}PFP}}^{(\ell)}(t_\ell)
 =\sum_{S\in\mathcal S_\ell}q_S^{(\ell)}\boldsymbol b_S^{\mathsf T}
 V_S e^{\Lambda_S t_\ell}V_S^{-1}\boldsymbol p_S^{(\ell)}(0^+),
 \label{eq:app-direct-spectral-t50}
\end{equation}
Its first crossing of $1/2$ defines $t_{50,\mathrm{dir}}^{(\ell)}$.

Equation~\eqref{eq:app-direct-spectral-t50} defines the directly propagated
curve $P_{\mathrm{s\text{-}PFP},\mathrm{dir}}^{(\ell)}(t_\ell)$.  The corresponding
MFPT reconstruction used for the two loading directions is
\begin{equation}
 P_{\mathrm{s\text{-}PFP},\mathrm{MFPT}}^{(\ell)}(t_\ell)
 =\begin{cases}
  1-\displaystyle\sum_{S\in\mathcal S_1}q_S^{(1)}e^{-t_1/T_S^{(1)}},&\ell=1,\\[2pt]
  \displaystyle\sum_{S\in\mathcal S_2}q_S^{(2)}e^{-t_2/T_S^{(2)}},&\ell=2.
 \end{cases}
 \label{eq:app-mfpt-basin-curves}
\end{equation}
Figure~\ref{fig:app-direct-mfpt-curves} compares these full curves for the
largest directly propagated paired protocol, $40\to41\to42$.

\begin{smfullfigure}[!htbp]
\centering
\includegraphics[width=0.8\columnwidth]{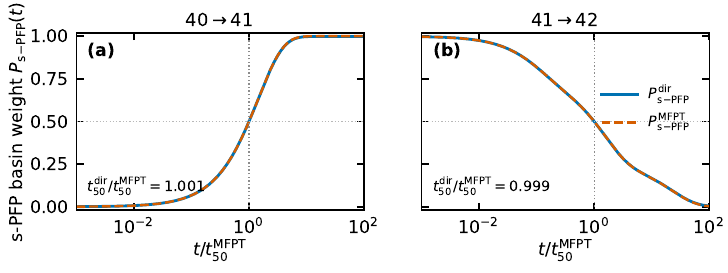}
\caption{\label{fig:app-direct-mfpt-curves}
Full-curve validation of the MFPT reconstruction at $\eta=0.90$ and even base
size $N_0=40$.  Solid blue curves are the directly propagated s-PFP basin
weights from Eq.~\eqref{eq:app-direct-spectral-t50}. Dashed orange curves are
Eq.~\eqref{eq:app-mfpt-basin-curves}.  Panels (a) and (b) show the
$40\to41$ growth and $41\to42$ decay, respectively.  Vertical and horizontal
dotted lines mark $t=t_{50}^{(\ell)}$ and basin weight $1/2$.}
\end{smfullfigure}

Over the displayed interval
\(10^{-3}\leq t/t_{50}^{(\ell)}\leq10^2\), the maximum absolute
direct--MFPT deviations are \(1.68\times10^{-3}\) and
\(1.85\times10^{-3}\) for the first and second loadings, respectively,
with
\(t_{50,\rm dir}^{(1)}/t_{50}^{(1)}=1.0007\) and
\(t_{50,\rm dir}^{(2)}/t_{50}^{(2)}=0.9992\).
Thus the reconstruction captures the full basin-weight transfer, rather than
only its equal-weight crossing. The remaining discrepancy is a finite-size
nonexponential correction.
This validation directly concerns the basin weights.  The Dicke-level and
Husimi-$Q$ reconstruction in Fig.~\ref{fig:parity-quench-collapse}(a)
additionally uses the metastable separation assumed in Sec.~VI.B: rapid
intrabasin relaxation establishes the conditional target profile before the
rare interbasin escape.

\begin{figure}[!htbp]
\centering
\includegraphics[width=0.4\columnwidth]{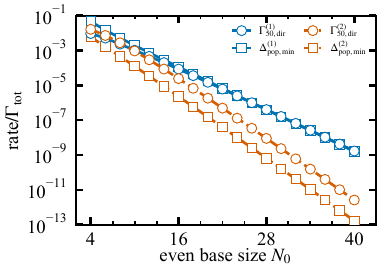}
\caption{\label{fig:app-rate-gap-validation}
Rate--gap comparison for the paired protocols of
Fig.~\ref{fig:end-loading-t50-validation}:
$\Gamma_{50,\mathrm{dir}}^{(\ell)}
=\ln2/t_{50,\mathrm{dir}}^{(\ell)}$ (circles) and
$\Delta_{{\rm pop},\min}^{(\ell)}=\min_S\Delta_{{\rm pop},S}$ (squares).  The first
loading is blue and solid. The second is orange and dashed.}
\end{figure}

The comparison in Fig.~\ref{fig:app-rate-gap-validation} uses
$\Delta_{{\rm pop},\min}^{(\ell)}=\min_S\Delta_{{\rm pop},S}$, with every sector
gap obtained from the blockwise Jacobi representation of
\hyperref[app:population-jacobi]{Sec.~V.B}. For an integer sector this
includes both the transient negative block and the recurrent nonnegative block.
$\Delta_{{\rm pop},S}$ is the smallest nonzero decay rate in their combined
spectrum.
Figure~\ref{fig:end-loading-t50-validation} compares the log-domain construction with
this direct finite-size propagation over their common resolved range.

Figure~\ref{fig:end-loading-t50-validation} shows that the sector-MFPT
construction tracks the directly propagated equal-weight time for both steps
throughout the resolved range. The residual first-step discrepancy at the
smallest sizes is the nonexponential correction removed as metastability
develops.  Figure~\ref{fig:app-rate-gap-validation} additionally shows that a
minimum gap and a median switching time need not select the same sector.

Let $N$ be the even base size and $S_2=(N+2)/2$ the maximal sector after the
second loading.  Applying the second unpolarized loading to the two
conditional first-stage s-PFP sectors gives the exact three-sector weights
$q_{S_2}^{(2)}=(N+3)/[4(N+1)]$, $q_{S_2-1}^{(2)}=1/2$, and
$q_{S_2-2}^{(2)}=(N-1)/[4(N+1)]$.
Across the displayed sequence, the maximal sector is the slowest one,
$\Delta_{\rm s}=\Delta_{{\rm pop},S_2}=\Delta_{{\rm pop},\min}^{(2)}$. Write
$\Delta_{\rm m}=\Delta_{{\rm pop},S_2-1}$ and
$\Delta_{\rm f}=\Delta_{{\rm pop},S_2-2}$ for the middle and fastest sectors.
After intrabasin mixing, the decaying s-PFP weight is
\begin{equation}
 P_{\mathrm{s\text{-}PFP}}^{(2)}(t_2)\simeq
 q_{S_2}^{(2)}a_{\rm s}e^{-\Delta_{\rm s}t_2}
 +\frac12a_{\rm m}e^{-\Delta_{\rm m}t_2}
 +q_{S_2-2}^{(2)}a_{\rm f}e^{-\Delta_{\rm f}t_2},
\label{eq:app-reverse-metastable-sector-mixture}
\end{equation}
where the $a_j$ are spectral-overlap amplitudes. In the scale-separated
metastable limit, rapid intrabasin mixing before escape gives
$a_j=1+o(1)$ for the leading survival mode. This is the same killed-generator
separation derived explicitly for the forward process in
Eqs.~\eqref{eq:app-loading-killed-spectrum}--
\eqref{eq:app-loading-survival-exponent}. The identical one-dimensional
argument applies to each reverse transient block. The intrabasin separation does not imply infinite separation between
the three sector escape times. Put $M=(N+2)/2$ and $a=Mz_\eta$.
On the negative half-ladder a common reversible-weight convention is
$w_{-n}=a^{-2n}\Gamma(n)^2$. Thus
$w_{-S}/w_{-(S-1)}=(S-1)^2/a^2\to z_\eta^{-2}$.
In the exact reverse MFPT sum, the accumulated mass is localized near the
south endpoint and the resistance near $m=-a+O(\sqrt N)$.
Adjacent sectors have asymptotically equal rates in the latter window
and equal limiting geometric normalization factors at their endpoints.
Consequently $T_{\rm s}/T_{\rm m}\to z_\eta^{-2}$ and
$T_{\rm m}/T_{\rm f}\to z_\eta^{-2}$. Using $T_j\Delta_j\to1$,
the limiting threshold equation for $x=\Delta_{\rm m}t_{50}^{(2)}$ is
$\tfrac14e^{-z_\eta^2x}+\tfrac12e^{-x}
+\tfrac14e^{-x/z_\eta^2}=\tfrac12$.
All three sectors therefore contribute to the asymptotic prefactor.
At $\eta=0.90$, the limiting values are $x\to0.6596$ and
$t_{50}^{(2)}\Delta_{\rm m}/\ln2\to0.9516$.
The slowest sector still controls the longest tail.
Across the displayed second-loading sequence, from preloading size $5$ to
$41$, $\Delta_{\rm m}/\Delta_{\rm s}$ grows from $3.95$ to $15.53$, while
$\Gamma_{50,\mathrm{dir}}^{(2)}/\Delta_{\rm s}$ grows from $2.78$ to $15.49$
whereas $\Gamma_{50,\mathrm{dir}}^{(2)}/\Delta_{\rm m}$ is close to unity over this finite range,
changing from $0.704$ to $0.997$.  Thus the widening orange separation is the expected distinction
between a global long-time gap and a finite-weight median, not a failure of
the MFPT-based calculation.  For the first loading, by contrast, the slowest
half-integer sector has weight
$1/2+1/[2(N+1)]$ and controls the threshold asymptotically, explaining the
 convergence of the blue curves.  These sector-selection differences affect
finite-size prefactors and the longest tail, but the proof below shows that all
sectors participating in a given loading share its leading exponential scale.

\paragraph{\textbf{Forward switching exponent.---}}
The leading forward exponent can be obtained analytically.  Fix
$\eta>\eta_c^{\mathrm{o}}$ and write
$S_1=N_1/2=(N_0+1)/2$ for the maximal spin after the first loading.  Its two
occupied sectors have $S=S_1-\delta_S$ with $\delta_S\in\{0,1\}$, so
$S/S_1=1+O(N_0^{-1})$.  The positive mode remains at
$m_{+,S}=S_1z_\eta+O(1)$, and the exact-MFPT bottleneck is at
$m_{\rm b}^{(S)}=-S_1z_\eta+O(1)$. The Stirling and reflection estimates of
\hyperref[app:parity-thermodynamic-limits]{Sec.~III} give
\begin{equation}
\begin{aligned}
 \ln\sum_{r=m_{\rm b}^{(S)}}^{S}w_{r,S}
 &=2S_1z_\eta+o(N_0),\smjoinrow
 \ln w_{m_{\rm b}^{(S)},S}&=-2S_1z_\eta+o(N_0),
\end{aligned}
 \label{eq:app-loading-mfpt-barrier-weights}
\end{equation}
whereas
$W_{m_{\rm b}^{(S)}\to m_{\rm b}^{(S)}-1}^{(S,1)}
=\Gamma_{\rm tot}e^{o(N_0)}$.
The bottleneck term in Eq.~\eqref{eq:app-log-mfpt-construction} therefore has
exponent $4S_1 z_\eta=2(N_0+1)z_\eta$, whose ratio to $N_0$ tends to
$2z_\eta$.  All other summands have no larger exponential order, and their
number is only $O(N_0)$, yielding
\begin{equation}
 T_S^{\rm s}=\Gamma_{\rm tot}^{-1}e^{2z_\eta N_0+o(N_0)},
 ~ S\in\{S_1,S_1-1\}.
 \label{eq:app-loading-mfpt-exponent}
\end{equation}
The endpoint used in $T_S^{\rm s}$ lies inside the s-PFP basin.
First arrival at the bottleneck is a different stopping time: a trajectory
there has a nonvanishing probability of returning to the SSO basin.
Writing $i=m_{+,S}$, $b=m_{\rm b}^{(S)}$, and $s=-S$, the strong Markov
property gives $T_{i\to s}=T_{i\to b}+T_{b\to s}$.
The exact MFPT sum for $T_{i\to b}$ retains the part of the resistance peak
on the SSO side of $b$. Its Gaussian width is $O(\sqrt{N_0})$, and the
one-site displacement of $b$ from the saddle is negligible on this scale.
Thus $T_{i\to b}/T_{i\to s}\to1/2$ and
$T_{b\to s}/T_{i\to s}\to1/2$.
Both stopping times have the exponent in
Eq.~\eqref{eq:app-loading-mfpt-exponent}; the unconditional time from the
bottleneck to the south endpoint is not subexponential.
To pass from the MFPT to a fixed quantile without assuming
an exact finite-size exponential law, let
$0<\kappa_{1,S}<\kappa_{2,S}\leq\cdots$ be the decay rates of the population
generator killed at the basin boundary.  Within the SSO basin the
large-deviation potential has a single minimum and no further extensive
barrier. After the escape edge is excluded, every resistance--mass product is
$e^{o(N_0)}$.  The one-dimensional Hardy bound in
Eq.~\eqref{eq:app-hardy-gap-bounds}, applied to this restricted chain, therefore
gives an intrabasin mixing time $e^{o(N_0)}$.
Together with
Eq.~\eqref{eq:app-loading-mfpt-exponent}, this gives the metastable spectral
separation
\begin{equation}
\hspace{-1.5em}
\begin{aligned}
\kappa_{1,S}&=\Gamma_{\rm tot}e^{-2z_\eta N_0+o(N_0)},\smjoinrow
\kappa_{2,S}^{-1}&=\Gamma_{\rm tot}^{-1}e^{o(N_0)},~
\frac{\kappa_{1,S}}{\kappa_{2,S}}\longrightarrow0.
\end{aligned}
\label{eq:app-loading-killed-spectrum}
\end{equation}
This hierarchy realizes the standard metastable separation between rapid
intrabasin mixing and exponentially slow escape
\cite{Macieszczak2016,Macieszczak2021,BrownMacieszczakJack2024}.
The spectral expansion of the killed finite Markov generator then yields,
uniformly after the intrabasin mixing time,
\begin{equation}
\begin{aligned}
 \Sigma_S(t)&=[1+o(1)]e^{-\kappa_{1,S}t}
 +O(e^{-\kappa_{2,S}t}).
\end{aligned}
 \label{eq:app-loading-survival-exponent}
\end{equation}
The leading amplitude is $1+o(1)$ because escape during the mixing stage has
exponentially small probability.

This separation fixes every nontrivial survival quantile on the exponential
scale.  Indeed, for arbitrary fixed $\epsilon>0$, define
$t_\pm=\Gamma_{\rm tot}^{-1}
\exp[(2z_\eta\pm\epsilon)N_0]$.  Equations
\eqref{eq:app-loading-killed-spectrum} and
\eqref{eq:app-loading-survival-exponent} give
$\Sigma_S(t_-)\to1$ and $\Sigma_S(t_+)\to0$.  Both occupied-sector weights
have nonzero $O(1)$ limits, and their number is finite, so the same limits hold
for the physical first-passage survival probability
$\Sigma^{(1)}(t)=\sum_S q_S^{(1)}\Sigma_S(t)$.  Therefore the time $t_r$ defined
by $\Sigma^{(1)}(t_r)=1-r$ for any fixed $0<r<1$ obeys
\(\Gamma_{\rm tot}t_r=e^{2z_\eta N_0+o(N_0)}\).
It remains to connect first passage to the instantaneous basin weight used in
the definition of $t_{50,\mathrm{dir}}^{(1)}$.
Choose a point a fixed macroscopic distance south of the bottleneck, inside
the attraction basin of the south endpoint. The one-dimensional splitting
probability, obtained by summing the edge resistances, gives a probability
$1/2+o(1)$ of reaching this point before returning to the SSO core after
first hitting the bottleneck. Failed attempts return to the SSO core;
successful transition paths take only subexponential time.
The resulting renewal of escape attempts changes the transfer rate by a
finite factor and preserves the exponent $2z_\eta$.
After relaxation inside the south basin, detailed balance and the stationary
mass ratio $B_{\rm SSO}^{(S)}/B_{\rm s\text{-}PFP}^{(S)}
=e^{-N_0\Delta g(\eta)+o(N_0)}$ imply that reverse basin-to-basin escape
is negligible on the forward switching scale. This statement concerns
equilibrated basin transfer, not the immediate recrossings at the saddle.
The physical basin-weight crossing and the first-boundary-passage median
therefore have the same exponential scale, although their ratio need not
tend to one.
Taking $r=1/2$ therefore proves for the exact
threshold time
\begin{equation}
 \Gamma_{\rm tot}t_{50,\mathrm{dir}}^{(1)}
 =e^{2z_\eta N_0+o(N_0)}.
 \label{eq:app-loading-t50-exponent}
\end{equation}
At the value $\eta=0.90$ used in Fig.~\ref{fig:parity-quench-collapse},
$2z_\eta=2/\sqrt{19}\simeq0.4588$.

\paragraph{\textbf{Reverse switching exponent.---}}
For fixed $\eta>\eta_c^{\mathrm{o}}$, define
$I_{\rm rev}(z)=-\ln z-1+z>0$. Let $S_2=(N_0+2)/2$ be the maximal spin after
the second loading.  Its three
occupied sectors can be written as $S=S_2-\delta_S$, with
$\delta_S\in\{0,1,2\}$, so their south endpoints differ by only finitely many
lattice spacings.  On the transient negative half-ladder, introduce reversible
weights $\widetilde w_{m,S}^{(2)}$ by
\begin{equation}
\hspace{-1.5em}
\begin{aligned}
\widetilde w_{-S,S}^{(2)}=1,\frac{\widetilde w_{m+1,S}^{(2)}}{\widetilde w_{m,S}^{(2)}}
=\left(\frac{S_2z_\eta}{m+1}\right)^2,~ -S\leq m<-1.
\end{aligned}
\label{eq:app-reverse-transient-weight-ratio}
\end{equation}
The ratio follows directly from the exact rates in
Eq.~\eqref{eq:app-loading-trajectory-rates}. The tilde distinguishes these
transient-chain weights from a stationary distribution, which has no support
at $m<0$ for integer $S$.  The backward recurrence in
Eq.~\eqref{eq:app-reverse-loading-mfpt} is equivalently
\begin{equation}
 T_S^0=\sum_{k=-S}^{-1}
 \frac{\displaystyle\sum_{r=-S}^{k}\widetilde w_{r,S}^{(2)}}
 {W_{k\to k+1}^{(S,2)}\widetilde w_{k,S}^{(2)}}.
 \label{eq:app-reverse-mfpt-weight-sum}
\end{equation}
This representation makes the reverse barrier explicit.  The weights decrease
from the south boundary to their minimum at
$m_{\rm b}^{(S)}=-S_2z_\eta+O(1)$ and then increase toward $m=0$. At the
minimum, Eq.~\eqref{eq:app-reverse-transient-weight-ratio} gives
\begin{equation}
\begin{aligned}
 \ln\frac{\widetilde w_{m_{\rm b}^{(S)},S}^{(2)}}
 {\widetilde w_{-S,S}^{(2)}}=2\sum_{j=-S+1}^{m_{\rm b}^{(S)}}
 \ln\frac{S_2z_\eta}{|j|}=2S_2\int_{z_\eta}^{1}\ln\frac{z_\eta}{x}\,dx+o(N_0)=-N_0I_{\rm rev}(z_\eta)+o(N_0),
\end{aligned}
\label{eq:app-reverse-barrier-action}
\end{equation}
where replacing $S/S_2$ by $1$ changes only the $o(N_0)$ term.  Near the
south endpoint the ratios are uniformly bounded above by a constant strictly smaller than unity, so
$\sum_{r=-S}^{m_{\rm b}^{(S)}}\widetilde w_{r,S}^{(2)}=e^{o(N_0)}$ under the
normalization in Eq.~\eqref{eq:app-reverse-transient-weight-ratio}.  Moreover,
$W_{m_{\rm b}^{(S)}\to m_{\rm b}^{(S)}+1}^{(S,2)}
=\Gamma_{\rm tot}e^{o(N_0)}$.  The bottleneck term in
Eq.~\eqref{eq:app-reverse-mfpt-weight-sum} therefore has exponent
$N_0I_{\rm rev}(z_\eta)$. No other term has a larger exponential order, and
there are only $O(N_0)$ terms.  Hence, uniformly for all three occupied
sectors,
\begin{equation}
 T_S^0=\Gamma_{\rm tot}^{-1}
 e^{N_0I_{\rm rev}(z_\eta)+o(N_0)},
 ~S\in\{S_2,S_2-1,S_2-2\}.
 \label{eq:app-reverse-mfpt-exponent}
\end{equation}

The first-stage conditional s-PFP distributions loaded into these sectors are
localized within a subextensive south-boundary layer.  Replacing a south-endpoint
initial condition by any such distribution changes
Eq.~\eqref{eq:app-reverse-mfpt-exponent} only by an $e^{o(N_0)}$ factor.  The
same one-dimensional Hardy bound used above gives subexponential mixing inside
the negative basin and a killed-spectrum separation with principal rate
$\Gamma_{\rm tot}e^{-N_0I_{\rm rev}(z_\eta)+o(N_0)}$.  Because the edge
$0\to-1$ is absent, first arrival at $m=0$ is irreversible and the surviving
negative-half-ladder probability is exactly the s-PFP weight.  Every fixed
survival quantile therefore has the exponent in
Eq.~\eqref{eq:app-reverse-mfpt-exponent}.

Finally, all three sector weights displayed above have nonzero limits.  A finite
mixture of their survival laws consequently retains the common exponent,
irrespective of which sector fixes its subexponential prefactor. Thus
\begin{equation}
  \hspace{-1em}
 \Gamma_{\rm tot}t_{50,\mathrm{dir}}^{(2)}
 =e^{N_0I_{\rm rev}(z_\eta)+o(N_0)},~
 \ln\frac{t_{50,\mathrm{dir}}^{(2)}}{t_{50}^{(2)}}=o(N_0).
 \label{eq:app-reverse-loading-t50-exponent}
\end{equation}
All three sectors contribute to the asymptotic equal-weight prefactor
described above, while sharing the same large-deviation exponent.  At $\eta=0.90$,
$I_{\rm rev}(z_\eta)=-\ln(1/\sqrt{19})-1+1/\sqrt{19}
\simeq0.7016$.

\subsubsection{\textbf{VI.D. Exact sector-resolved quantum-jump trajectories}}
\label{app:sector-resolved-stochastic-dynamics}
The autonomous sector population blocks proved in
\hyperref[app:intersector-coherence-closure]{Sec.~VI.A} admit an exact
continuous-time Markov-chain unraveling, generated with the Gillespie
algorithm \cite{Gillespie1977}. Its first-passage observables resolve
metastable switching \cite{BrownMacieszczakJack2024,Menczel2026}.

At loading step $\ell$, a trajectory is initialized in two stages.  First a
sector $S\in\mathcal S_\ell$ is sampled with probability $q_S^{(\ell)}$, and
then its initial Dicke level is sampled from $p_{m,S}^{(\ell)}(0)$. For fixed $S$ and $m$,
Eq.~\eqref{eq:app-loading-trajectory-rates} supplies the allowed jumps, with
missing boundary rates set to zero.

For the $10\to11$ trajectories in
Fig.~\ref{fig:parity-quench-collapse}(c), both $S=11/2$ and $9/2$ sectors have
$m_{\rm b}^{(S)}=-3/2$.  Hence the sector-dependent partition in
Eq.~\eqref{eq:app-loading-basin-partition} gives the common plotted boundary
$m_z^{\rm b}=-3/11$, with no sector average.  The first visit to
$m_{\rm b}^{(S)}$ is the first-passage event marked in panel (c). It differs
from the south-endpoint MFPT of
\hyperref[app:loading-switching-time]{Sec.~VI.C} but shares its asymptotic
exponent.

For a current level $m$, define
$R_m=W_{m\to m+1}^{(S,\ell)}+W_{m\to m-1}^{(S,\ell)}$.  The Gillespie update
is: (i) draw independent $u_1,u_2\in(0,1)$; (ii) advance time by
$\delta t=-\ln u_1/R_m$; and (iii) choose $m\to m+1$ when
$u_2<W_{m\to m+1}^{(S,\ell)}/R_m$, otherwise choose $m\to m-1$.  Iterating
these steps gives a piecewise-constant trajectory with
$m_z(t)=2m(t)/N_\ell$.

Panel (c) displays eight trajectories selected uniformly by first-passage-time
rank from 64 realizations.  All 64 cross before
$t\Gamma_{\rm tot}=5000$. Color-matched markers record first entry, while the
curves retain subsequent relaxation and recrossings.

Trajectory averaging recovers the deterministic multisector dynamics without
sector postselection:
\begin{equation}
\begin{aligned}
\Pr[S,m(t_\ell)=m]&=q_S^{(\ell)}p_{m,S}^{(\ell)}(t_\ell),~P_\alpha^{(\ell)}(t_\ell)&=\sum_Sq_S^{(\ell)}
\mathbb E_S\!\left[\mathbf1_{\mathcal B_{\alpha,\ell}^{(S)}}(m_{t_\ell})\right],~\frac{2\langle\hat S_z(t_\ell)\rangle_\ell}{N_\ell}
=\sum_Sq_S^{(\ell)}\frac{2\mathbb E_S[m_{t_\ell}]}{N_\ell}.
\end{aligned}
\label{eq:app-trajectory-ensemble-averages}
\end{equation}

\subsubsection{\textbf{VI.E. One-spin removal and the reverse parity switch}}
\label{app:spin-removal-switch}

\paragraph{\textbf{Exact trace-out map.---}}
We now consider the complementary physical operation in which one spin is
removed without resolving or postselecting its internal state.  Let the
pre-removal state lie in the maximal representation $S=N/2$ and be Dicke
diagonal,
$\hat\rho_N=\sum_{m=-S}^{S}p_m^{(N)}|S,m\rangle\langle S,m|$.
Permutation symmetry makes the choice of the removed constituent irrelevant,
and the instantaneous post-removal state is
$\hat\rho_{N-1}(0^+)=\operatorname{Tr}_{j}\hat\rho_N$.
The Clebsch--Gordan decomposition with respect to that constituent is
\begin{equation}
\begin{aligned}
 |S,m\rangle={}&
 \sqrt{\frac{S+m}{2S}}
 \left|S-\frac12,m-\frac12\right\rangle|\!\uparrow\rangle
 +\sqrt{\frac{S-m}{2S}}
 \left|S-\frac12,m+\frac12\right\rangle|\!\downarrow\rangle .
\end{aligned}
\label{eq:app-removal-cg}
\end{equation}
Tracing the last factor removes the cross terms and leaves the squared
Clebsch--Gordan amplitudes.  With
$S'=S-\tfrac12=(N-1)/2$ and
$\mu=-S',\ldots,S'$, the exact population map is therefore
\begin{equation}
 p_{\mu}^{(-)}(0^+)
 =\frac{S+\mu+\tfrac12}{2S}p_{\mu+1/2}^{(N)}
 +\frac{S-\mu+\tfrac12}{2S}p_{\mu-1/2}^{(N)}.
\label{eq:app-removal-population-map}
\end{equation}
Equation~\eqref{eq:app-removal-population-map} is normalized and, unlike an
independent-spin loading, occupies only the maximal $S'$ representation:
tracing one constituent from a fully symmetric state leaves the remaining
$N-1$ constituents fully symmetric.  The removal map is thus not the inverse
of the loading map, but it requires no multisector average in the setting
considered here.

\paragraph{\textbf{Instantaneous representation-parity quench.---}}
Equation~\eqref{eq:app-removal-population-map} gives
\begin{equation}
 \langle\hat S_z'\rangle_{0^+}
 =\frac{N-1}{N}\langle\hat S_z\rangle_{0^-},
 \qquad
 \frac{2\langle\hat S_z'\rangle_{0^+}}{N-1}
 =\frac{2\langle\hat S_z\rangle_{0^-}}{N}.
\label{eq:app-removal-polarization-invariance}
\end{equation}
Hence the normalized polarization is exactly unchanged at the removal event.
What changes discontinuously is the representation lattice and the generator
governing the subsequent evolution.  Removing one spin from an odd system
creates an integer Dicke ladder containing $\mu=0$, deletes the
$0\to-1$ nonlinear-loss edge, and makes the negative half-ladder transient.
Removing one spin from an even system instead produces a half-integer ladder,
eliminates the sampled zero, and reconnects the SSO and south-boundary
regions.  The macroscopic switch therefore occurs during relaxation under
$\mathcal L_{N-1}$, not at $t=0^+$.

\paragraph{\textbf{Reverse switching dynamics and time scales.---}}
After removal we propagate Eq.~\eqref{eq:app-removal-population-map} with the
exact maximal-$S'$ population generator obtained from
Eq.~\eqref{eq:app-loading-trajectory-rates} by setting the physical size to
$N-1$.  The basin partitions, MFPT reconstruction, and direct
quantum-jump construction are then the single-sector specializations of
Secs.~VI.B--VI.D.  If $N_{\mathrm{r}}$ denotes the remaining particle number,
their leading switching scales, with $I_{\rm rev}(z)=-\ln z-1+z$, are
\begin{equation}
 \Gamma_{\rm tot}t_{50}^{\mathrm{o\to e}}
 =e^{N_{\rm r}I_{\rm rev}(z_\eta)+o(N_{\rm r})},
 \qquad
 \Gamma_{\rm tot}t_{50}^{\mathrm{e\to o}}
 =e^{2z_\eta N_{\rm r}+o(N_{\rm r})}.
\label{eq:app-removal-switching-exponents}
\end{equation}
Thus the same forward and reverse activation exponents found for loading
reappear in the opposite temporal order under consecutive removals.

At $\eta=0.90$, the exact steady populations give
\begin{equation}
\begin{aligned}
 1025\to1024:\quad
 -0.9999&=m_z(0^+)\longrightarrow
 m_z^{\rm ss}=0.2289,\\
 1024\to1023:\quad
 0.2289&=m_z(0^+)\longrightarrow
 m_z^{\rm ss}=-0.9999.
\end{aligned}
\label{eq:app-removal-large-size-polarization}
\end{equation}
The corresponding log-domain MFPT reconstruction yields
$\log_{10}(\Gamma_{\rm tot}t_{50}^{(1)})=311.153$ and
$\log_{10}(\Gamma_{\rm tot}t_{50}^{(2)})=204.272$.
Direct finite-size propagation gives
$\Gamma_{\rm tot}t_{50}^{11\to10}=1.2136\times10^3$ and
$\Gamma_{\rm tot}t_{50}^{10\to9}=1.7172\times10^2$.

\begin{smfullfigure}[!htbp]
\centering
\includegraphics[width=0.6\columnwidth]{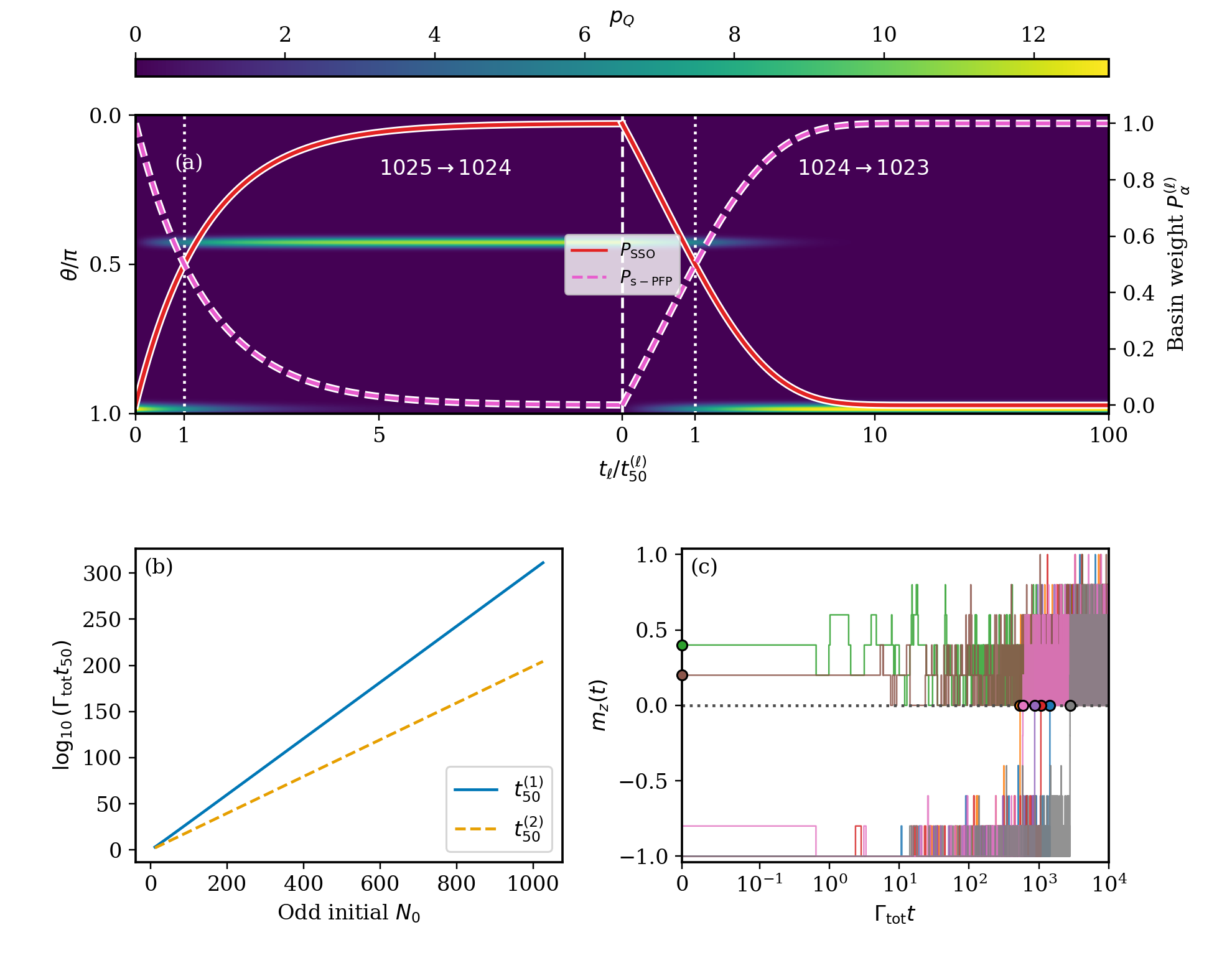}
\caption{\label{fig:app-spin-removal-switch}
Reverse parity switching under state-unresolved one-spin removal at
\(\eta=0.90\).  (a) MFPT-based metastable reconstruction for
\(1025\to1024\to1023\).  The background is the polar Husimi-$Q$ density, with
solid red and dashed magenta curves showing the SSO and s-PFP basin weights.  The
white dashed line separates the two stage-local time axes, and white dotted
lines mark \(t_\ell=t_{50}^{(\ell)}\).  (b) Log-domain MFPT switching times
versus the odd initial size \(N_0\): blue solid, the first
s-PFP-to-SSO removal and orange dashed for the second SSO-to-s-PFP removal.
(c) Eight independently sampled exact quantum-jump trajectories for
\(11\to10\), generated with random seed 20260907. Circles mark first entry
into the nonnegative half-ladder (at \(t=0\) for samples initially there);
entry from the negative half-ladder occurs at \(m=0\). The deleted
\(0\to-1\) edge prevents subsequent return to the negative half-ladder.}
\end{smfullfigure}

This construction proves the removal protocol only for a pre-removal state in
the maximal-spin representation.  Removing a constituent from a general
lower-spin or multisector state can populate multiple descendant
representations and requires the corresponding Clebsch--Gordan and
multiplicity bookkeeping. No such broader claim is made here.

\subsection{VII. Generalized nonlinear jumps, zero-preserving robustness, and edge restoration}
\label{app:general-mechanism}

\subsubsection{\textbf{VII.A. Ordered jumps and lattice--zero matching}}
\label{app:ordered-jump-lattice-zero}

The graph mechanism extends beyond $\hat S_-\hat S_z$ to
\begin{equation}
 \hat L_{\sigma,F}^{(q)}
 =c_{\sigma,q}(S)\hat S_\sigma^{\,q}F(\hat S_z/S),
 ~ \sigma=\pm1,
 \label{eq:app-general-ordered-jump}
\end{equation}
where the function on the right acts first and $c_{\sigma,q}(S)$ fixes the
desired large-$S$ normalization. For $q=1$, its action on a Dicke state gives
\begin{equation}
\begin{aligned}
 W_{m\to m+\sigma}^{(\sigma,F)}=|c_{\sigma,1}(S)|^2
 \left|F\!\left(\frac mS\right)\right|^2
 \smjoincontinuation (S-\sigma m)(S+\sigma m+1).
\end{aligned}
\label{eq:app-general-ordered-rate}
\end{equation}
Thus a zero $F(m_0/S)=0$ deletes the directed edge
$m_0\to m_0+\sigma$ only when $m_0$ belongs to the finite representation
lattice. The operator ordering matters: placing $F(\hat S_z/S)$ on the left
would evaluate it at the arrival site and shift the deleted edge.

Deleting an edge in one jump channel is not by itself a statement about the
full Liouvillian.  Within the population-closed class considered here, the
deleted bond changes recurrent support only if no other active jump and no
coherent matrix element reconnects the two sides.  For example, a transverse
term proportional to $\hat S_x=(\hat S_++\hat S_-)/2$ couples adjacent Dicke
levels and bypasses an isolated zero of the nonlinear loss.  Outside a closed
population description, one must therefore verify the invariant and recurrent
subspaces of the complete generator, rather than infer them from a zero of a
single $\hat L_\mu$.

\subsubsection{\textbf{VII.B. Multistep jumps and \texorpdfstring{$N\bmod q$}{N mod q}}}
\label{app:multistep-jumps}

For a genuine \(q\)-step jump in a spin-$S$ representation, introduce
\(j=m+S\in\{0,1,\ldots,2S\}\).
In the maximal representation $S=N/2$, its endpoint $2S=N$ converts the
representation-lattice dependence into a particle-number dependence.
The ordered channel in Eq.~\eqref{eq:app-general-ordered-jump} changes \(j\)
by \(\sigma q\). Consequently it preserves \(r=j\bmod q\), and the Dicke
graph splits into
\begin{equation}
\begin{aligned}
\mathcal C_r^{(S)}
&=\{j=r+kq:0\leq r+kq\leq 2S\}
\smjoinrow
&r=0,1,\ldots,q-1.
\end{aligned}
\label{eq:app-qstep-residue-classes}
\end{equation}
The endpoint $2S$ fixes each class size and produces a possible
$2S\bmod q$ dependence.  A zero $F(x_0)=0$ deletes a bond only when
$m_0=Sx_0$ lies on the Dicke lattice, in the class
$r_0=(m_0+S)\bmod q$.  Different subsequences can therefore sample the zero in
different classes, or sample versus miss it, producing distinct recurrent
graphs. The endpoint dependence is governed by $N\bmod q$, but zero matching
also requires $j_0=N(1+x_0)/2\in\mathbb Z$ and $r_0=j_0\bmod q$.
For example, $q=2$ and $x_0=0$ distinguish $N=4k$ from $N=4k+2$ by
the class containing the zero, even though both have $N\bmod2=0$.
The combined subsequence classification need not be $N\bmod q$.

With active step sizes $q_1,q_2,\ldots$, only $j\bmod d$ remains conserved,
where $d=\gcd(q_1,q_2,\ldots)$. Any one-step channel gives $d=1$.  Exact
residue classes therefore require every channel to preserve them.  This is
only the kinematic ingredient: a dissipative phase distinction further
requires the deleted bonds to reorganize recurrent support and the surviving
regions to compete exponentially.

More explicitly, let $\mathcal A_\nu$ be the macroscopic branches admitted by
representation subsequence $\nu$ within one specified irreducible recurrent
class, and let $g_i$ be their stationary rate functions in a common
normalization convention. When the minimum is unique, the selected branch is
\begin{equation}
 i_\nu^*=\operatorname*{arg\,min}_{i\in\mathcal A_\nu}g_i.
 \label{eq:app-general-selected-branch}
\end{equation}
This minimization cannot by itself compare disconnected recurrent classes.
Their stationary mixture is $p_\infty=\sum_r\alpha_r\pi^{(r)}$, with
$\alpha_r$ fixed by the initial state and absorption probabilities.
Each class has an independent normalization; cross-class selection requires
specifying these weights, including any exponential dependence on size.
Different recurrent supports, $\mathcal A_\nu\ne\mathcal A_{\nu'}$, establish
only different branch eligibility.  They produce distinct thermodynamic
phases only when $i_\nu^*\ne i_{\nu'}^*$ over a finite parameter interval.  In
the model of the Letter, the jump zero removes the s-PFP branch from the even
stationary support, while the SSO--s-PFP exchange of exponential weights makes
that additional branch dominant only along the odd subsequence.  The three
logically separate ingredients are therefore common macroscopic candidates,
representation-dependent accessibility, and a subsequence-specific global
branch exchange.

\begin{smfullfigure}[!t]
\centering
\includegraphics[width=0.7\columnwidth]{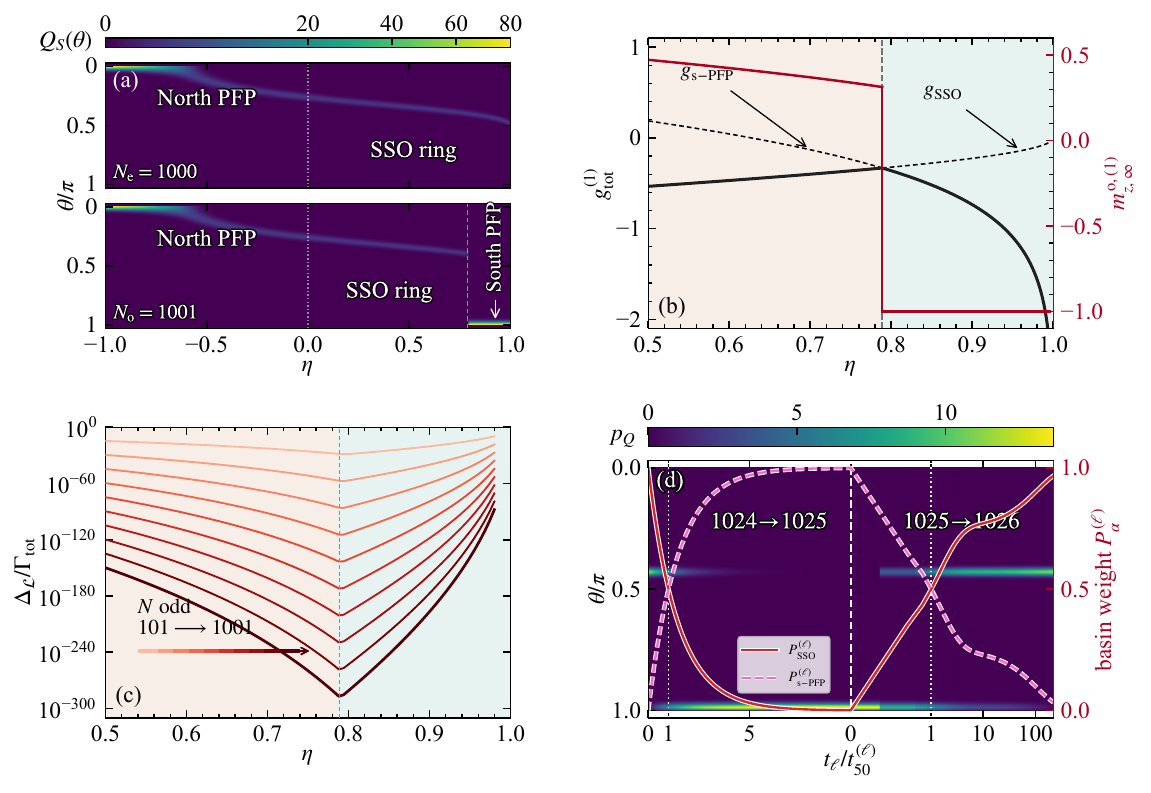}
\caption{\label{fig:app-generalized-loss-robustness}
Zero-preserving robustness under $F_{\chi=1}(z)=z(1+z^2)$.
(a) Polar Husimi-$Q$ densities for $N=1000$ (top) and $1001$ (bottom), with gray
dotted/dashed lines marking $\eta=0$/$\eta_c^{\mathrm{o},(1)}$.
(b) Branch rates (black dashed), $g_{\rm tot}^{(1)}$ (black solid), and
$m_{z,\infty}^{\mathrm{o},(1)}$ (red).  (c) Exact full gaps for odd
$N=101,201,\ldots,1001$ (light to dark), with the gray dashed line marking
$\eta_c^{\mathrm{o},(1)}$.  (d) Unpolarized $1024\to1025\to1026$ loading at
$\eta=0.9$: heat map for the polar Husimi-$Q$ density, solid/dashed curves for SSO/s-PFP
weights, white dotted lines for $t_{50}^{(\ell)}$, and a white dashed line for the loading
boundary.}
\end{smfullfigure}

\subsubsection{\textbf{VII.C. Zero-preserving deformation of the nonlinear loss}}
\label{app:deformed-loss}

To separate this graph mechanism from the special Bessel--Gamma form of the
\hyperref[eq:unified-population]{main-text solution}, retain the linear gain
of \hyperref[eq:lindblad]{the main text} and consider
\begin{equation}
\begin{aligned}
 F_\chi(z)=z(1+\chi z^2),~ \hat L_-^{(\chi)} =\sqrt{\frac{\Gamma_{\rm tot}(1+\eta)}{2S}}\hat S_-F_\chi(\hat S_z/S),
\end{aligned}
 \label{eq:app-deformed-loss}
\end{equation}
Equation~\eqref{eq:app-deformed-loss} defines the static maximal-representation
family, for which $S=N/2$. In the multisector loading of
Fig.~\ref{fig:app-generalized-loss-robustness}(d), the physical value
$S_{{\rm phys},\ell}=N_\ell/2$ instead fixes the Kac normalization and the
intensive argument of the nonlinear function in every occupied spin-$S$ block:
\begin{equation}
 \hat L_{-,S}^{(\chi,\ell)}
 =\sqrt{\frac{\Gamma_{\rm tot}(1+\eta)}{2S_{{\rm phys},\ell}}}\,
 \hat S_-^{(S)}F_\chi\!\left(\frac{\hat S_z^{(S)}}{S_{{\rm phys},\ell}}\right).
 \label{eq:app-deformed-loading-loss}
\end{equation}
The linear gain is normalized by the same $S_{{\rm phys},\ell}$, while the
sector value $S$ fixes only the Dicke ladder and its matrix elements, as in
\hyperref[app:finite-uniqueness-proof]{Sec.~I.C}.
For $\chi>-1$, this deformation retains
$z=0$ as its only zero on the physical interval $z\in[-1,1]$ while changing
every nonzero loss rate. Below we use $\chi=1$. Zero current gives
\begin{equation}
 \frac{p_{m+1}}{p_m}
 =\frac{z_\eta^2}{
 \left[\dfrac{m+1}{S}
 \left(1+\chi\dfrac{(m+1)^2}{S^2}\right)\right]^2}.
 \label{eq:app-deformed-recurrence}
\end{equation}
For integer $S$, $F_\chi(0)=0$ again makes $m<0$ transient and leaves
$m\geq0$ as the unique closed class. For half-integer $S$, the zero is missed
and the entire ladder remains irreducible. The ratio is used only on bonds with nonzero stationary populations;
at the deleted edge one must retain the cross-multiplied zero-current identity.
Equation
\eqref{eq:app-deformed-recurrence} is an exact finite product but, for
$\chi\ne0$, no longer has the truncated Bessel form of the main model.

Writing $m=Sz$, the large-deviation slope is
\begin{equation}
 A_\chi(z)
 =2\ln\!\frac{z_\eta}{|z|(1+\chi z^2)},
 ~
 z_*(1+\chi z_*^2)=z_\eta.
 \label{eq:app-deformed-rate-slope}
\end{equation}
The odd-sequence rate-coexistence point is determined by equality of the south
boundary and positive interior actions,
\begin{equation}
 \int_{-1}^{z_*}A_\chi(z)\,\mathrm{d}z=0.
 \label{eq:app-deformed-coexistence}
\end{equation}
For the finite deformation $\chi=1$, this gives
$z_{*,c}^{(1)}\simeq0.3129$, $z_{\eta,c}^{(1)}\simeq0.3436$, and
$\eta_c^{\mathrm{o},(1)}\simeq0.7888$.
At coexistence, the leading switching exponent is
\begin{equation}
 I_{\rm sw}^{(1)}
 =2\int_0^{z_{*,c}^{(1)}}
 \ln\!\frac{z_{\eta,c}^{(1)}}{z(1+z^2)}\,\mathrm{d}z
 \simeq0.6645,
 \label{eq:app-deformed-switching-exponent}
\end{equation}
so both basin-to-basin MFPTs and the population gap satisfy
$T_{\rm MFPT}=\Gamma_{\rm tot}^{-1}e^{NI_{\rm sw}^{(1)}+o(N)}$ and
$\Delta_{\rm pop}=\Gamma_{\rm tot}e^{-NI_{\rm sw}^{(1)}+o(N)}$.
Figure~\ref{fig:app-generalized-loss-robustness}(a,b) uses
Eq.~\eqref{eq:app-deformed-recurrence} at $N=1000,1001$ and
Eqs.~\eqref{eq:app-deformed-rate-slope}--
\eqref{eq:app-deformed-coexistence}, respectively.  Panel (c) evaluates
Eq.~\eqref{eq:end-block-gap-minimum} over all coherence blocks at
$N=101,201,\ldots,1001$ and 81 uniform $\eta\in[0.50,0.98]$, using the
\hyperref[app:gap-sturm-computation]{Sec.~V.B} solver for $\nu=0$.
All 810 minima are $\nu=0$ (blocks validated at small $N$).  Panel (d) uses
\hyperref[app:cg-multisector-dynamics]{Secs.~VI.A} and
\hyperref[app:loading-switching-time]{VI.C} and, at $N_0=1024$, the
\hyperref[app:loading-large-size-reconstruction]{Sec.~VI.B} MFPT
reconstruction (not direct propagation).

\begin{smfullfigure}[!t]
\centering
\includegraphics[width=0.8\columnwidth]{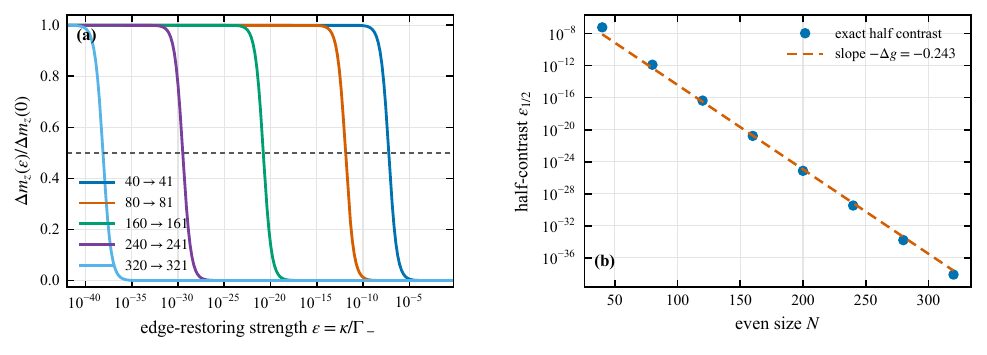}
\caption{\label{fig:app-zero-lifting-crossover}
Edge-restoration crossover at \(\eta=0.90\).
(a) Exact adjacent-size polarization contrast for even-\(N\) and odd-\(N+1\),
normalized by its value at the exact zero, versus
\(\epsilon=\kappa/\Gamma_-\).  The gray dashed line marks half contrast.
(b) Exact half-contrast scale \(\epsilon_{1/2}\) (symbols). The dashed guide
has slope \(-\Delta g(0.90)=-0.243\), as predicted by
Eq.~\eqref{eq:app-edge-restoring-crossover}.}
\end{smfullfigure}

\subsubsection{\textbf{VII.D. Restoring the deleted edge: finite-size crossover}}
\label{app:edge-restoration-crossover}

The robustness in \hyperref[app:deformed-loss]{Sec.~VII.C} is conditional on preserving
the sampled zero.  To restore the missing edge without shifting the nonlinear
zero, we add a weak Kac-normalized collective loss to the maximal-spin model,
\begin{equation}
 \dot{\hat\rho}=\mathcal L_0[\hat\rho]
 +\mathcal D[\hat L_{\rm rec}]\hat\rho,
 \qquad
 \hat L_{\rm rec}=\sqrt{\frac{2\kappa}{N}}\,\hat S_-,
 \label{eq:app-edge-restoring-channel}
\end{equation}
where \(\mathcal L_0\) is the Liouvillian of the Letter and \(S=N/2\).
Writing \(\epsilon=\kappa/\Gamma_-\), the added channel changes the downward
rate but leaves the upward rate unchanged.  Cancellation of the common Dicke
matrix element across each bond gives the exact stationary recurrence
\begin{equation}
 \frac{p_{m+1}^{(\epsilon)}}{p_m^{(\epsilon)}}
 =\frac{\Gamma_+}
 {\kappa+\Gamma_-[(m+1)/S]^2}
 =\frac{z_\eta^2}{\epsilon+[(m+1)/S]^2}.
 \label{eq:app-edge-restoring-recurrence}
\end{equation}
In particular, the formerly deleted bond now satisfies
\begin{equation}
 \frac{p_0^{(\epsilon)}}{p_{-1}^{(\epsilon)}}
 =\frac{z_\eta^2}{\epsilon}=\frac{\Gamma_+}{\kappa}.
 \label{eq:app-restored-edge-ratio}
\end{equation}
Thus every \(\epsilon>0\) reconnects the integer-spin ladder.  The exact zero
is an algebraic selection rule of the engineered \(\hat S_z\) factor, not a
generic symmetry protection against additional loss channels.

The crossover follows directly from the branch weights.  For
\(\eta>\eta_c^{\mathrm{o}}\), define the positive rate difference
\begin{equation}
 \Delta g(\eta)=g_{\rm SSO}-g_{\rm s\text{-}PFP}
 =-\bigl[1+z_\eta+\ln z_\eta\bigr]>0.
 \label{eq:app-edge-restoring-rate-difference}
\end{equation}
In the joint crossover regime $\epsilon\to0$ with
$-\ln\epsilon/N\to c\in(0,\infty)$, the restored bond supplies
the factor \(\epsilon\), whereas the two macroscopic branch shapes retain
their unperturbed leading actions away from the bond.  Consequently,
\begin{equation}
 \ln\frac{\mathcal Z_{\rm s\text{-}PFP}^{\rm e}(\epsilon)}{\mathcal Z_{\rm SSO}^{\rm e}(\epsilon)}=\ln\epsilon+N\Delta g(\eta)+o(N),\qquad \ln\epsilon_c(N)=-N\Delta g(\eta)+o(N).
\label{eq:app-edge-restoring-crossover}
\end{equation}
For any fixed \(\epsilon>0\), the denominator in
Eq.~\eqref{eq:app-edge-restoring-recurrence} is smooth on the intensive Dicke
lattice.  The half-spacing difference between the even and odd lattices then
vanishes as \(N\to\infty\), so both parities approach the same perturbed rate
function and the strict stationary parity splitting disappears.  Finite
systems nevertheless retain the exact-zero contrast when
\(\epsilon\ll\epsilon_c(N)\). Dynamically, the restored edge converts the
strict selection rule into a long-lived parity-sensitive metastable regime.

An observation must also preserve particle-number parity over the relevant
stage. If each of the \(N\) particles is lost independently at rate
\(\gamma_1\), the no-loss probability over \(t_{\rm obs}\) is
\(e^{-N\gamma_1t_{\rm obs}}\). A parity-resolved loading switch therefore
requires \(t_{50}\lesssim t_{\rm obs}\ll(N\gamma_1)^{-1}\), in addition to
the edge-restoration tolerance above.

Figure~\ref{fig:app-zero-lifting-crossover} confirms this scaling from the
exact stationary populations at \(\eta=0.90\).  The normalized adjacent-size
polarization contrast crosses over at progressively smaller \(\epsilon\) as
\(N\) increases [panel (a)], while the extracted half-contrast scale follows
the exponential size dependence predicted by
Eq.~\eqref{eq:app-edge-restoring-crossover} [panel (b)].

\FloatBarrier

\end{document}